\documentclass[useAMS,usenatbib]{mn2e}

\usepackage{graphicx}
\usepackage{amssymb}
\usepackage{subfig}             
\usepackage{amsmath}
\usepackage{wasysym}            
\usepackage{color}

\title[Variation of Area-to-Mass-Ratio of HAMR Space Debris Objects]{Variation
  of Area-to-Mass-Ratio of HAMR Space Debris Objects}
\author[C. Fr{\"u}h and T. Schildknecht]{C. Fr{\"u}h$^{1}$\thanks{E-mail:
frueh@aiub.unibe.ch (CF); thomas.schildknecht@aiub.unibe.ch
(TS)}\thanks{corresponding author} and T. Schildknecht$^{1}$\\
$^{1}$Astronomical Institute, University of Bern, Sidlerstrasse 5, 3012 Bern, Switzerland}
\begin{document}

\date{Accepted 2011 October 12. Received 2011 October 11; in original form
  2011 August 5, the definitve version is available at www.blackwell-synergy.com}

\pagerange{\pageref{firstpage}--\pageref{lastpage}} \pubyear{2011}

\maketitle

\label{firstpage}

\begin{abstract}
An unexpected space debris population has been detected in 2004
\citet{Schildknecht03,Schildknecht04a} with the unique properties of a very
high area-to-mass ratio (HAMR) \citet{Schildknecht05}. Ever since it has been
tried to investigate the dynamical properties of those objects further. 
The orbits of those objects are
heavily perturbed by the effect of direct radiation pressure. Unknown attitude
motion complicates orbit prediction. The area-to-mass ratio of the
objects seems to be not stable over time. Only sparse optical
data is available for those objects in drift orbits. \\
\\
The current work uses optical observations of five HAMR objects, observed over several years
and investigates the variation of their area-to-mass ratio and orbital parameters. A normalized orbit determination setup has been established and validated
with two low and two of the high ratio objects, to ensure, that comparable orbits over
longer time spans are determined even with sparse optical data. 
\end{abstract}


\section{Introduction}
The Astronomical Institute of the University of Bern (AIUB) detected high
area-to-mass ratio (HAMR) objects in GEO-like orbits in 2004 \citet{Schildknecht03,Schildknecht04a,Schildknecht05}. Since then, the
AIUB observes HAMR objects on a regular basis and keeps a small catalog of
HAMR and other space debris objects, which are not listed in the USSTRATCOM
catalog. The observations are performed with the one meter ESA Space Debris
Telescope (ESASDT), located on Tenerife, Spain, and the one meter Zimmerwald
Laser and Astrometry Telescope (ZIMLAT), located in Zimmerwald,
Switzerland. Additional observations for some objects, which were detected by
the AIUB, are provided by courtesy
of the Keldysh Institute of Applied Mathematics, Moscow, via the ISON network.\\
\\
Maintaining a catalogue of HAMR objects is especially challenging due to the
unique properties of these objects; the orbits are highly perturbed by direct
radiation pressure. Regular observations on
short time intervals are mandatory. In routine orbit determination for
catalogue maintenance, variations in the value of the effective area-to-mass ratio (AMR)
were detected, first investigations were performed \citet*{retoIAC08}.\\
\\
For the investigations presented in this paper orbits are determined with an enhanced version of the CelMech tool \citet*{Beutler05}. The
area-to-mass (AMR) value is determined as a scaling parameter of the direct
radiation pressure. The acceleration due to the direct radiation pressure is calculated as:
\begin{eqnarray}
\vec a_{\text{rad}}=\frac{C}{2}\cdot\frac{S}{c}\cdot\frac{AU^2}{|\vec r - \vec r_{\astrosun}|^2}\cdot\frac{A}{m}\cdot\frac{\vec r - \vec r_{\astrosun}}{|\vec r - \vec r_{\astrosun}|},
\end{eqnarray}
where $\vec r$ is the geocentric position of the satellite, $\vec r_{\astrosun}$ 
the geocentric position vector of the sun, $AU$ the astronomical unit, $A$ the effective cross section exposed to the radiation, $m$ the mass of
the satellite, and $c$ the speed of light. $C$ is the reflection coefficient.
The direct radiation pressure is determined under the assumption of a
spherically shaped object. In contrast to the calculation of the radiation
pressure acceleration by other sources (compare e.g. \citet*{vallado}),
the coefficient $C$ is divided by two in the formula above. A value for $C$
has to be chosen, by default, 2.0 is selected in the standard processing. This
corresponds to an assumption of full absorption. All AMR values presented in
this paper have to be interpreted as the effective area-to-mass ratio scaled
with $C/2=1$; the AMR values of other sources may be scaled with a different factor. It is assumed that the AMR is
constant over the orbital fit interval. A default value of 0.02
\,$\text{m}^2\text{kg}^{-1}$ is selected, which corresponds to an AMR value of
a standard GPS satellite, in case the AMR parameter is not estimated but kept
fixed in the orbit determination. For HAMR objects always an AMR value is
estimated. \\
\\
The shadow paths of the orbit are modeled, under the assumption of a
spherical earth on a mean circular orbit; the boundary between sunlit and
eclipsed part is assumed to be cylindrical, no distinction between penumbra and
umbra is made, earth atmosphere is neglected.\\
\\
For a long term investigation of the orbits and the AMR values, different
comparable orbits have to be determined. Only sparse observations are available, which are
unequally spaced in time. A normalized setup is developed, tested with two low AMR objects and two
of the HAMR objects of the AIUB
catalog and applied for the creation of comparable orbits for the
investigation of the HAMR objects. 

\section{Normalized Sparse Data Setup}

\subsection{The Method}

Four representative GEO objects from the internal catalogue of the AIUB were
chosen, they have been followed over longer time periods and are not listed in
the USSTRATCOM catalogue. Those objects are clearly space debris, since no maneuvers could be detected in the data. The AIUB did not have information what those objects actually were before becoming debris. From the apparent magnitude it can be concluded that those are all fragmentation pieces. They represent typical objects found in GEO surveys. Their properties are listed in Tab. \ref{prop}.\\
\setlength{\arrayrulewidth}{0.5pt}
\setlength{\doublerulesep}{0.6pt} 
\begin{table}
\begin{center}
\caption{\bf Internal name, eccentricity, inclination (deg), semi-major axis (km), area-to-mass ratio ($m^2/kg$) and apparent magnitude (mag) of the selected objects of the AIUB catalogue}
\begin{tabular}{lllllll}
\small NAME & \small Epoch &\small \textit{a} &\small \textit{e} &\small \textit{i} &\small AMR &\small Mag\\
\vspace{-0.4cm}\\
\hline
\vspace{-0.3cm}\\
\small 
\small E03174A&55208.0&41900&0.001&10.1&0.01&14.6\\
\small E06321D&55275.9&41400&0.035&7.00&2.29&15.3\\
\small E06327E&54470.1&40000&0.067&12.31&0.20&17.2\\
\small E08241A&55213.0&41600&0.041&13.26&1.24&16.1\\
\end{tabular}
\label{prop}
\vspace{-0.5cm}
\end{center}
\end{table}
\\
Two of the objects have low area to mass ratios, two objects qualify as HAMR
objects with an AMR value larger than 1$m^2/kg$. The optical angle-only observations are obtained with ZIMLAT (Zimmerwald, Switzerland),
and ESASDT (Tenerife, Spain), supplemented by some observations of the ISON
network provided by the Keldysh Institute of Applied Mathematics, Moscow,
Russia. The latter observations were obtained from different sites of the ISON
network, in these particular cases, all located in Eastern Europe.\\
\\
All orbits
were determined from two observation sets only, using a priori orbital
elements. A maximum of eight observations are allowed per set. An observation set may consist of more than
one tracklet. But the observations within the sets should not be distributed
over more than three days. \\
\begin{figure*}\textcolor{white}{[ht]}
  \centering
  \subfloat[]{\label{resE03174Aarc}\includegraphics[width=0.25\textwidth, height=0.13\textheight]{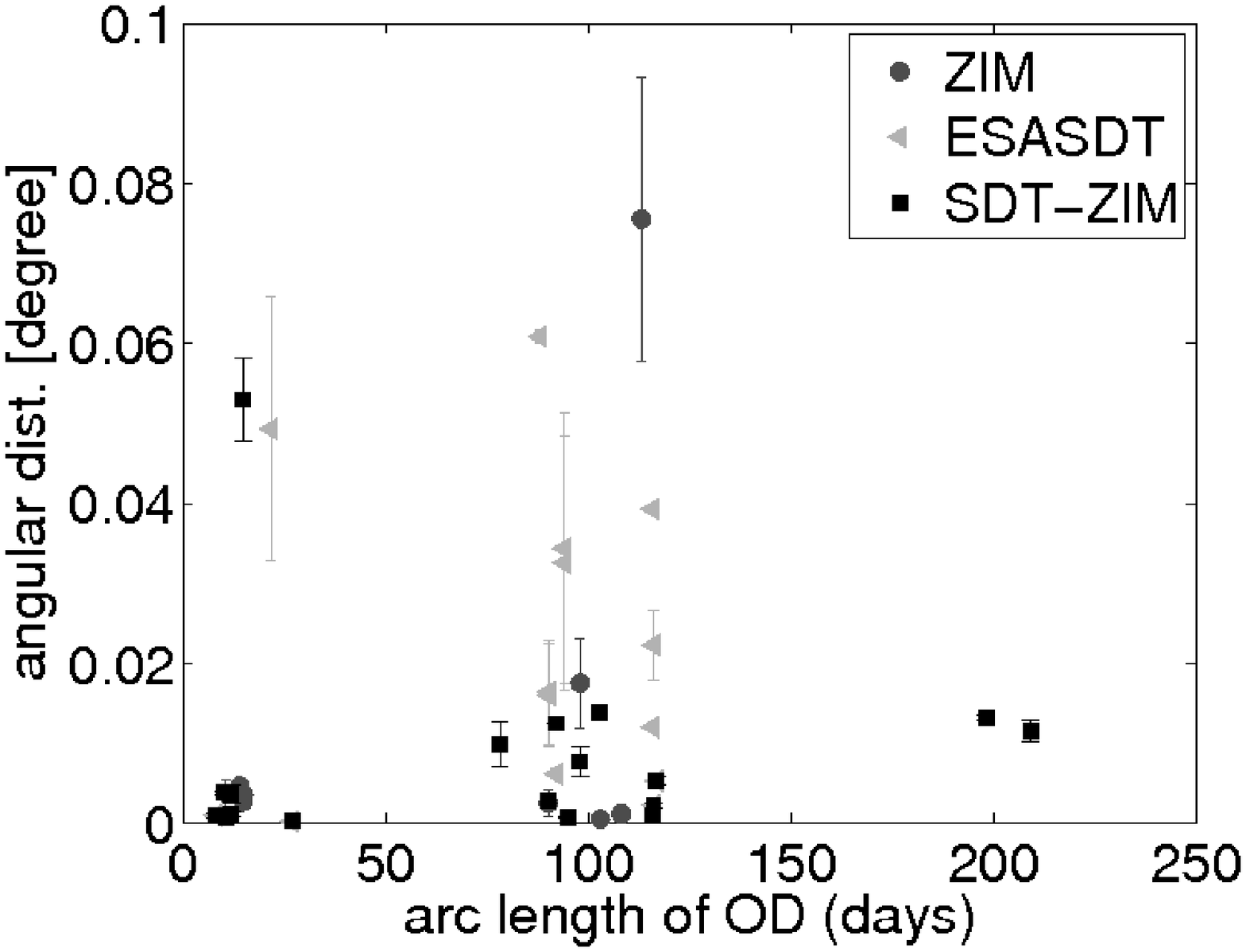}} 
  \subfloat[]{\label{resE06321Darc}\includegraphics[width=0.25\textwidth, height=0.13\textheight]{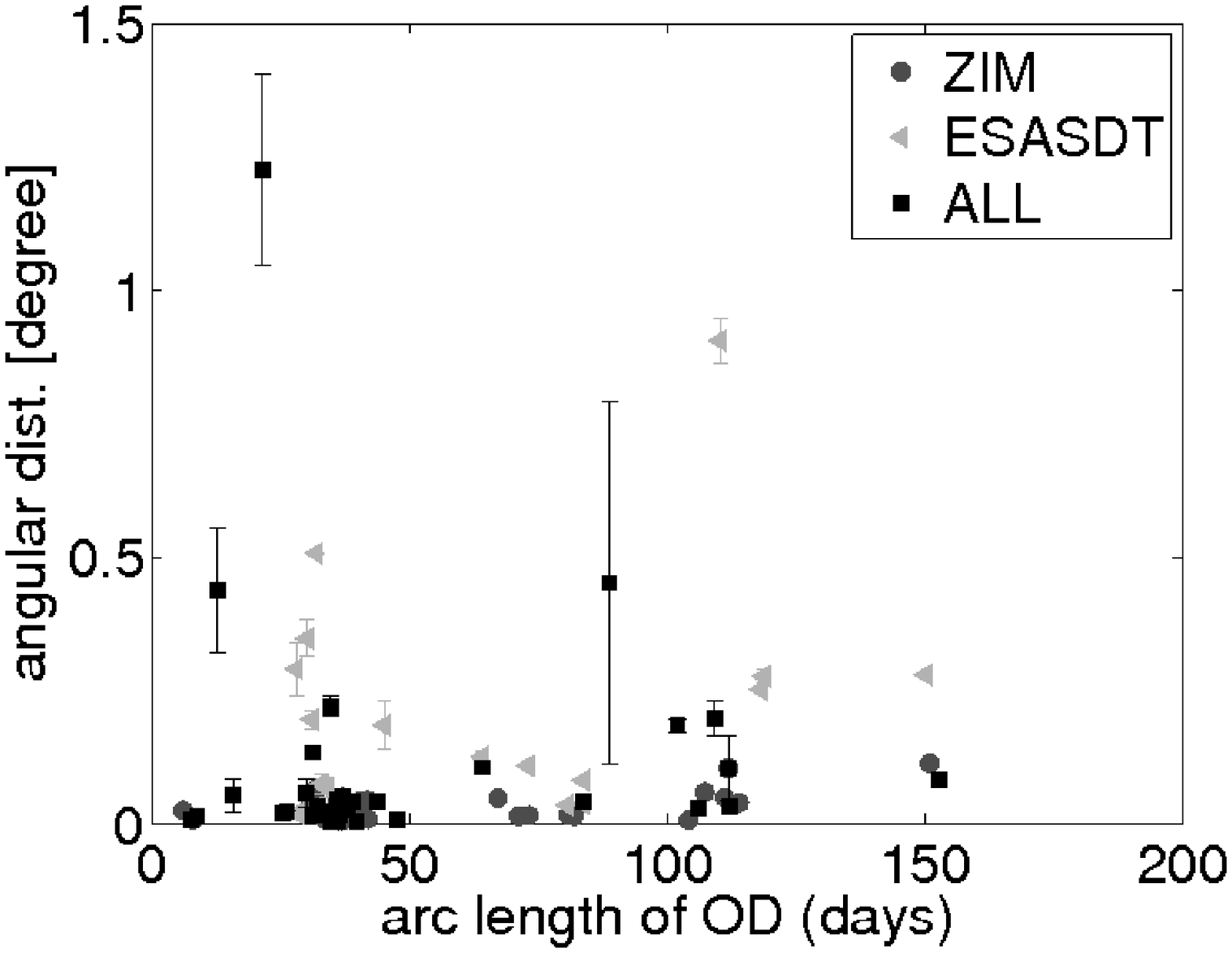}}
  \subfloat[]{\label{resE06327Earc}\includegraphics[width=0.25\textwidth, height=0.13\textheight]{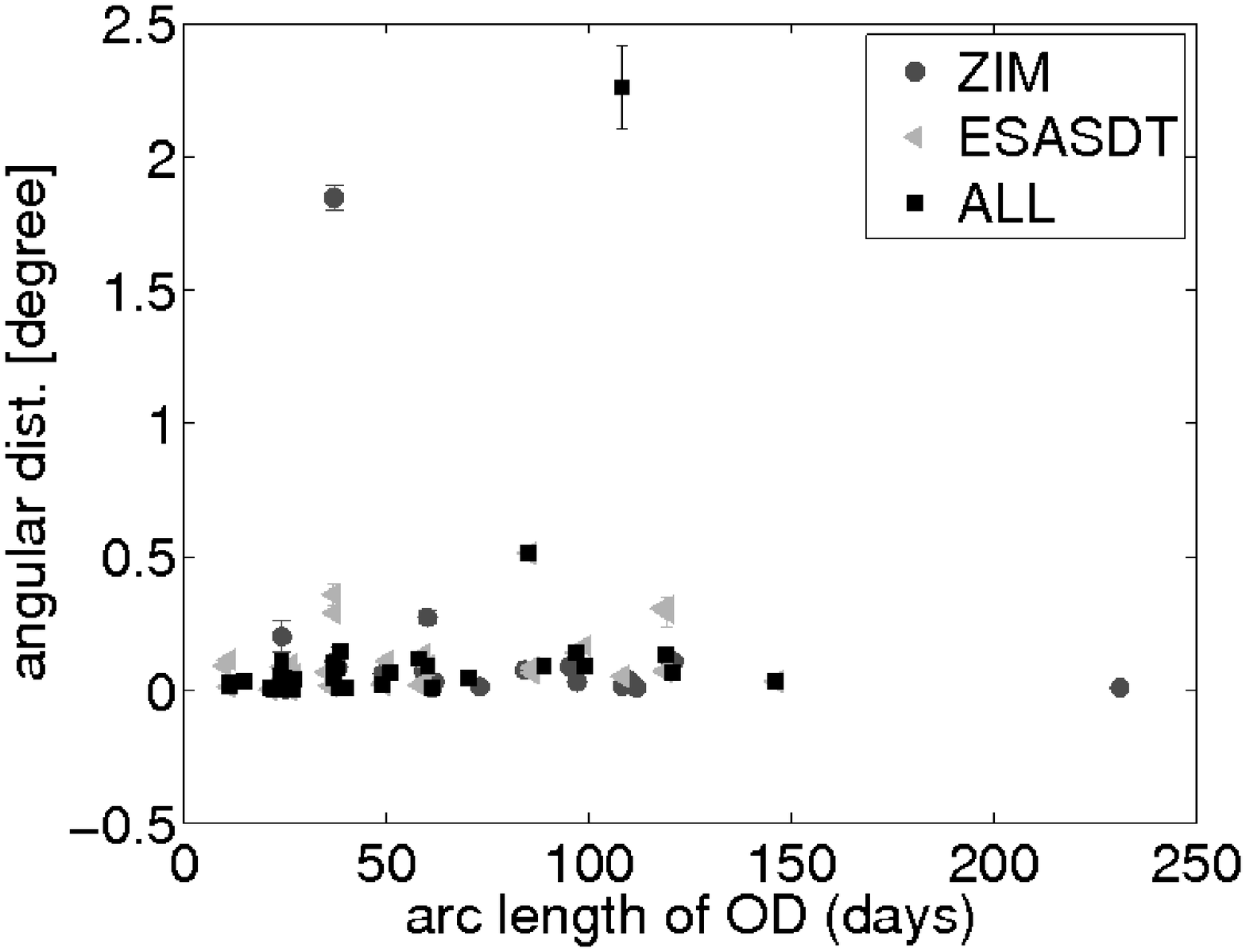}}
  \subfloat[]{\label{resE08241Aarc}\includegraphics[width=0.25\textwidth, height=0.13\textheight]{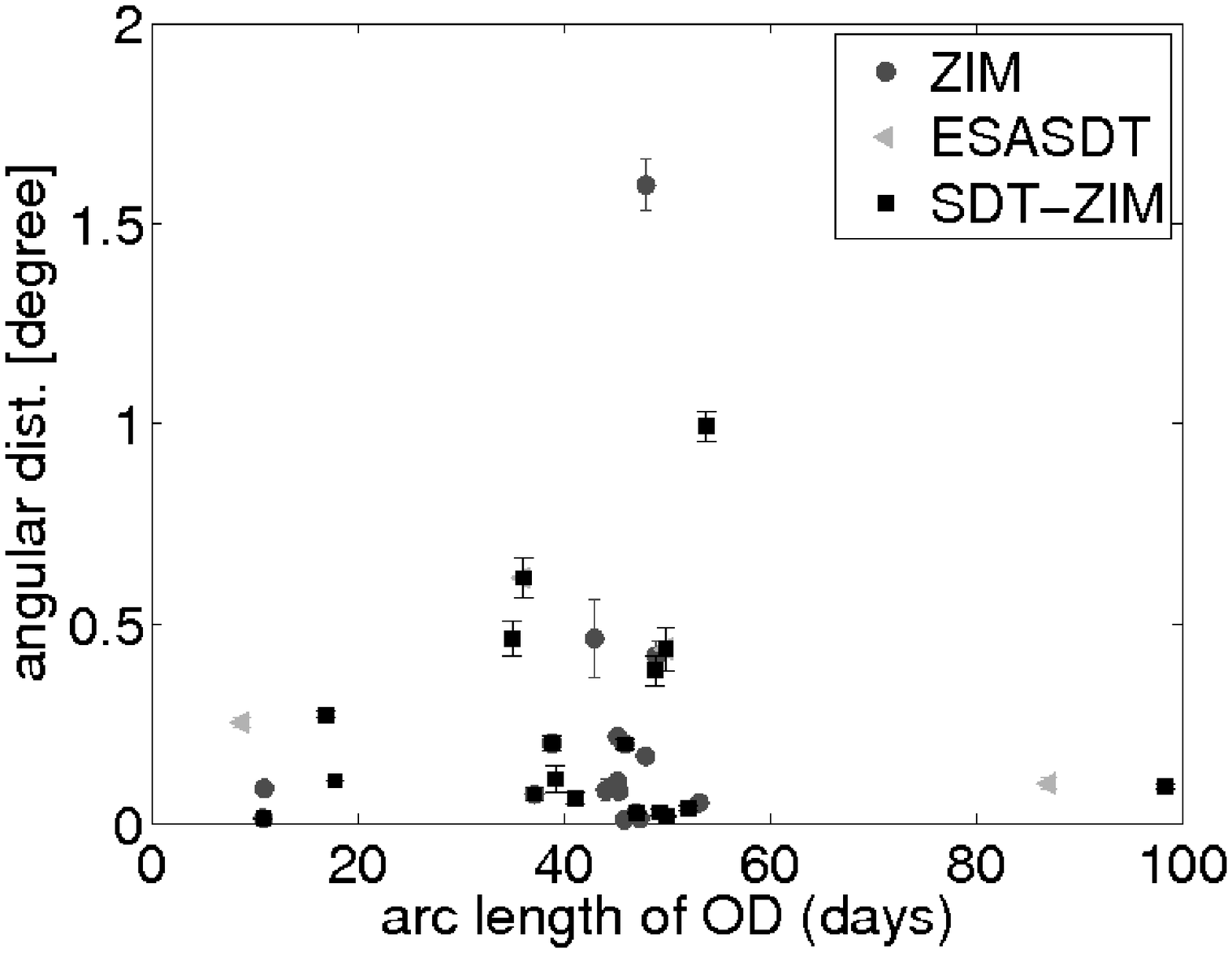}} 
  \caption{\bf Angular distances as a function of the time interval between the
    first and the last observation of the fit interval of orbit determination
    for object (a) E03174A, (b) E06321D, (c) E06327E
    and (d) E08241A.}
  \label{resE03174A}
\end{figure*}
\begin{figure*}\textcolor{white}{[h]}
  \centering
  \subfloat[]{\label{resresE03174A}\includegraphics[width=0.25\textwidth, height=0.13\textheight]{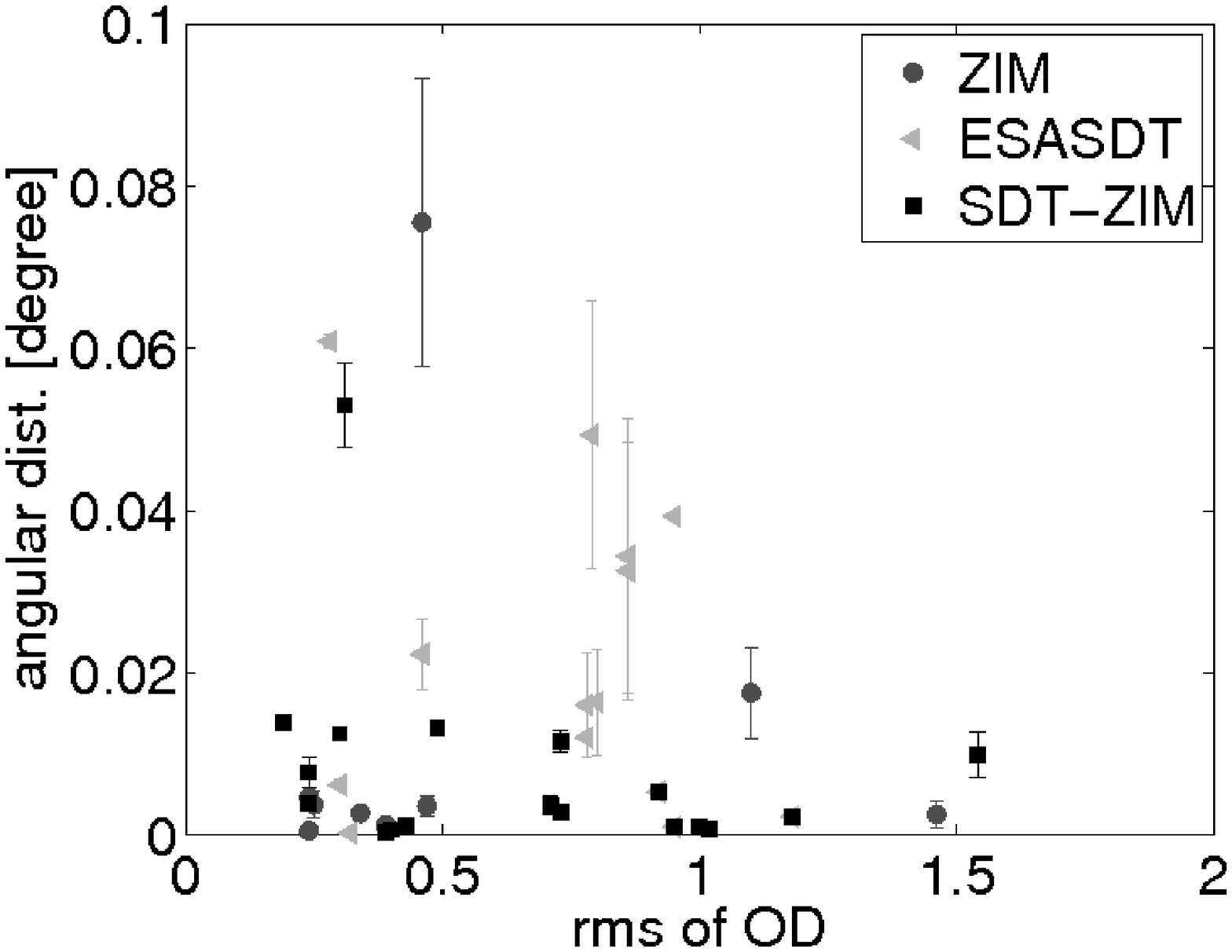}}  
  \subfloat[]{\label{resresE06321D}\includegraphics[width=0.25\textwidth, height=0.13\textheight]{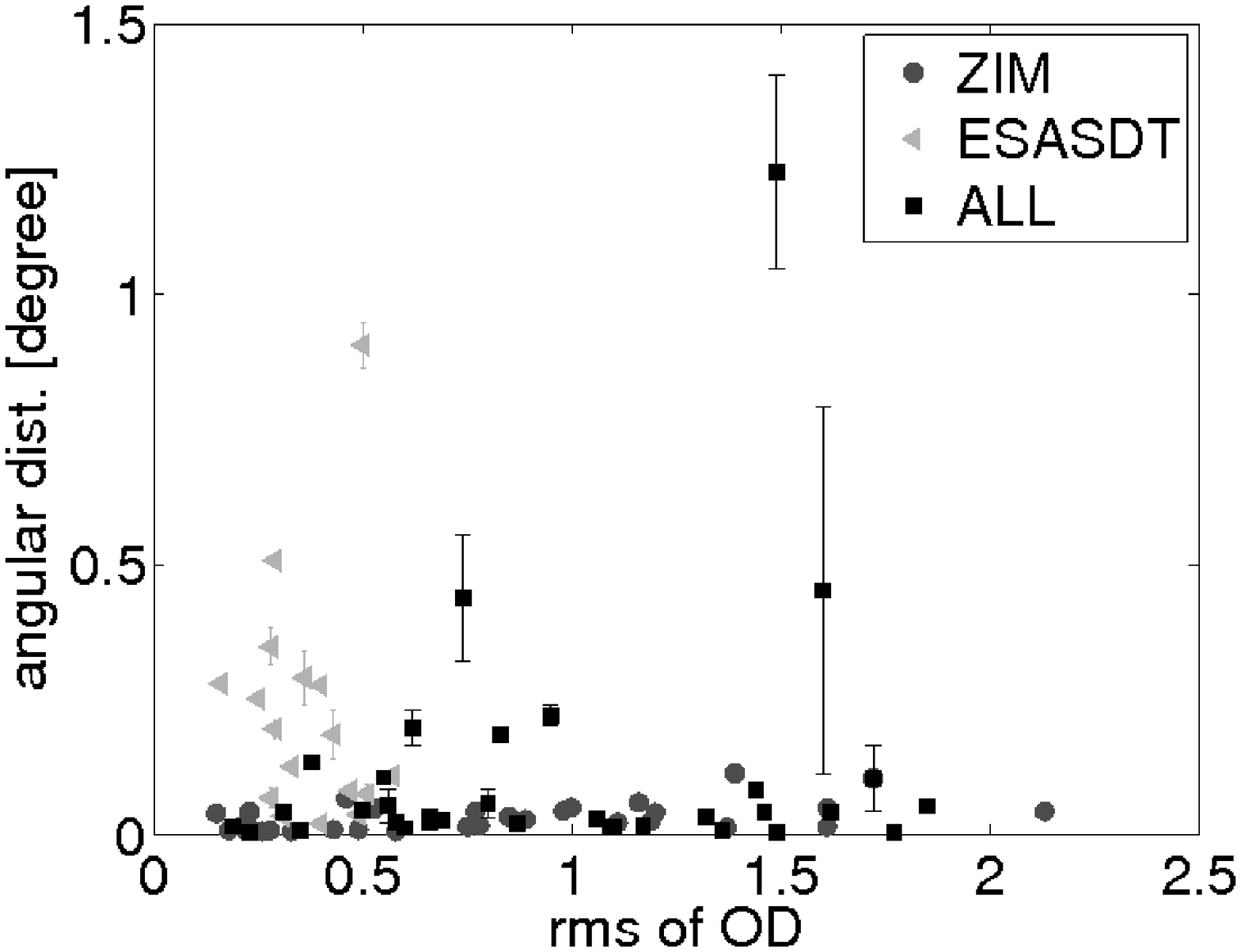}}  
  \subfloat[]{\label{resresE06327E}\includegraphics[width=0.25\textwidth, height=0.13\textheight]{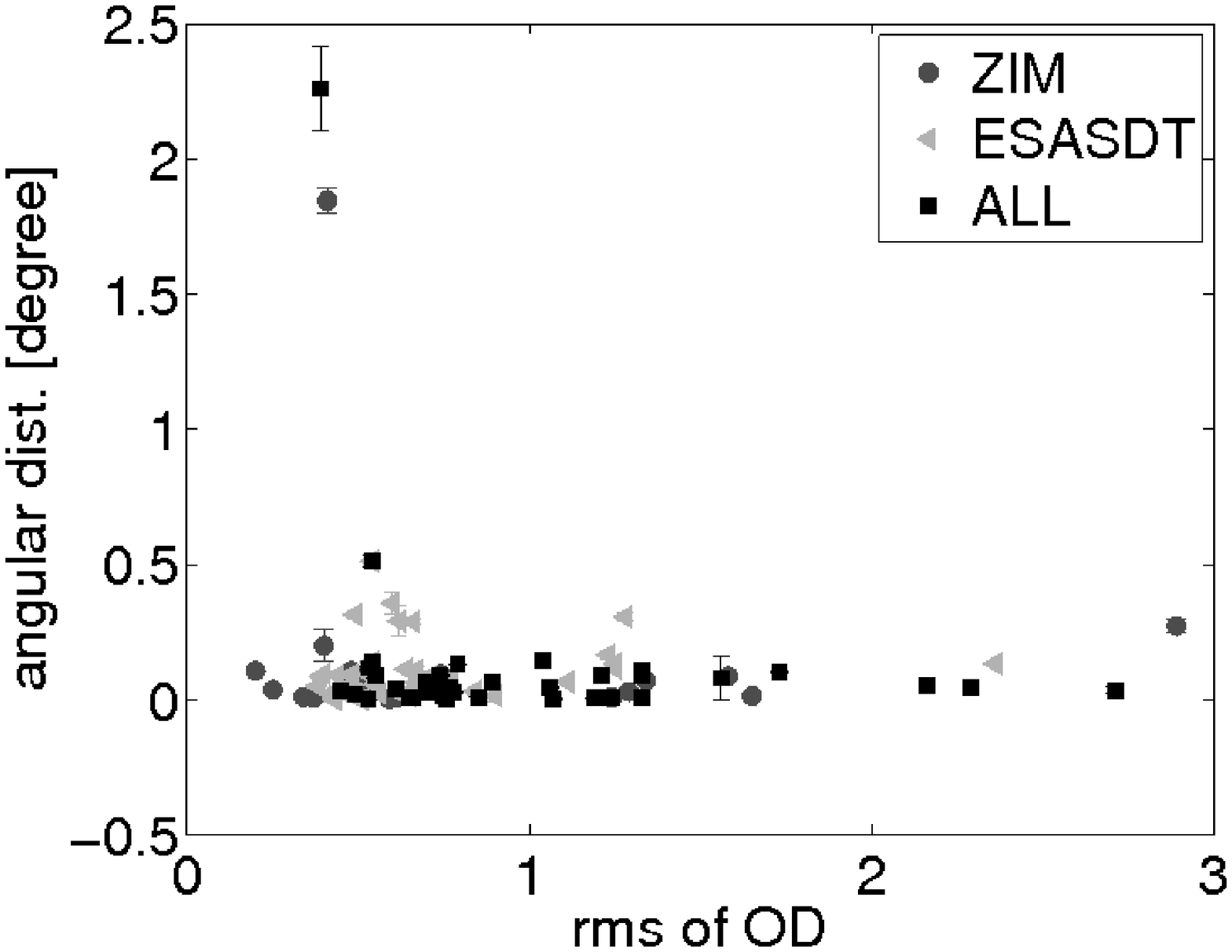}}
  \subfloat[]{\label{resresE08241A}\includegraphics[width=0.25\textwidth, height=0.13\textheight]{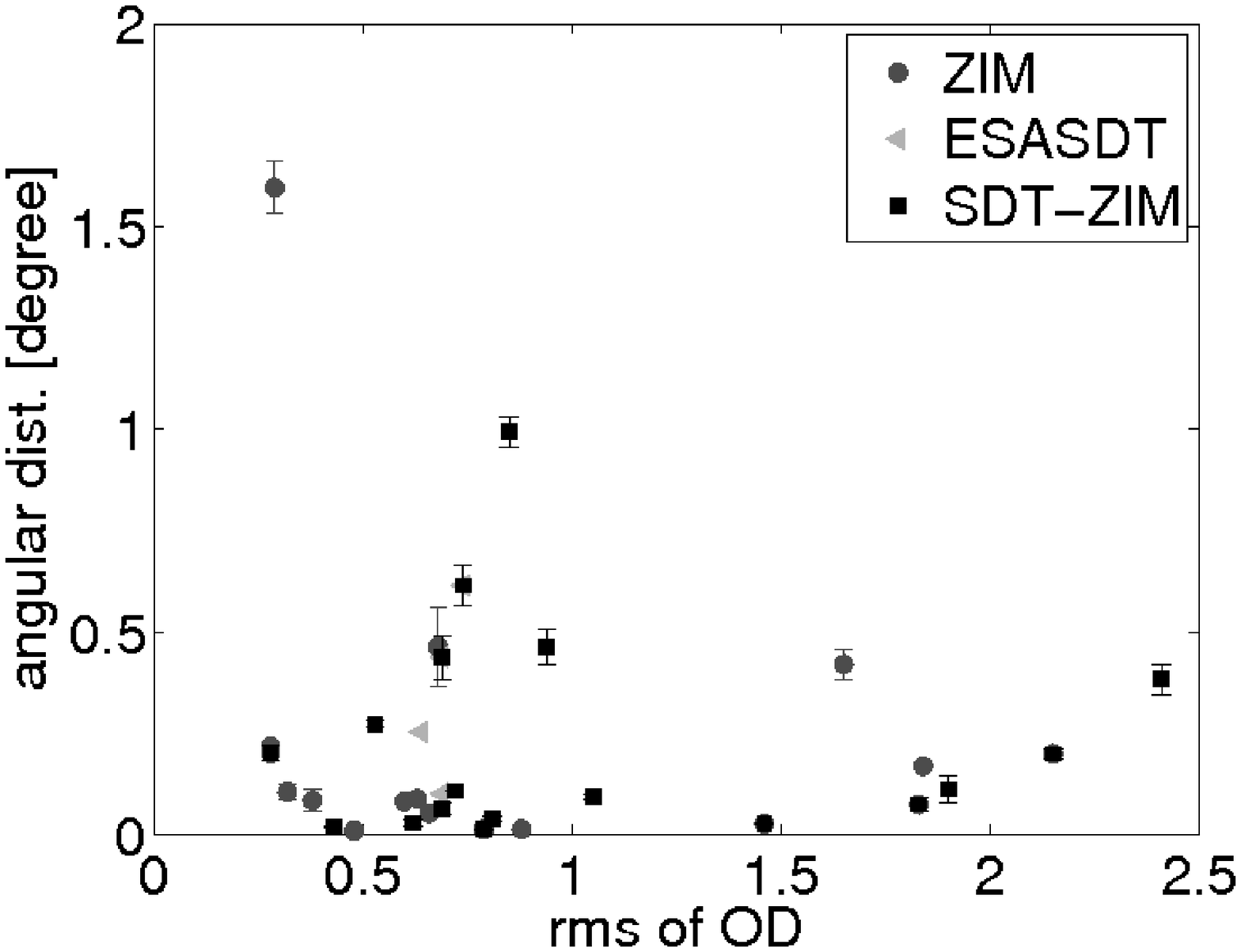}}
  \caption{\bf Root mean square of orbit determination as a function of the arc
    length of observations for object (a) E03174A, (b) E06321D, (c) E06327E
    and (d) E08241A.}
  \label{res}
\end{figure*}
\begin{figure*}\textcolor{white}{[h]}
  \centering
  \subfloat[]{\label{noobsE03174A}\includegraphics[width=0.25\textwidth, height=0.13\textheight]{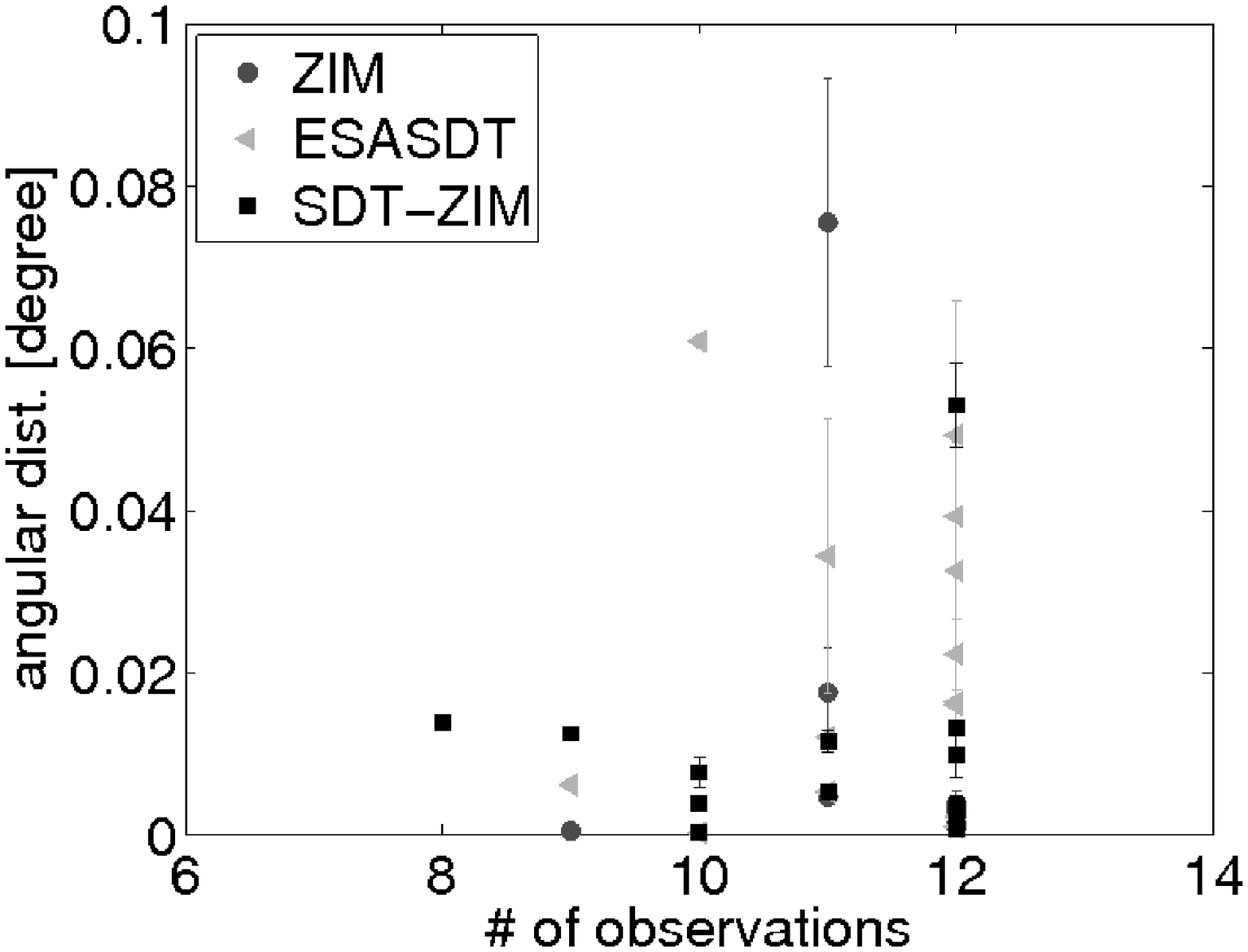}}  
  \subfloat[]{\label{noobsE06321D}\includegraphics[width=0.25\textwidth, height=0.13\textheight]{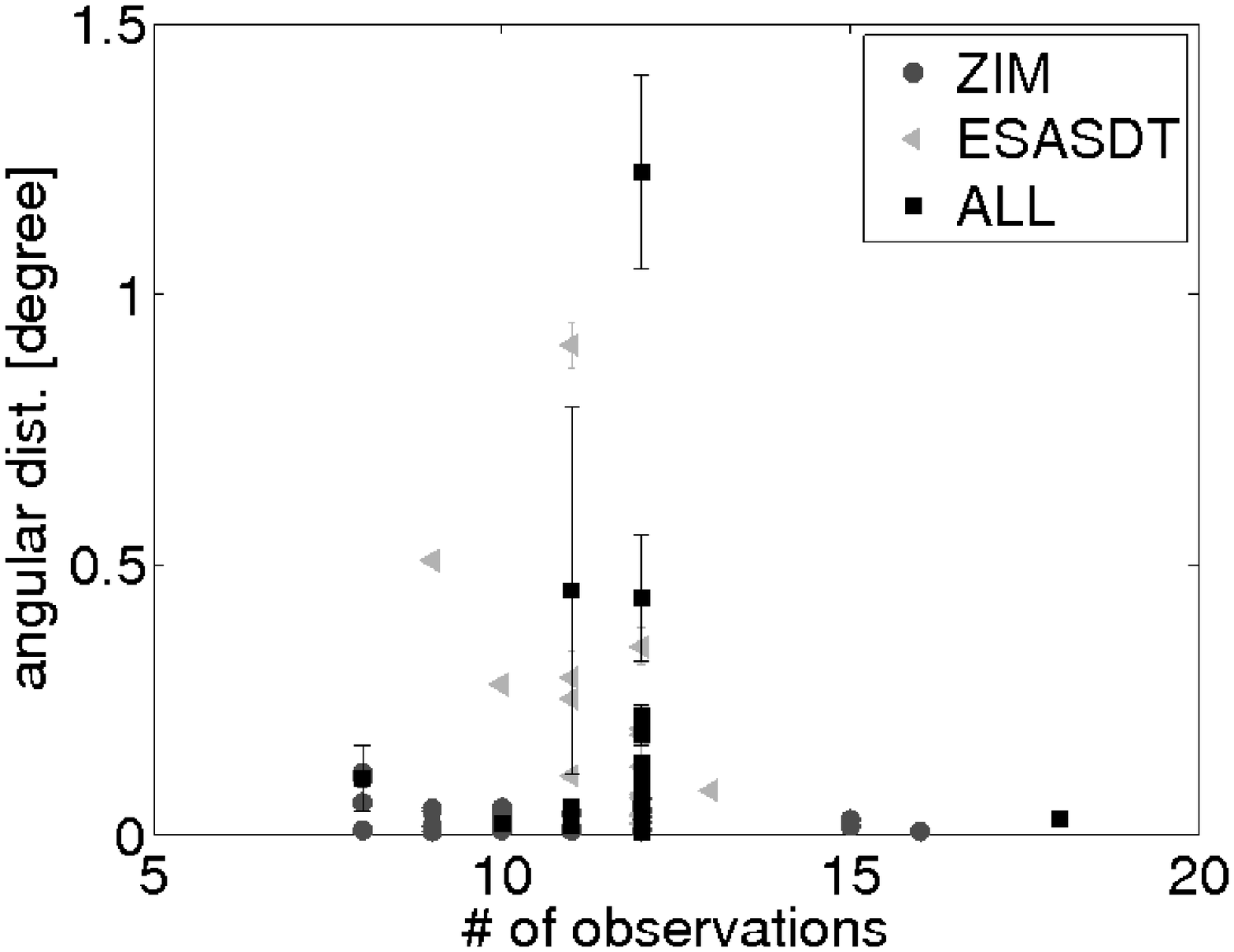}}
  \subfloat[]{\label{noobsE06327E}\includegraphics[width=0.25\textwidth, height=0.13\textheight]{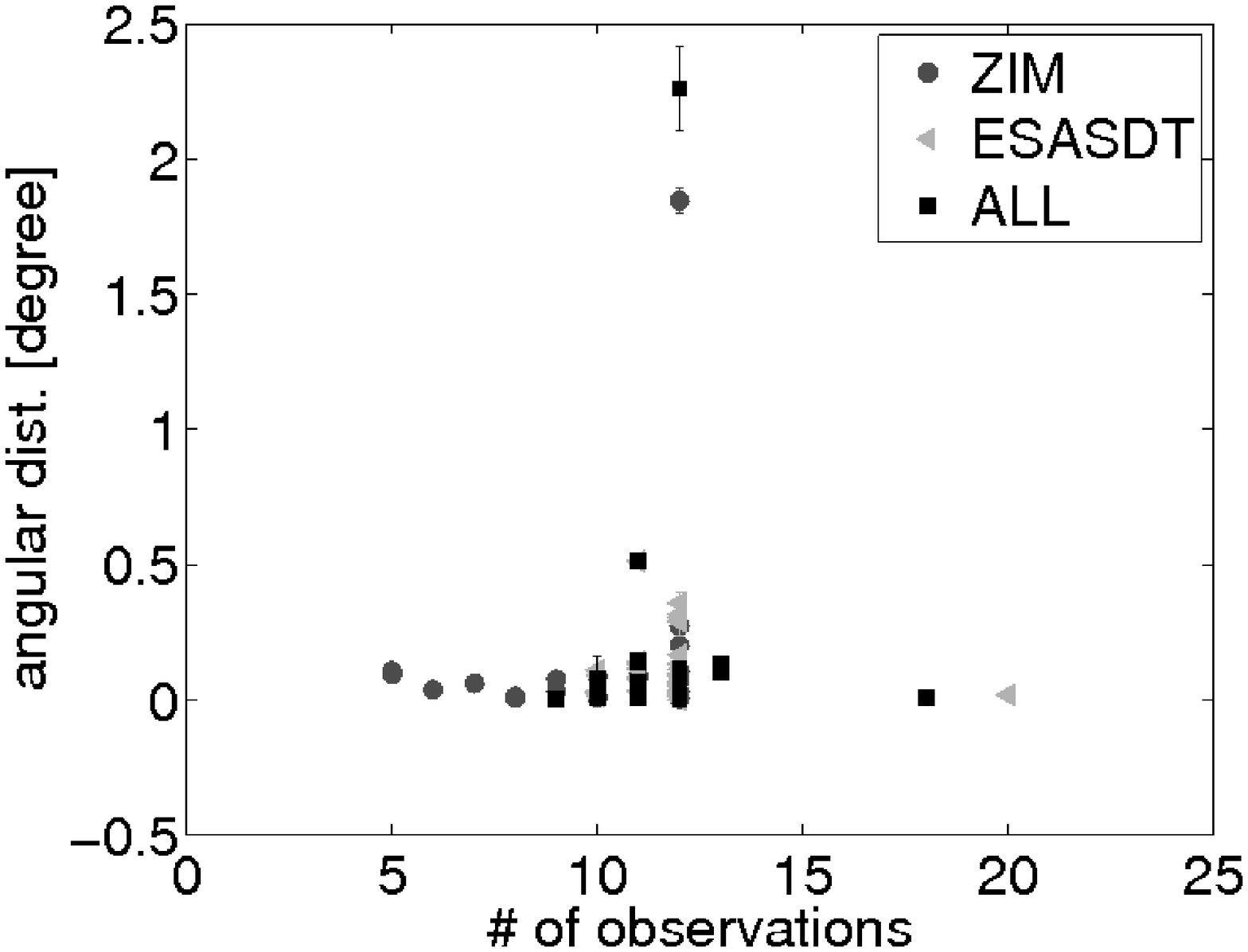}}
  \subfloat[]{\label{noobsE08241A}\includegraphics[width=0.25\textwidth, height=0.13\textheight]{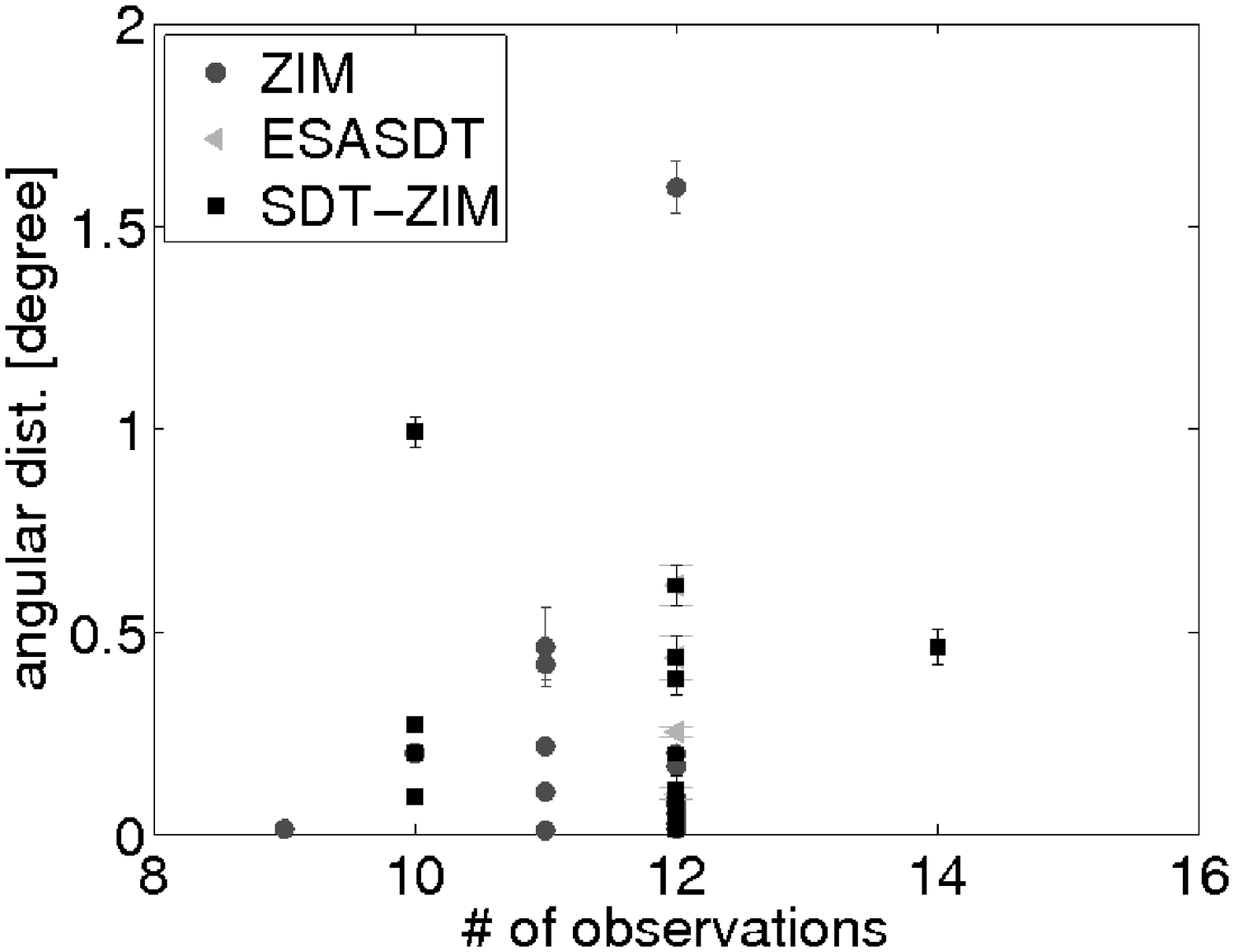}}
  \caption{\bf Angular distances as a function of the number of observations used
    for orbit determination for object (a) E03174A, (b) E06321D, (c) E06327E
    and (d) E08241A. }
  \label{noobs}
\end{figure*}
\begin{figure*}\textcolor{white}{[h]}
  \centering
  \subfloat[]{\label{anoarcE03174A}\includegraphics[width=0.25\textwidth, height=0.13\textheight]{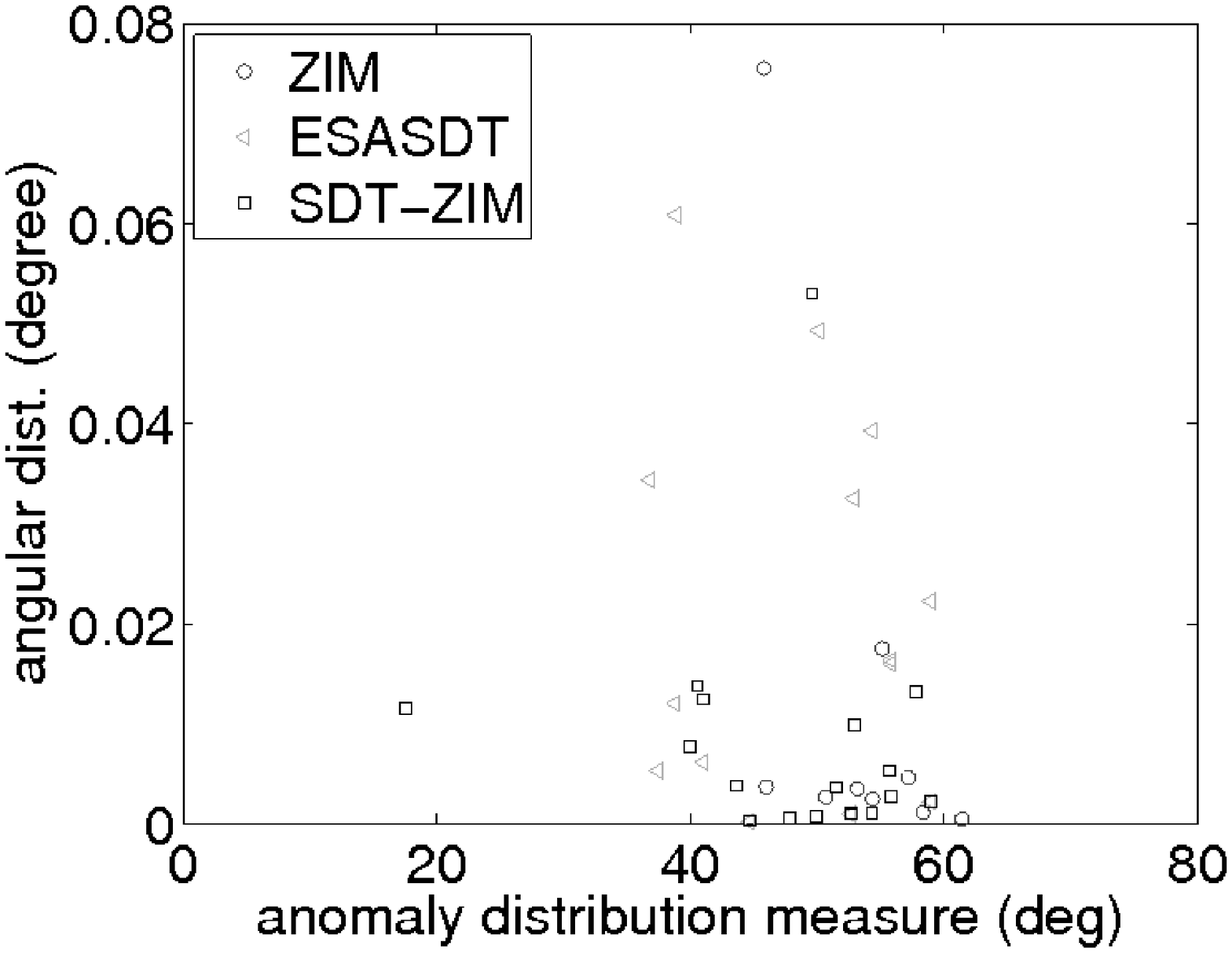}}  
  \subfloat[]{\label{anoarcE06321D}\includegraphics[width=0.25\textwidth, height=0.13\textheight]{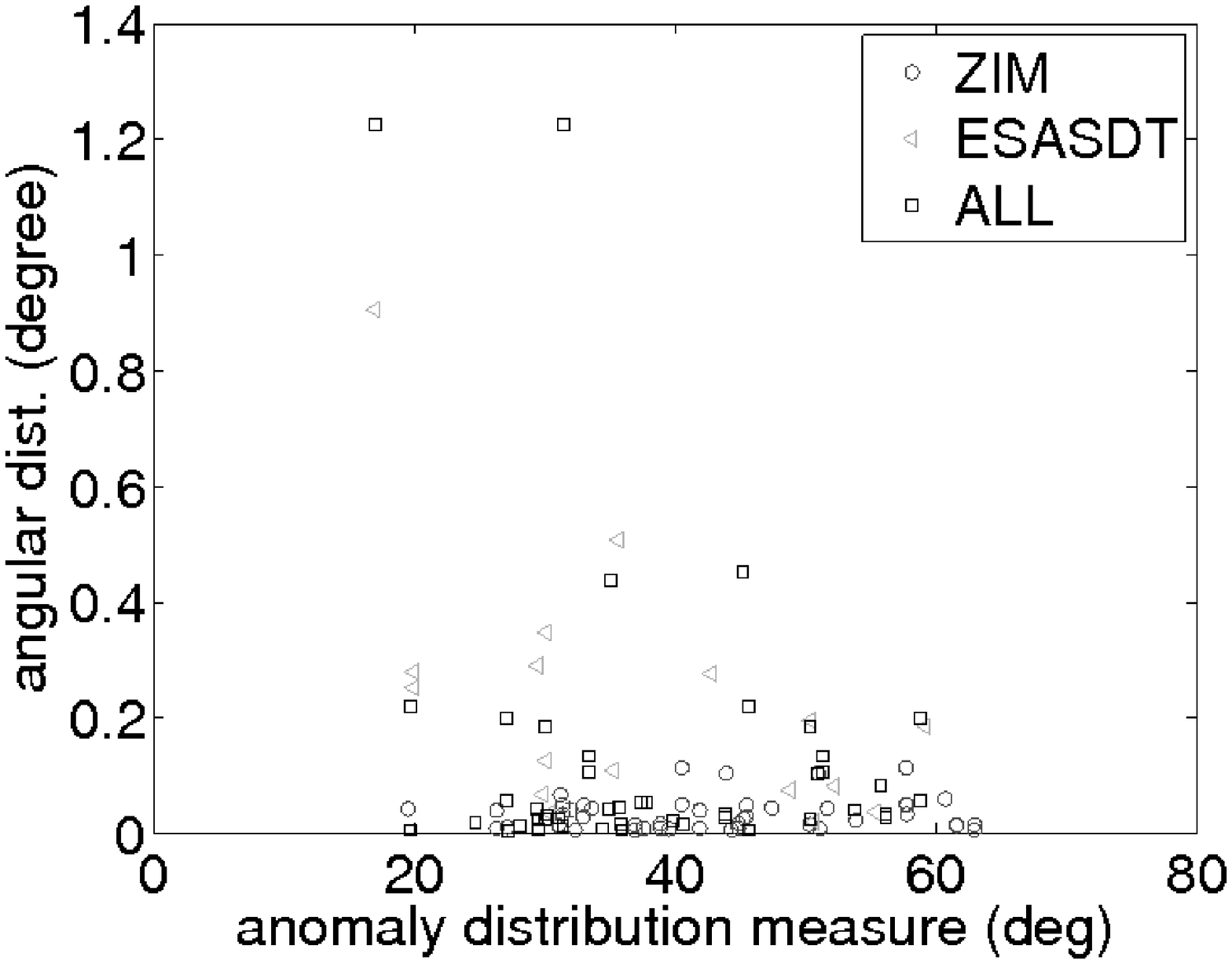}} 
  \subfloat[]{\label{anoarcE06327E}\includegraphics[width=0.25\textwidth, height=0.13\textheight]{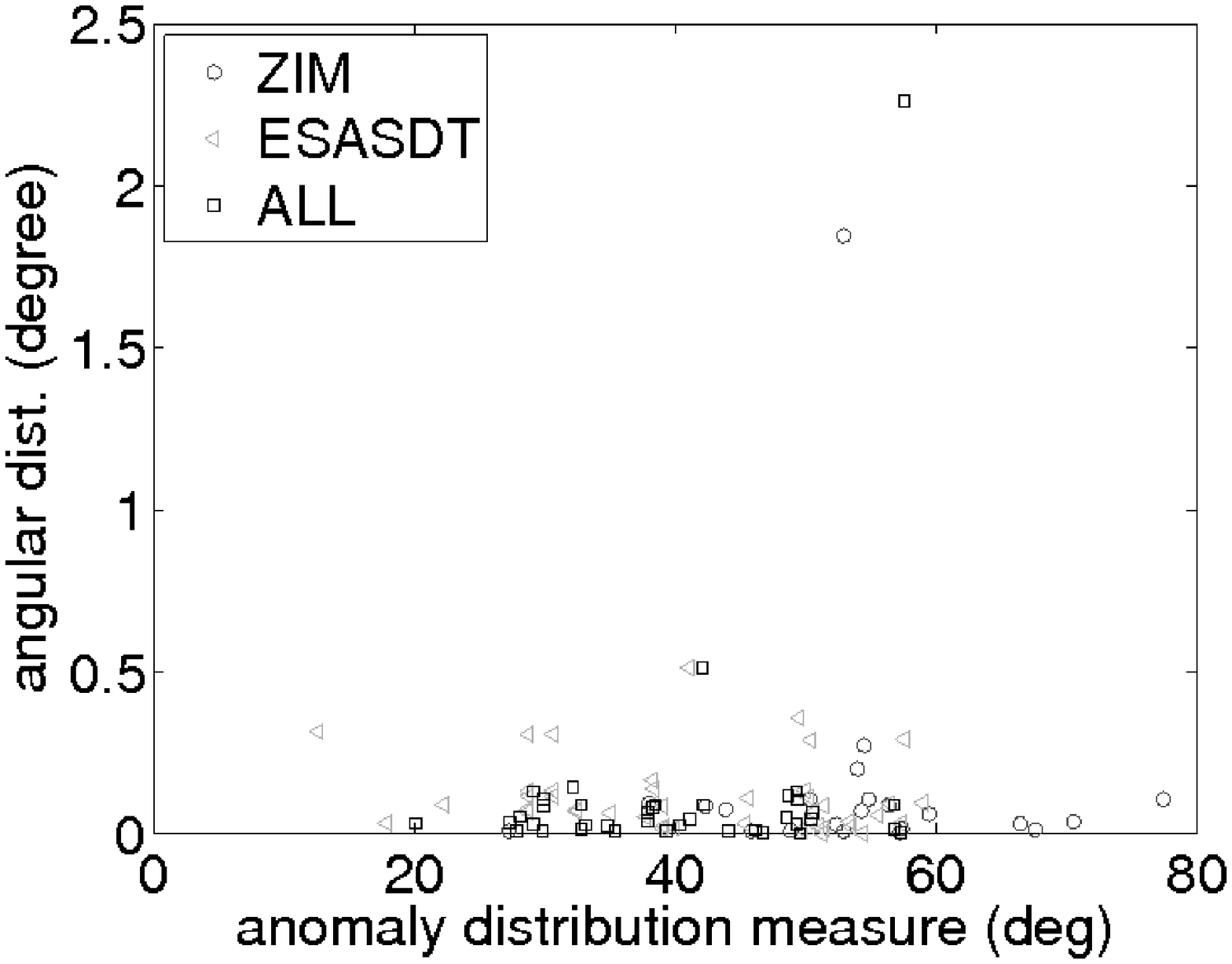}}
  \subfloat[]{\label{anoarcE08241A}\includegraphics[width=0.25\textwidth, height=0.13\textheight]{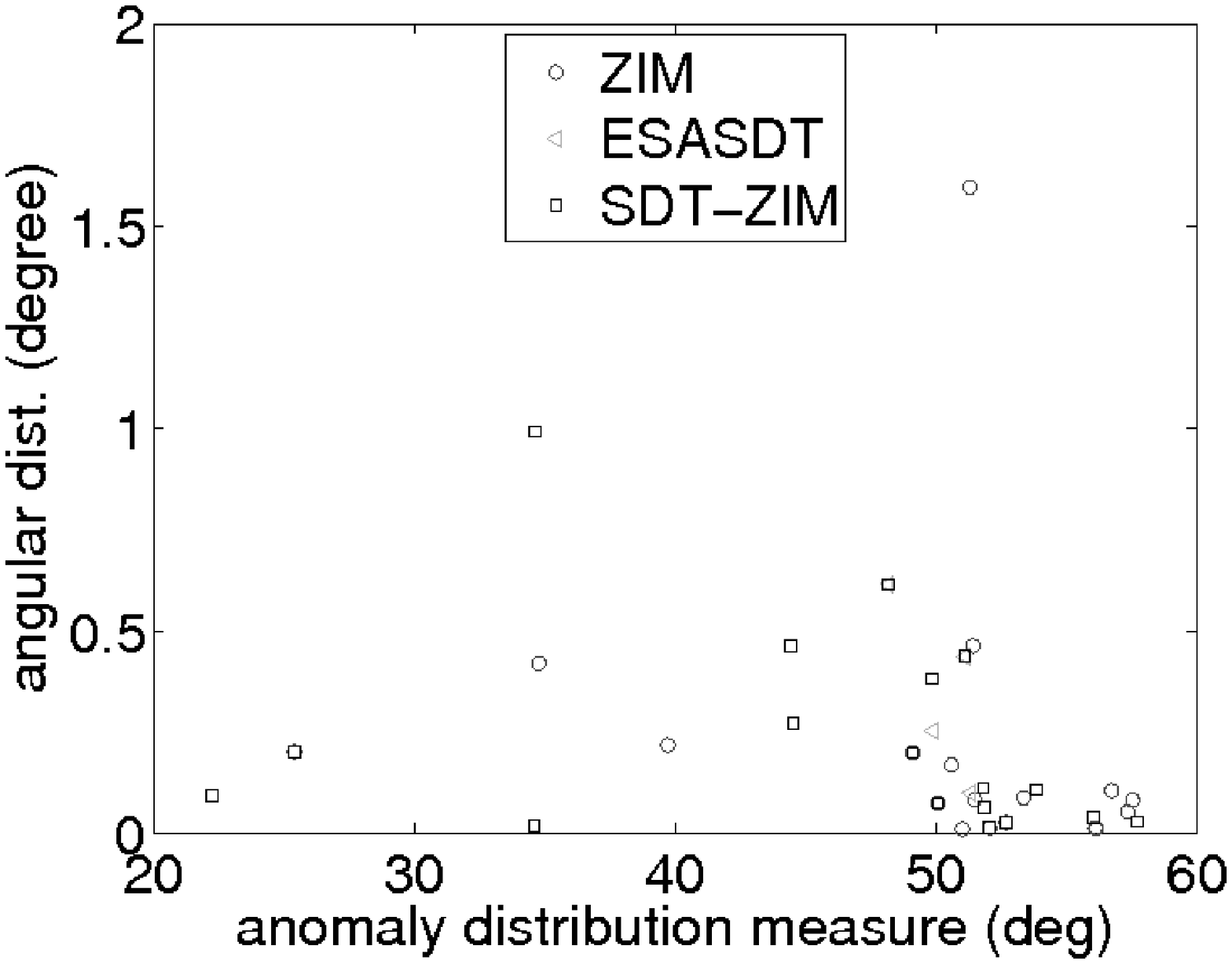}}
  \caption{\bf Angular distances as a function of anomaly distribution factor for object (a) E03174A, (b) E06321D, (c) E06327E
    and (d) E08241A.}
  \label{anoarc}
\end{figure*}
\begin{figure*}\textcolor{white}{[h]}
  \centering
  \subfloat[]{\label{covE03174A}\includegraphics[width=0.25\textwidth, height=0.13\textheight]{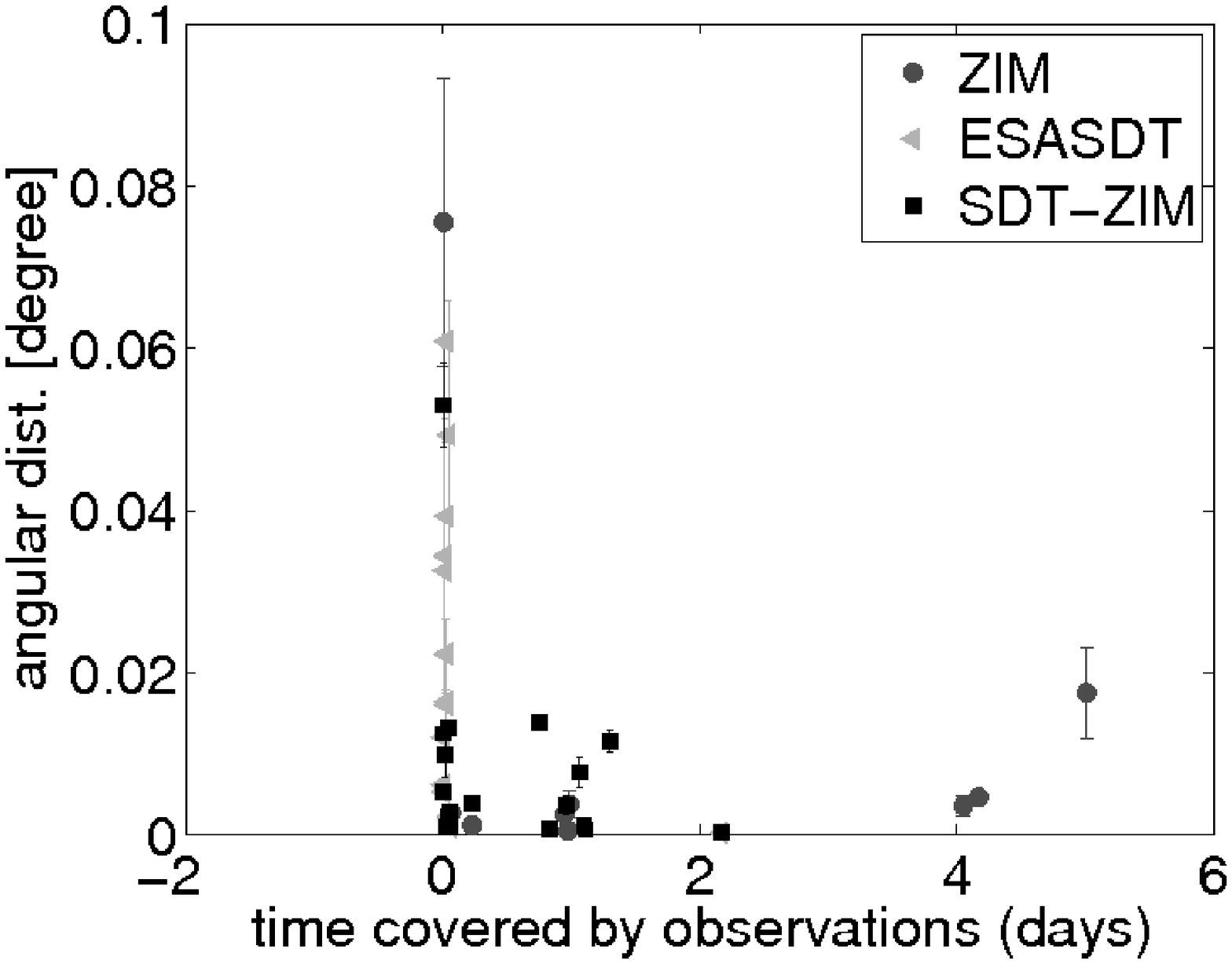}}  
  \subfloat[]{\label{covE06321D}\includegraphics[width=0.25\textwidth, height=0.13\textheight]{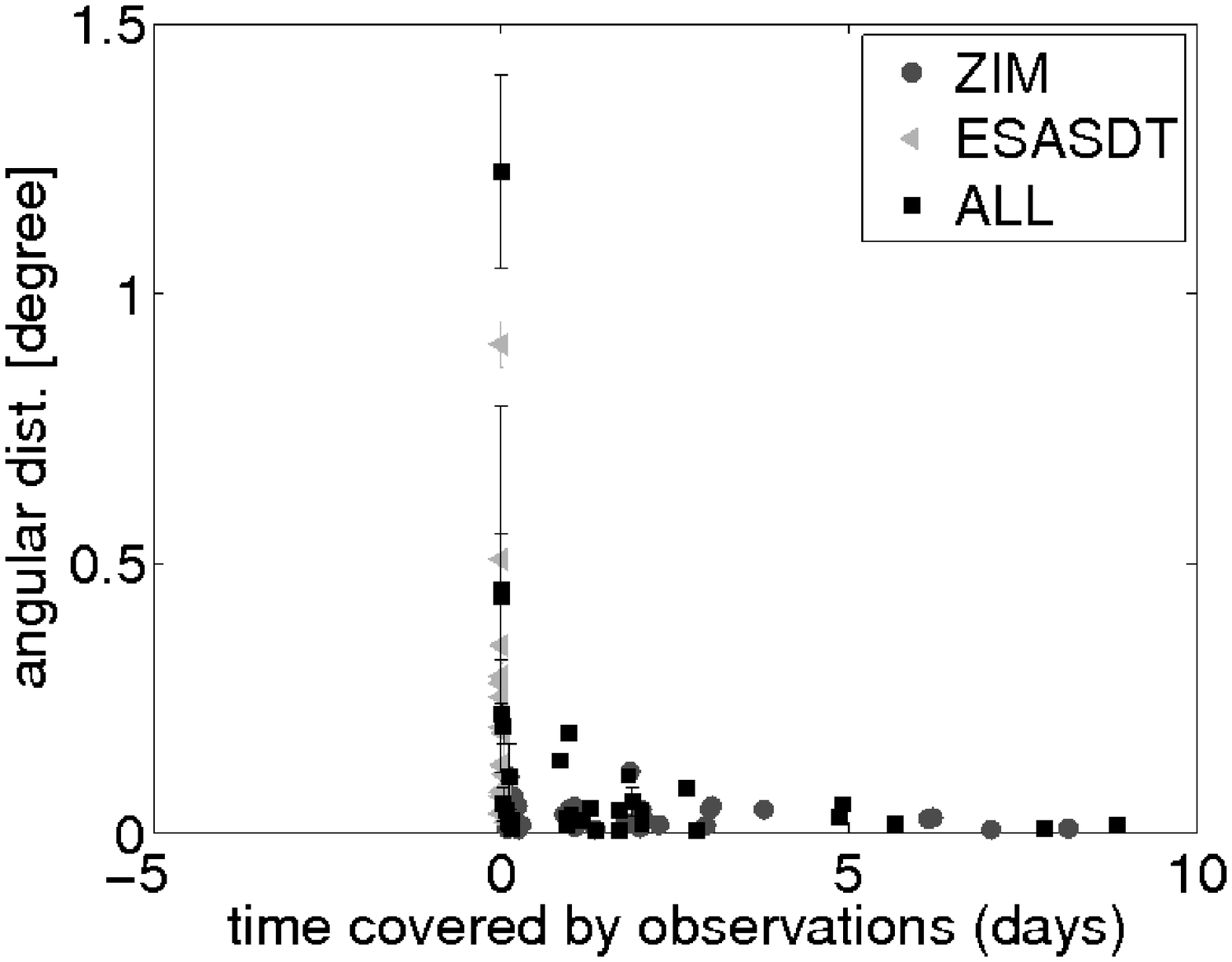}}
  \subfloat[]{\label{covE06327E}\includegraphics[width=0.25\textwidth, height=0.13\textheight]{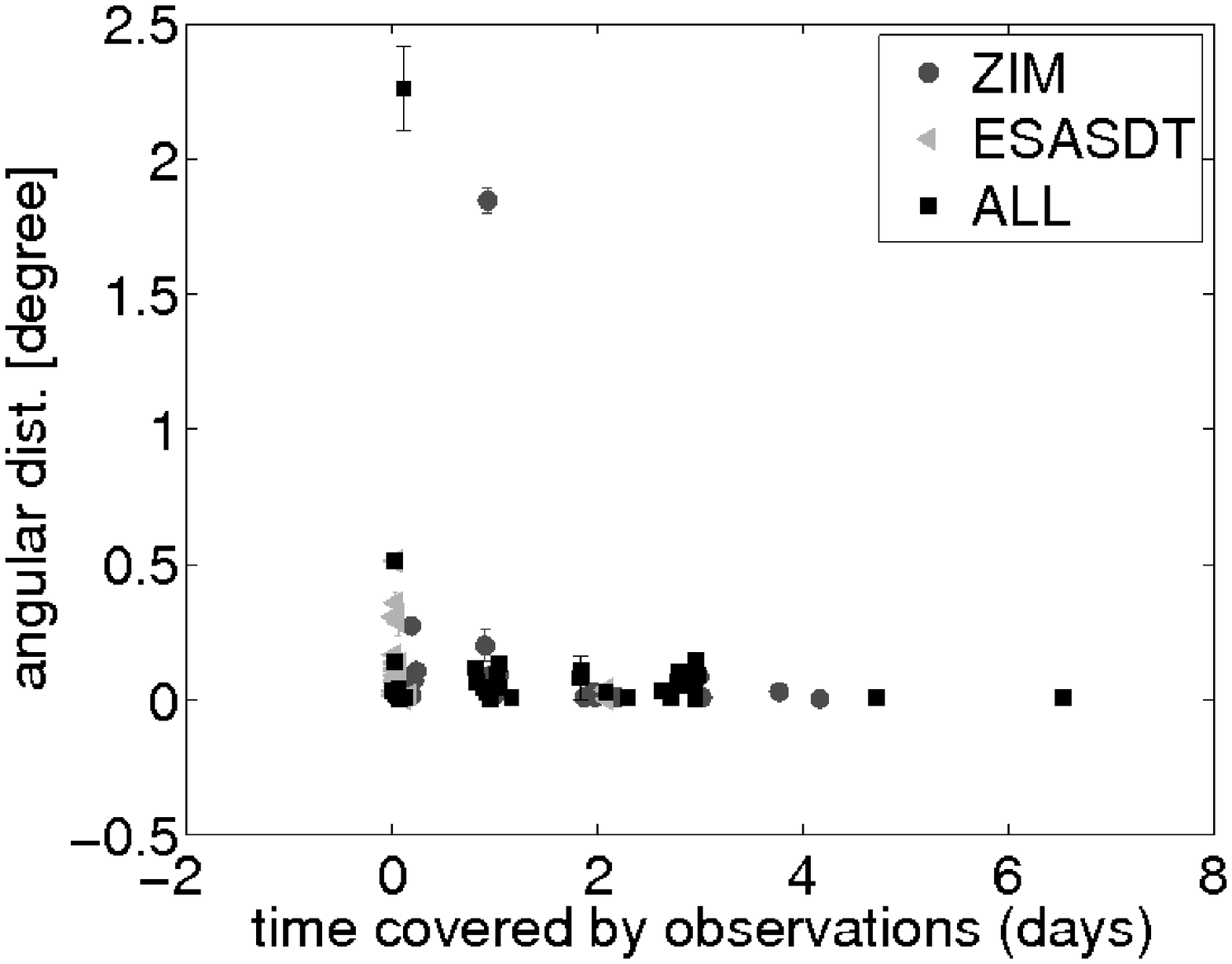}}
  \subfloat[]{\label{covE08241A}\includegraphics[width=0.25\textwidth, height=0.13\textheight]{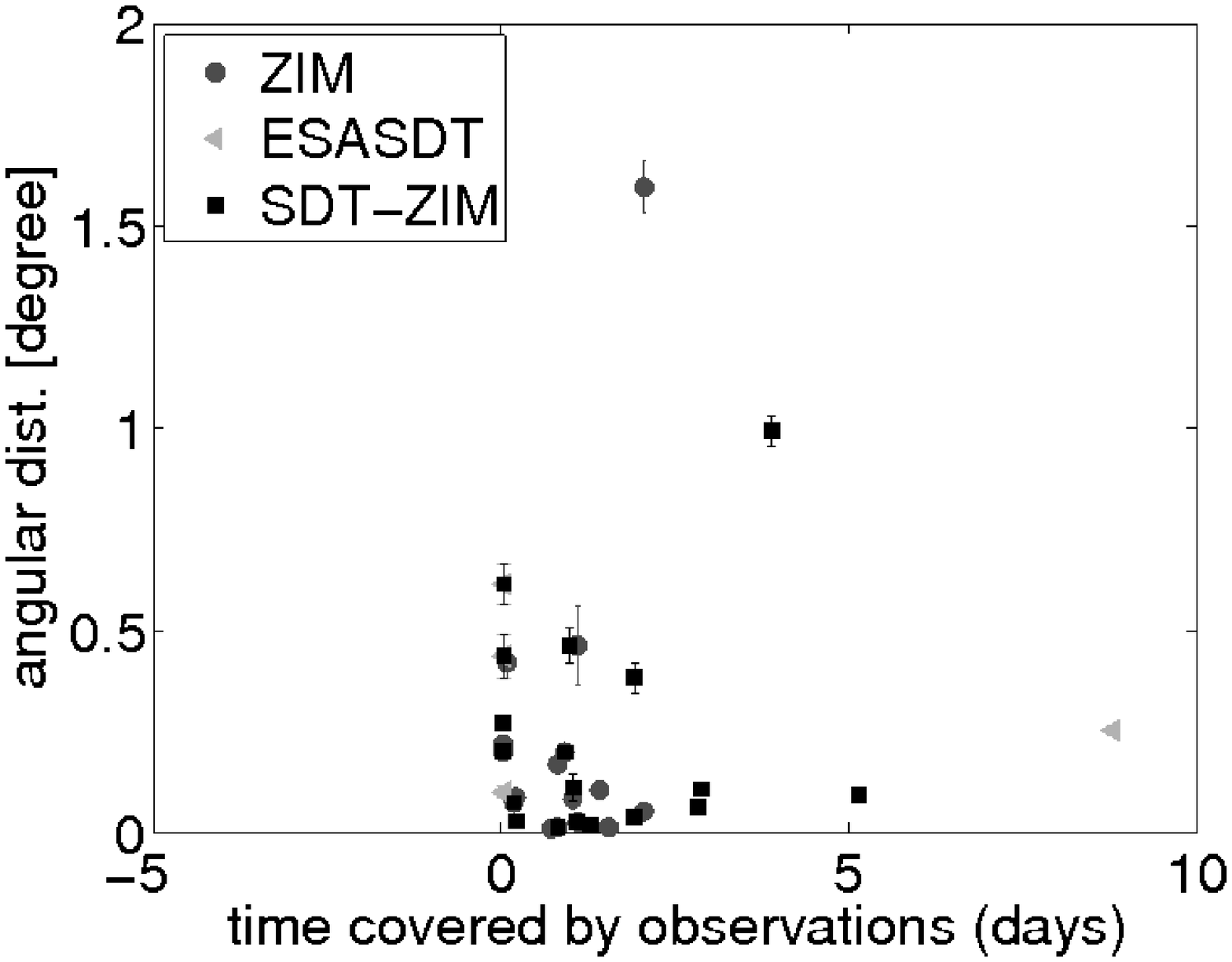}}
  \caption{\bf Angular distance as a function of time interval covered by
    observations used for orbit determination for object (a) E03174A, (b) E06321D, (c) E06327E and (d) E08241A.}
  \label{timecov}
\end{figure*}
\begin{figure*}\textcolor{white}{[h]}
  \centering
  \subfloat[]{\label{covnoE03174A}\includegraphics[width=0.25\textwidth, height=0.13\textheight]{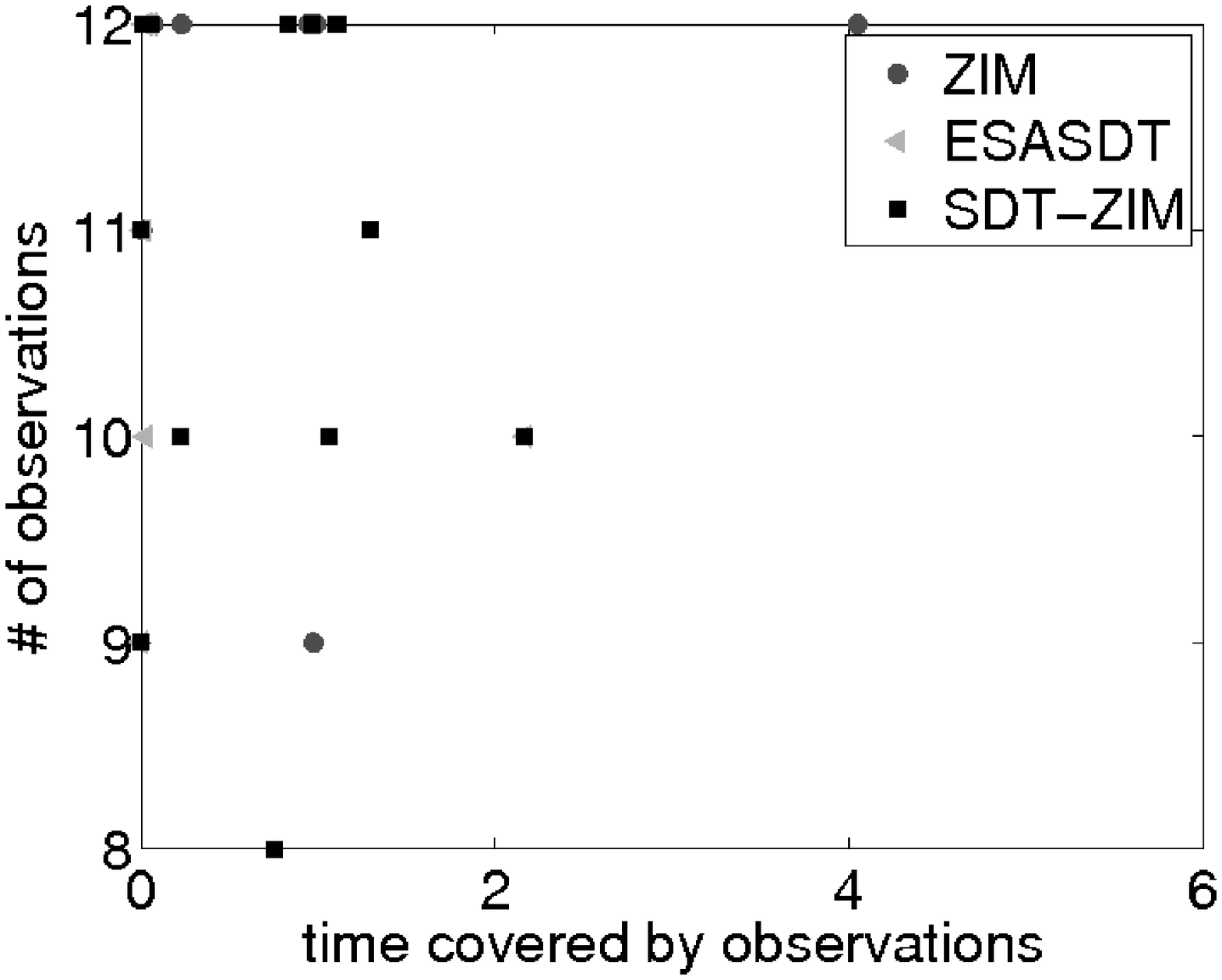}}  
  \subfloat[]{\label{covnoE06321D}\includegraphics[width=0.25\textwidth, height=0.13\textheight]{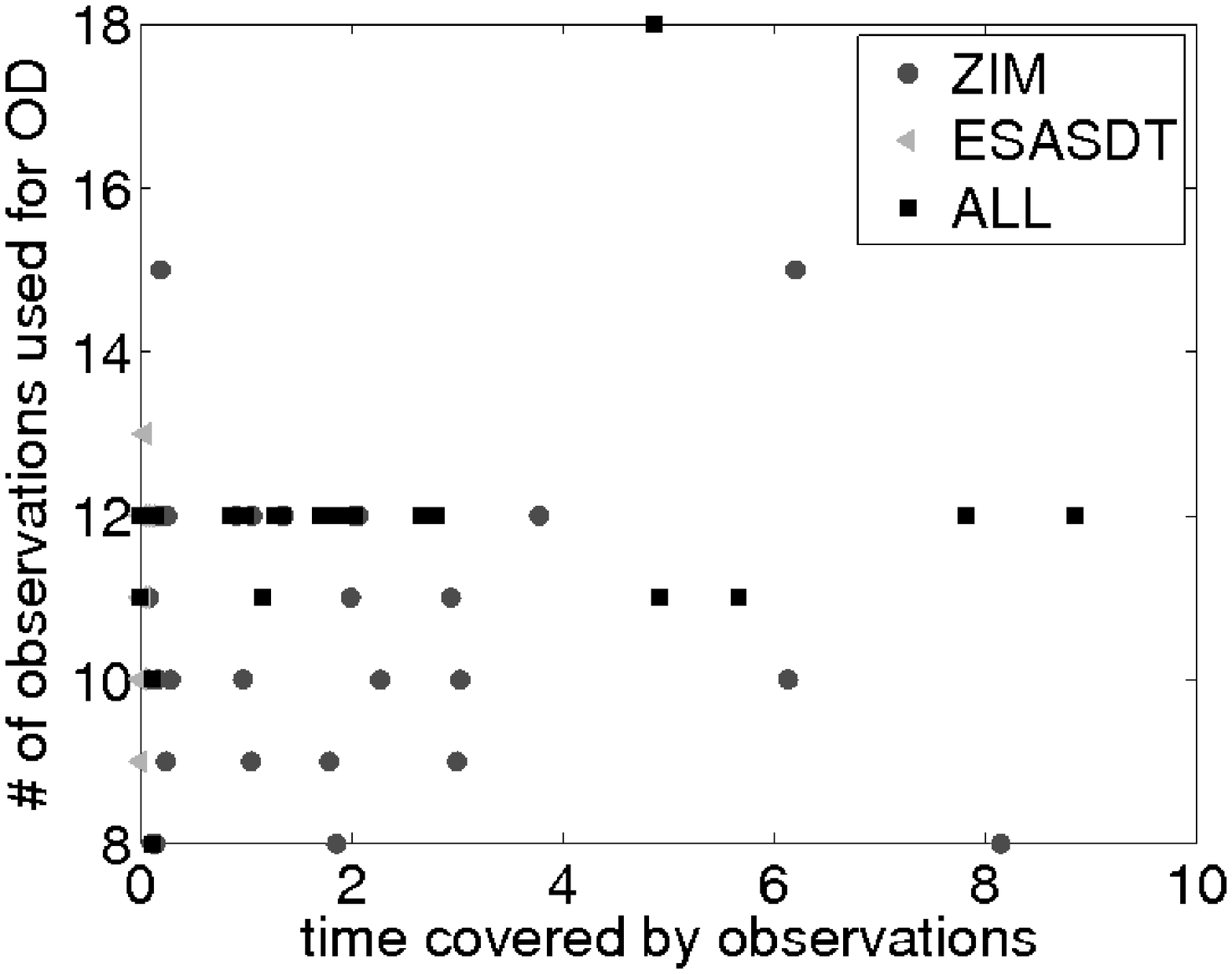}}
  \subfloat[]{\label{covnoE06327E}\includegraphics[width=0.25\textwidth, height=0.13\textheight]{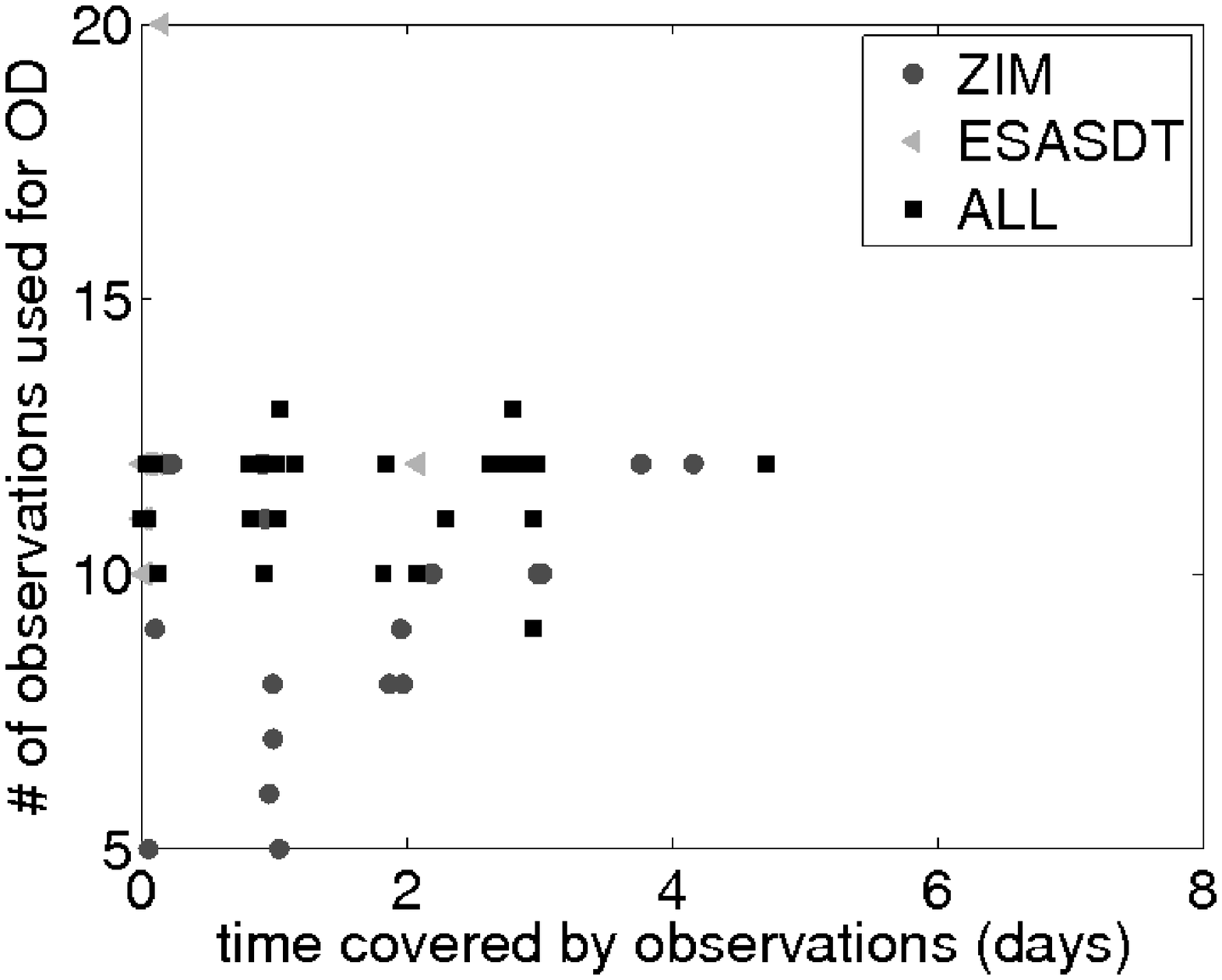}}
  \subfloat[]{\label{covnoE08241A}\includegraphics[width=0.25\textwidth, height=0.13\textheight]{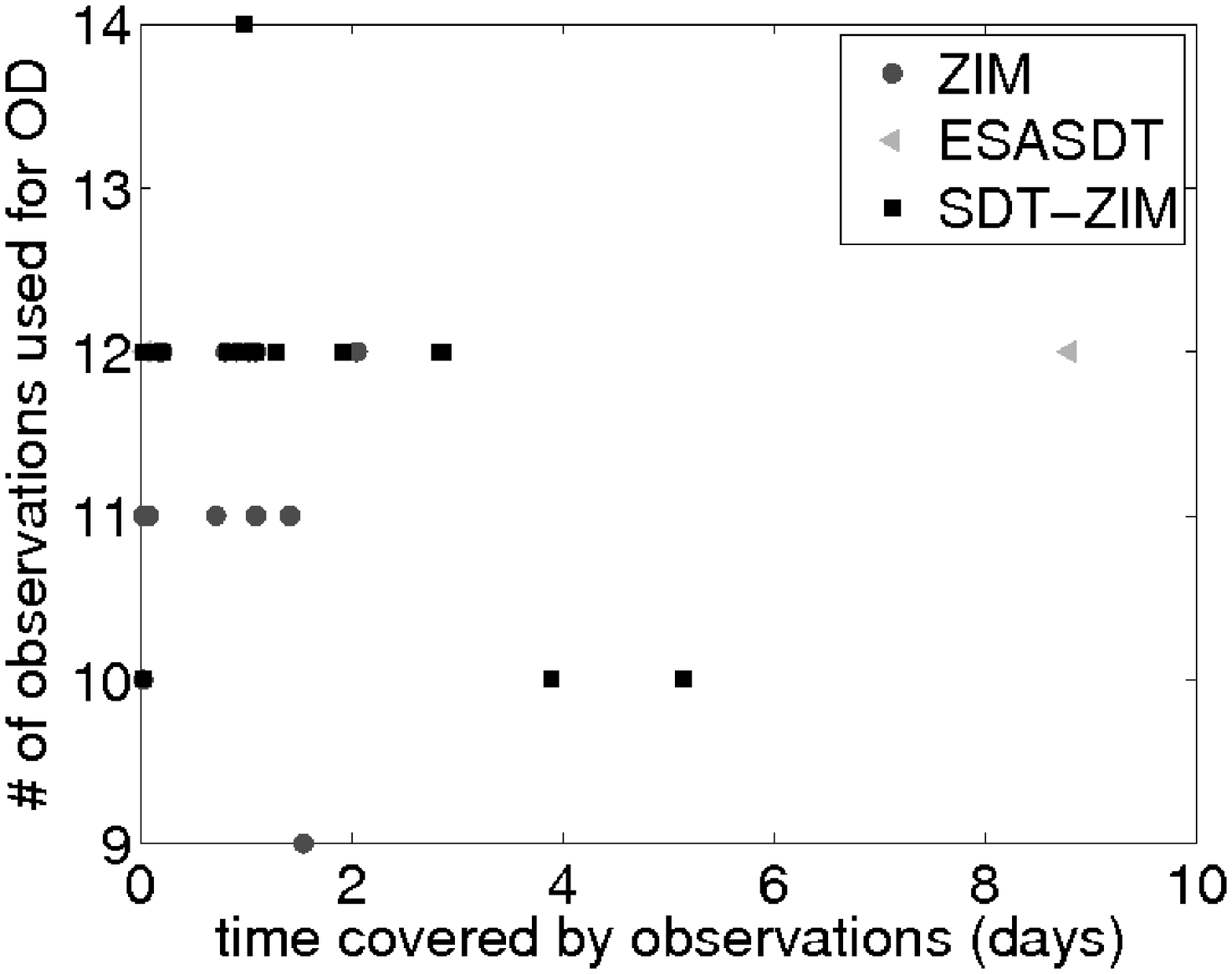}}
  \caption{\bf Time interval covered by the observations within the sets as a function of the number of observations used for orbit determination for object (a) E03174A, (b) E06321D, (c) E06327E and (d) E08241A.}
  \label{timecovnoobs}
\end{figure*}
\\
Orbits were determined for different spacings of two observation sets stemming
a) from one observation site only and b) from different sites. In the first
case, the observations either stem from ZIMLAT or from ESASDT only. In the
second case, not only the observations of ZIMLAT and ESASDT were combined
but also observations of the ISON network, if available. When observations from different sites are used in orbit
determination, the distribution is either that the first set of observations
stems from one site and the second from another, or that there are
observations from different sites at similar epochs used within the first
and/or the last set of observations or a mixture of those options. In the
figures the label \textit{ALL} is applied, when observations of ZIMLAT
(labeled \textit{ZIM}), the ESASDT and of the ISON network are combined; the
label \textit{SDT-ZIM} is applied, if only the observations of ZIMLAT and the
ESASDT are used. The distances between the observations and the ephemerides of the predicted
orbits of the four objects for a prediction interval of 50 days after the
last observation used for orbit determination were determined. The distances
were averaged and a mean value and standard deviation was calculated. Between six and 50 single distances between ephemerides and observations were averaged.\\
\\
The predicted ephemeris positions are compared to the optical angle-only observations, which were not used in orbit
 determination. Angular distances are determined on the celestial sphere. The
 observation used for the comparison stem from ZIMLAT and ESASDT and serve as
 ground truth. Calibration measurements with high accuracy ephemerides of
 Global Navigation Satellite System (GNSS)
 satellites provided by International GNSS Service (IGS) showed an accuracy of the measurements of
 ZIMLAT and ESASDT of below one arcsecond. That the further observations in
 fact belong to the same object is validated via an orbit determination with
 both the observations used in the original sparse data orbit determination
 and the observations, which they were compared to. An orbit determination
 with a root-mean-square of below two arcseconds is a reliable tool to
 associate observations of this accuracy of the same object to each other, as shown with
 cluster observations in \citet{Musci05b}.

\subsection{Results}

In Fig. \ref{resE03174A} the angular distance
between predicted and observed position are displayed as a function of the time
interval between the first and the last observation, which were used
in orbit determination. Displayed are the mean values and the standard
deviations of the angular distances of the single orbits. The mean value and
standard deviations are determined with the single angular distances of predicted position to observed ones, all within 50 days since orbit determination.\\
\\
Figure \ref{resE03174A} shows that the angular distances are in general very
small. The vast majority of the determined orbits even produce distances
smaller than 0.6 degrees. Except for the first object, each object also shows some outliers, with larger
angular distances. These larger distances also tend to show larger standard
deviations. The value of the angular distances seems to be, at least in this setup,
quite independent of how large the difference between the first and the last
observations of the fit interval is. Moreover, Figure \ref{resE03174A} also
shows that there is no significant difference in using observations only from
one observation site for orbit determination or using observations from
different sites. It could not be shown that the latter approach is more
advantageous for orbit determination. Different observation sites still have
advantages in terms of availability, weather conditions, which results in a
larger amount of observations, which are available. In
Fig. \ref{res}, the root mean square of the orbit determinations is shown,
which were used for the prediction, as a function of the angular distance. No
trend is visible, all orbits, which were determined had a small root mean
square of below three arcseconds.
\\
\\
In Fig. \ref{noobs}, the angular distances are
displayed as a function of the actual number of single observations that
entered orbit determination. It can be seen that no strong correlation is
visible between the actual number of observations used and the value for the
distances. \\
\\
To find a measure for the true anomaly distribution, an anomaly distribution measure
$f_{\text{ano}}$ was defined: It would be ideal to distribute all $n$
observations equally spaced with an angle of $2\pi/n$ between each observation. The deviation from
this ideal distribution is determined and normalized with the number of
observations. The smaller $f_{\text{ano}}$, the better distributed are the
observations in anomaly.
\begin{equation}
f_{\text{ano}}=\frac{1}{n}\sqrt{\sum_{i=1}^{n-1}\big(\frac{2\pi}{n}-(a_{i+1}-a_i)\big)^2+\big(\frac{2\pi}{n}-(a_1+2\pi-a_n)\big)^2},
\end{equation}
where as $n$ is the number of observations and $a_i$ with $i=1,..,n$ are the
anomalies of the single observations, in ascending anomaly order. The angular distances as a function of $f_{\text{ano}}$ are
displayed in Fig.\,\ref{anoarc}. There is no clear correlation between the
$f_{\text{ano}}$ and the distances, as it is expected for objects with
small eccentricities. Object E06327E, with the highest eccentricity of e=0.06, has the strongest correlation with $f_{\text{ano}}$.\\
\\
The crucial factor however, seems to be the time interval
covered by observations itself within the sets. In Fig \ref{timecov}, the angular distances
are displayed as a function of the time interval covered \textit{within} the
two sets used in the beginning and the end of the fit interval, without the
time gap in between the two sets. A strong correlation is
visible. Fig. \ref{timecovnoobs} shows that there is no strong correlation
between the number of used observations and the time interval covered within
the sets. For example for  the ESASDT observation strategy, primarily densely
spaced observations are available.
\\
\\
The investigation of
the data displayed in Fig.\,\ref{timecovnoobs} showed that a coverage of at
least 1.2 hours for both sets together seems to be necessary, in order to gain
an orbit which allows to safely re-detect the investigated objects in more
than 90 percent of all cases with a field of view of one square degree, that
is to have an accuracy of below 0.5 degrees.
\begin{figure*}\textcolor{white}{[ht]}
  \centering
  \subfloat[]{\includegraphics[width=0.25\textwidth, height=0.13\textheight]{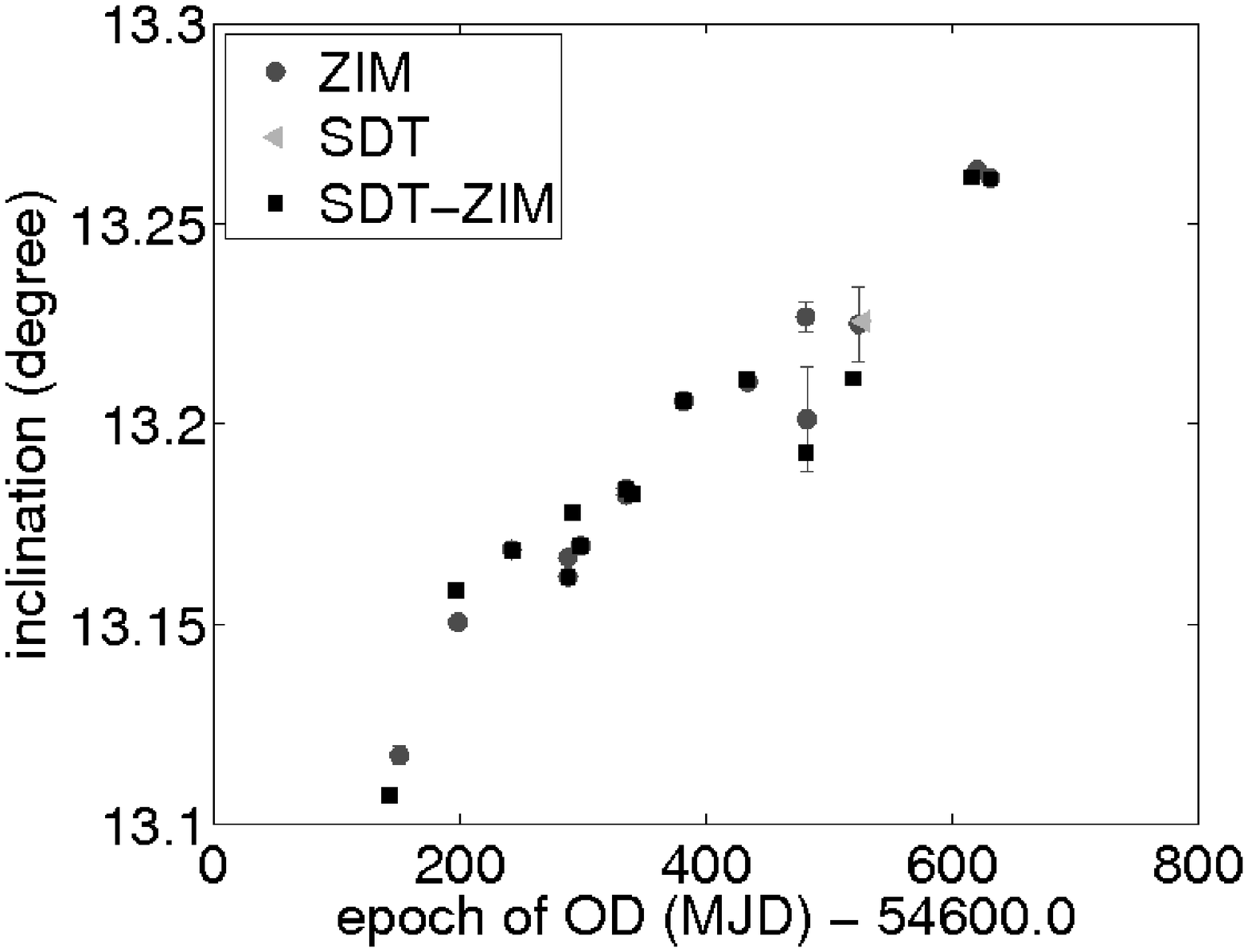}}  
  \subfloat[]{\includegraphics[width=0.25\textwidth, height=0.13\textheight]{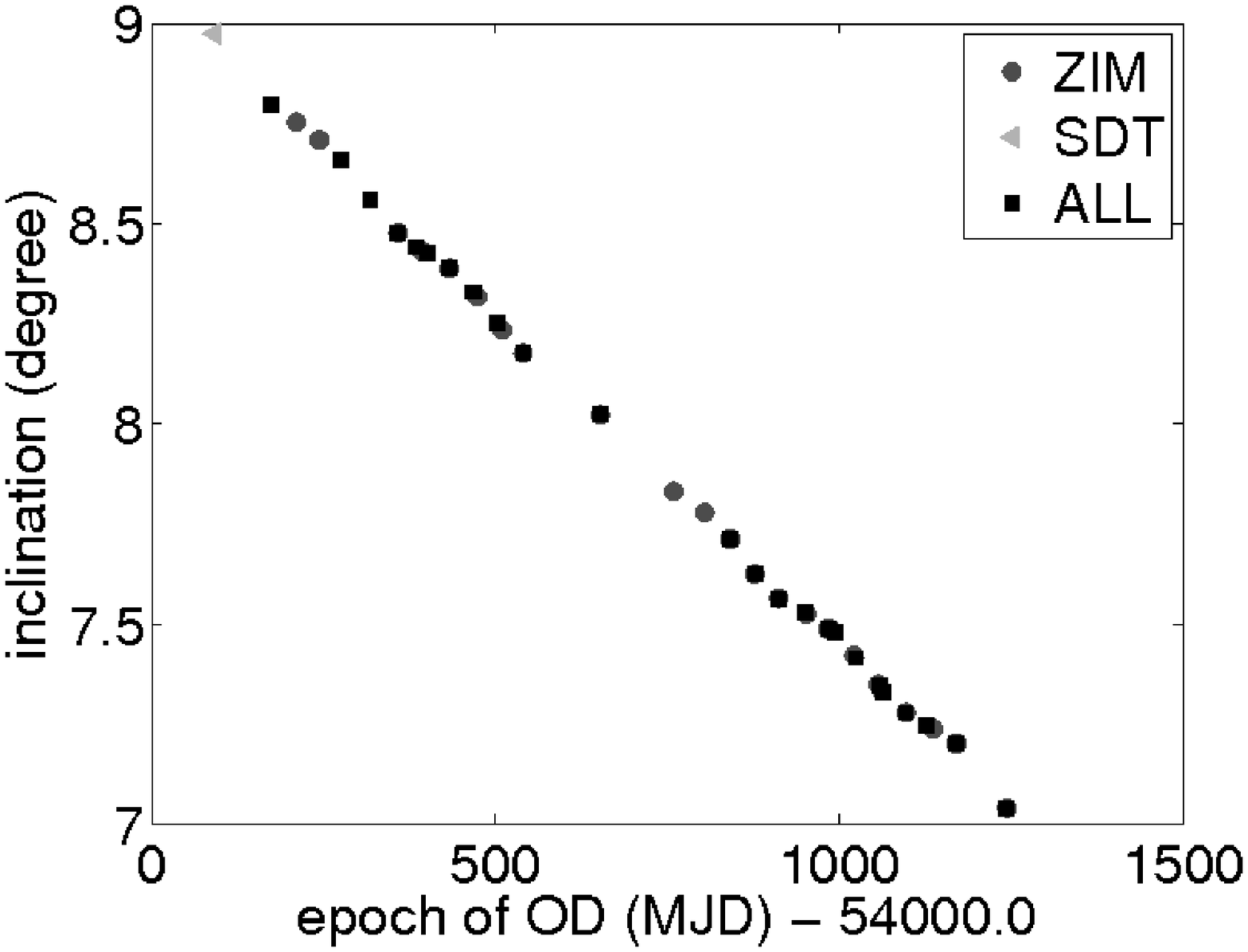}}
  \subfloat[]{\includegraphics[width=0.25\textwidth, height=0.13\textheight]{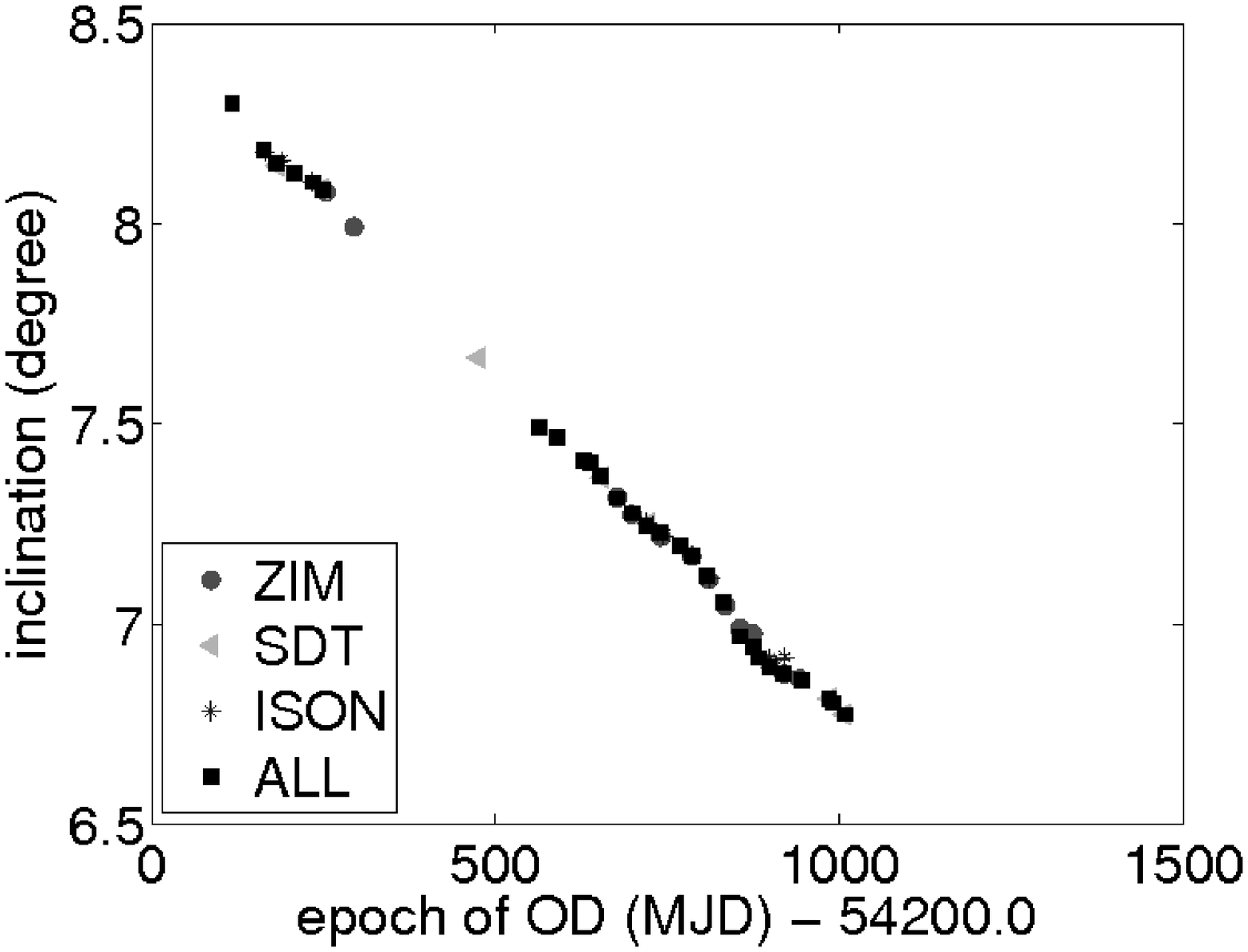}}\\
  \subfloat[]{\includegraphics[width=0.25\textwidth, height=0.13\textheight]{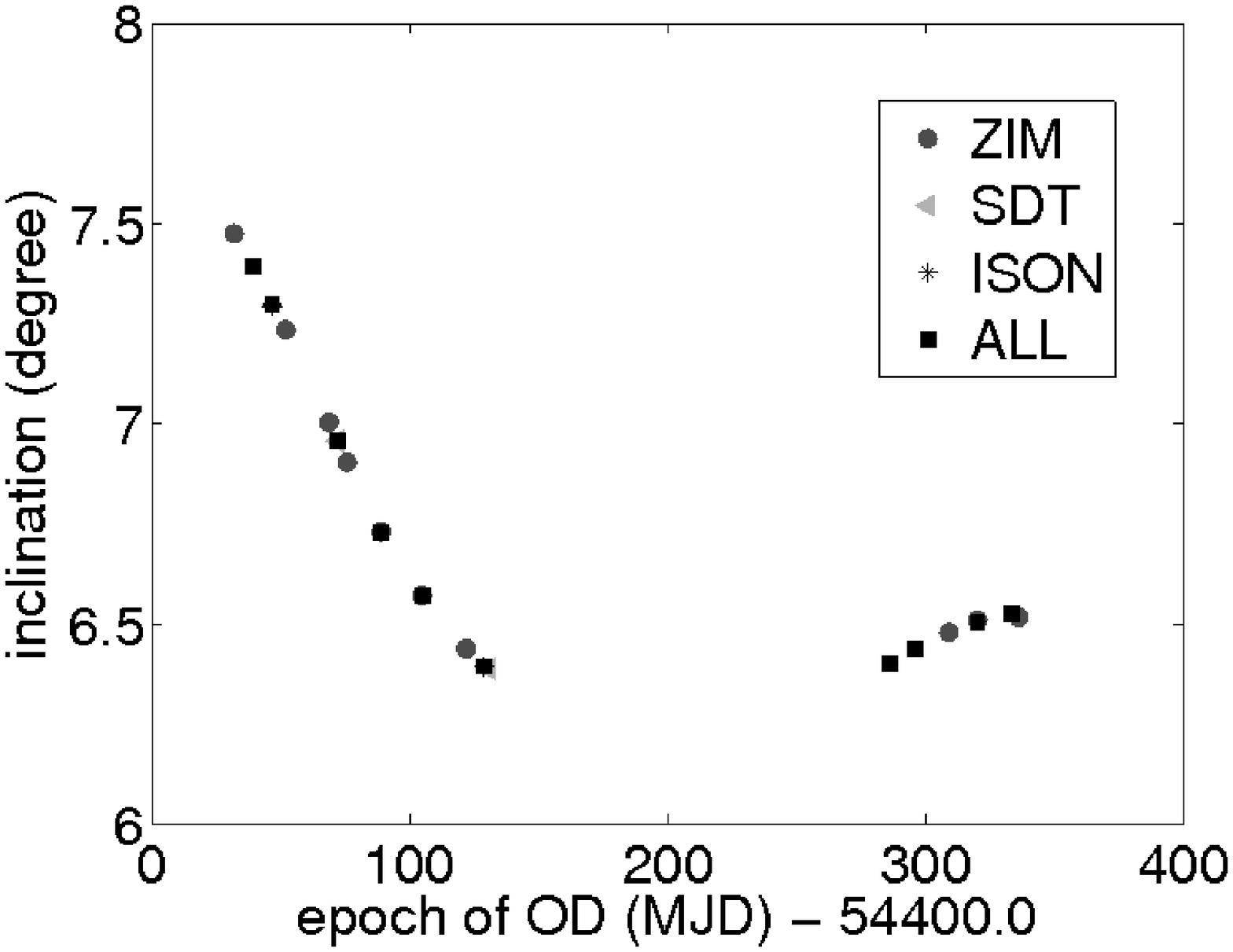}}
  \subfloat[]{\includegraphics[width=0.25\textwidth, height=0.13\textheight]{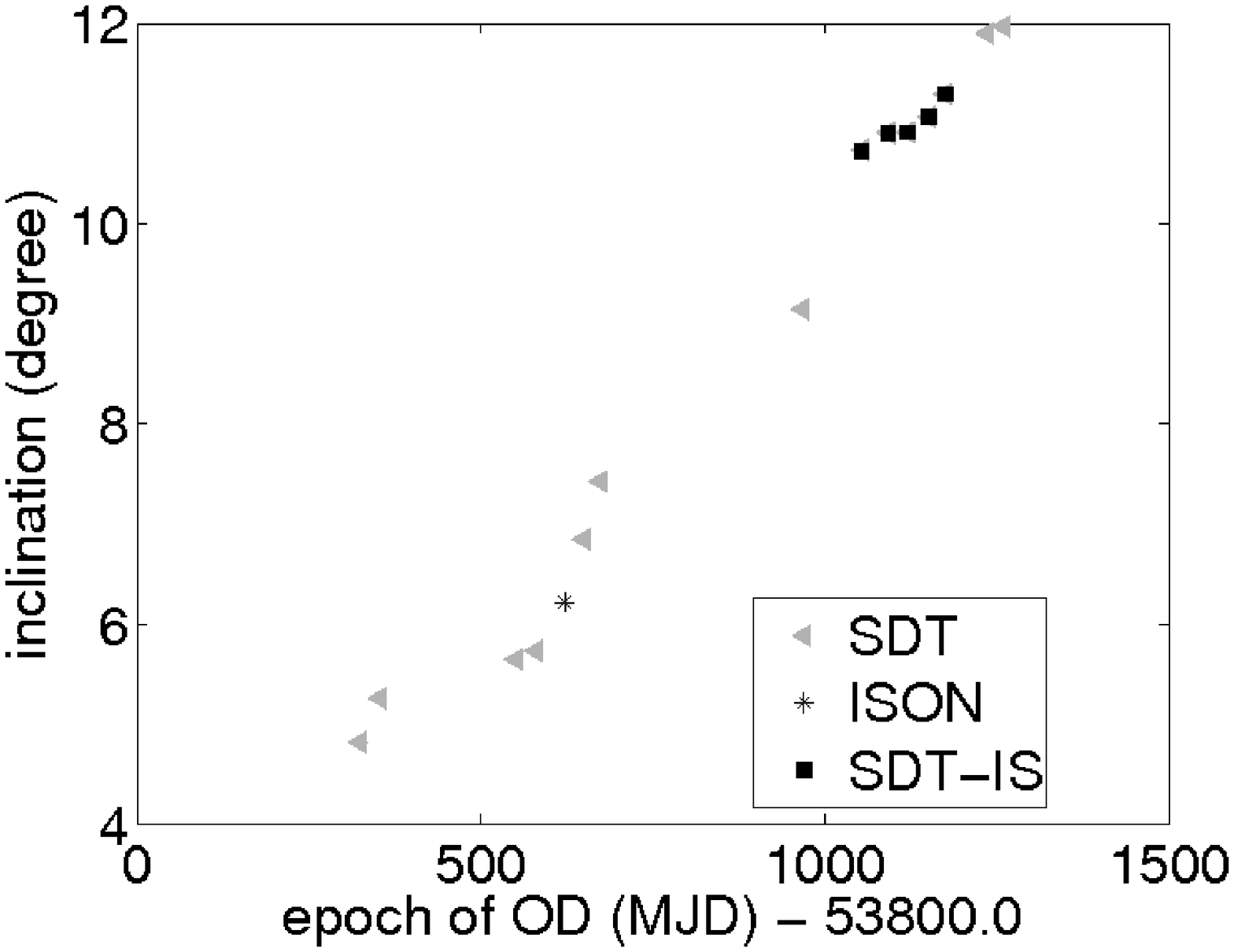}}
  \caption{Inclination as a function of time for orbits of the object (a)E08241A, (b) E06321D, (c) E07194A, (d) E07308B, (e) E06293A.}
  \label{inc}
\end{figure*} 
\begin{figure*}\textcolor{white}{[h]}
  \centering
  \subfloat[]{\includegraphics[width=0.25\textwidth, height=0.13\textheight]{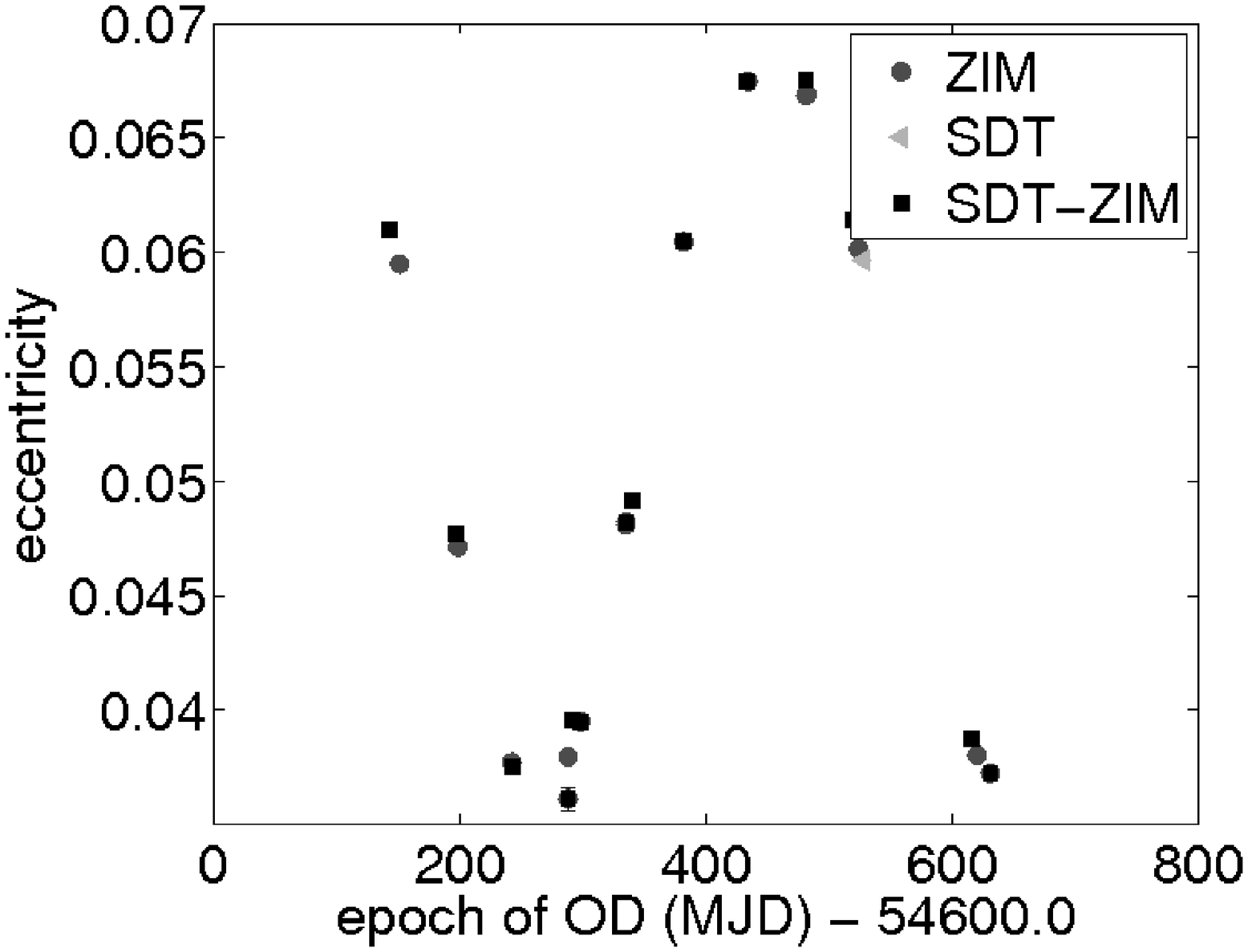}}  
  \subfloat[]{\includegraphics[width=0.25\textwidth, height=0.13\textheight]{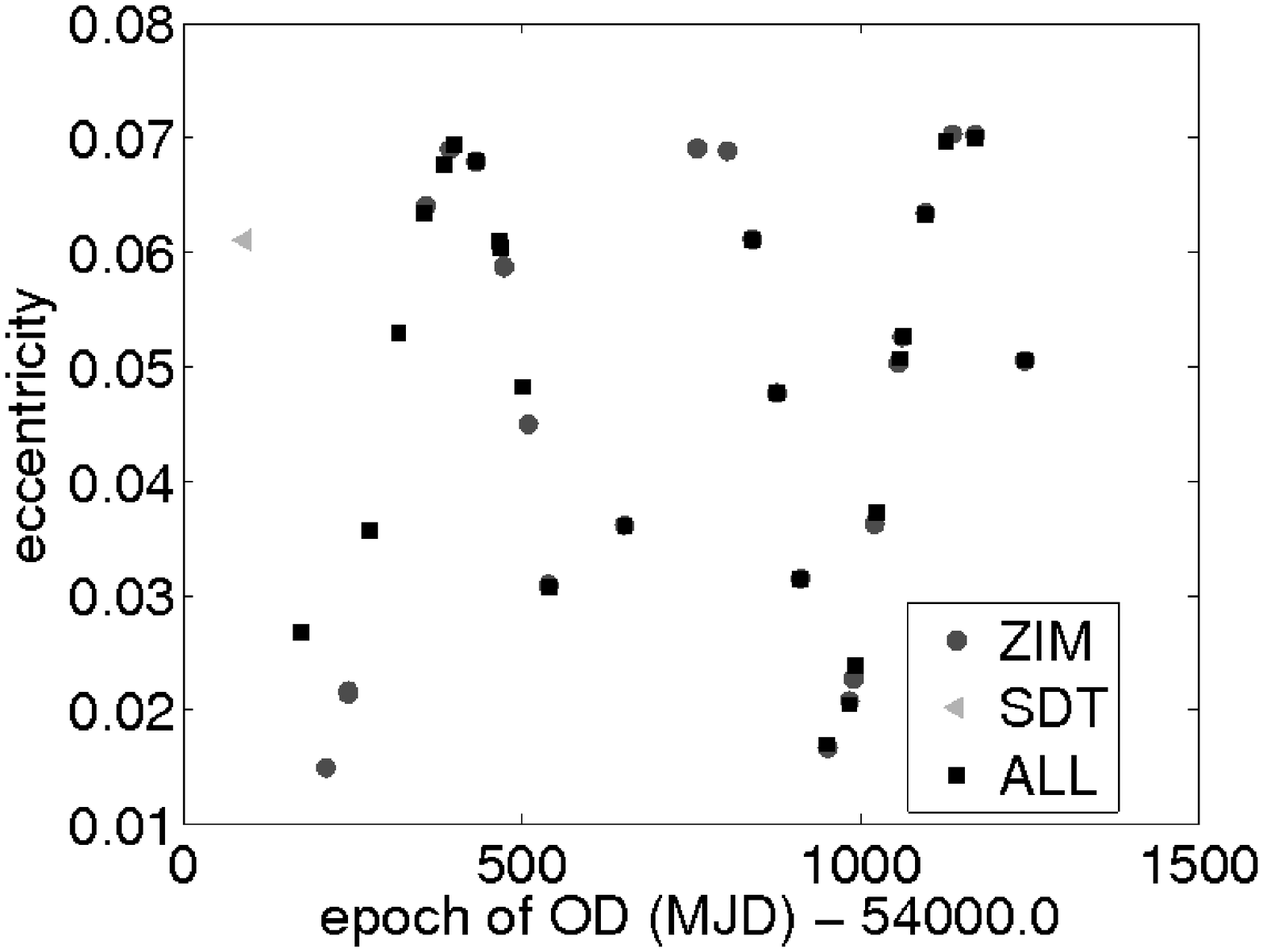}}
  \subfloat[]{\includegraphics[width=0.25\textwidth, height=0.13\textheight]{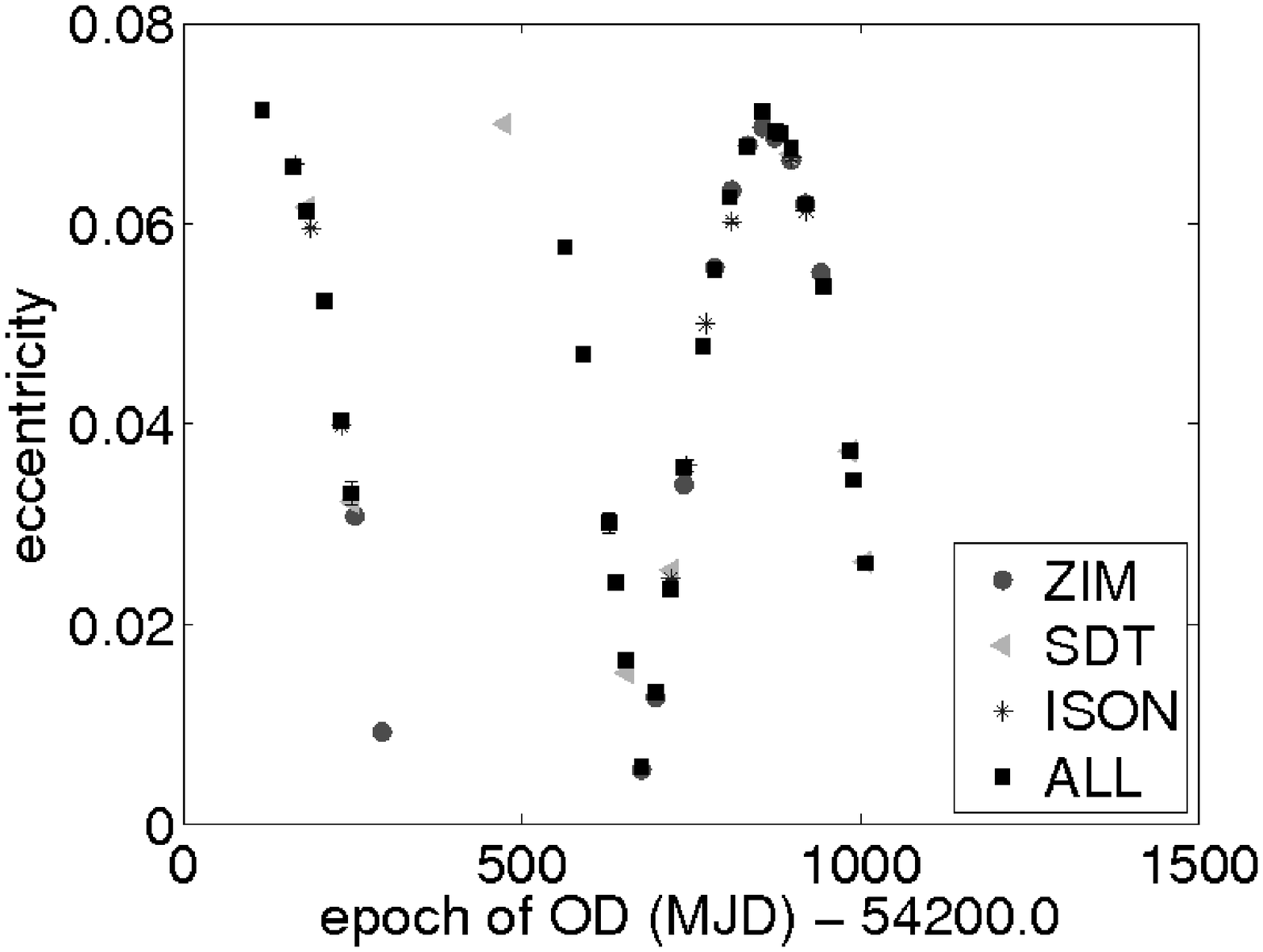}}\\
  \subfloat[]{\includegraphics[width=0.25\textwidth, height=0.13\textheight]{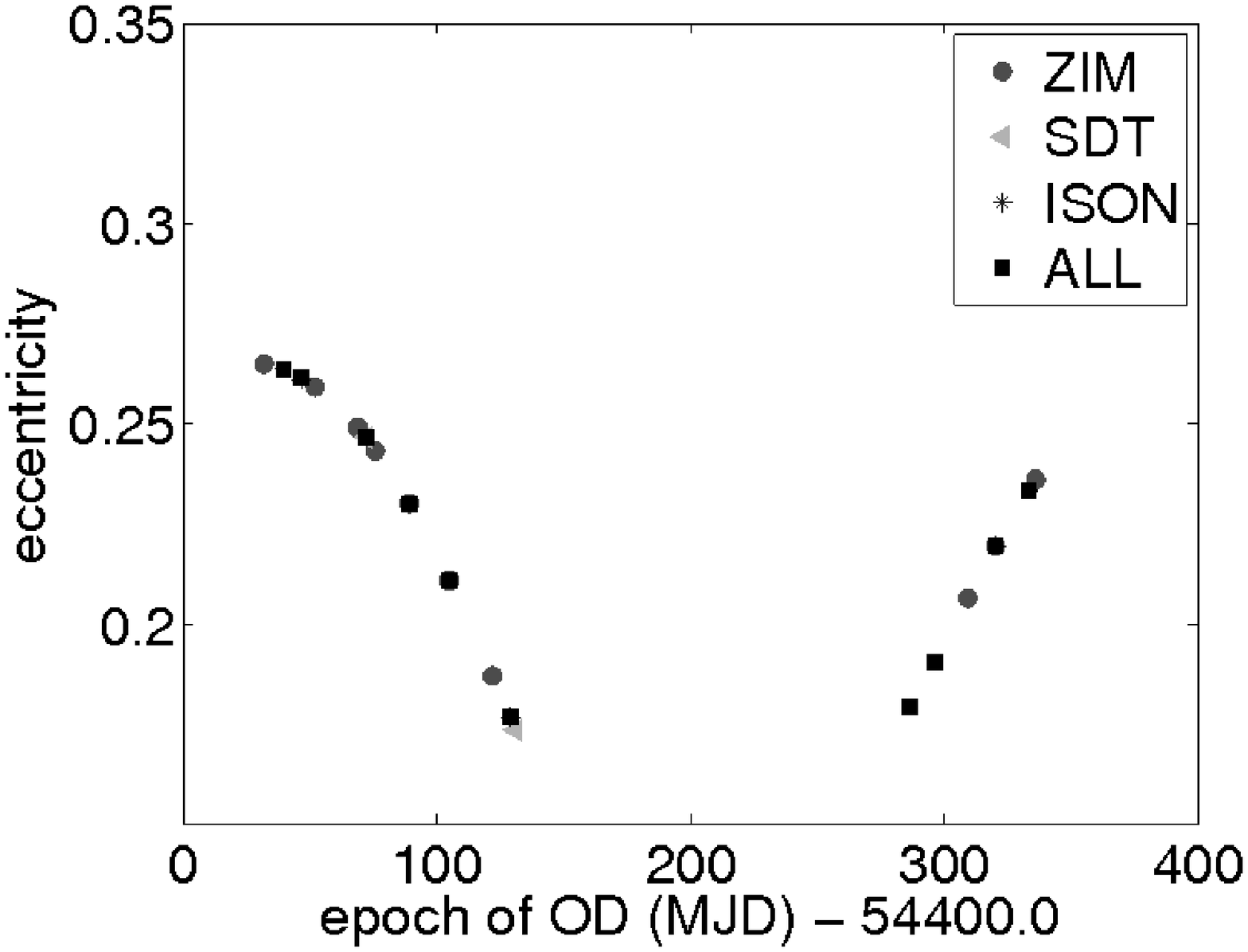}}
  \subfloat[]{\includegraphics[width=0.25\textwidth, height=0.13\textheight]{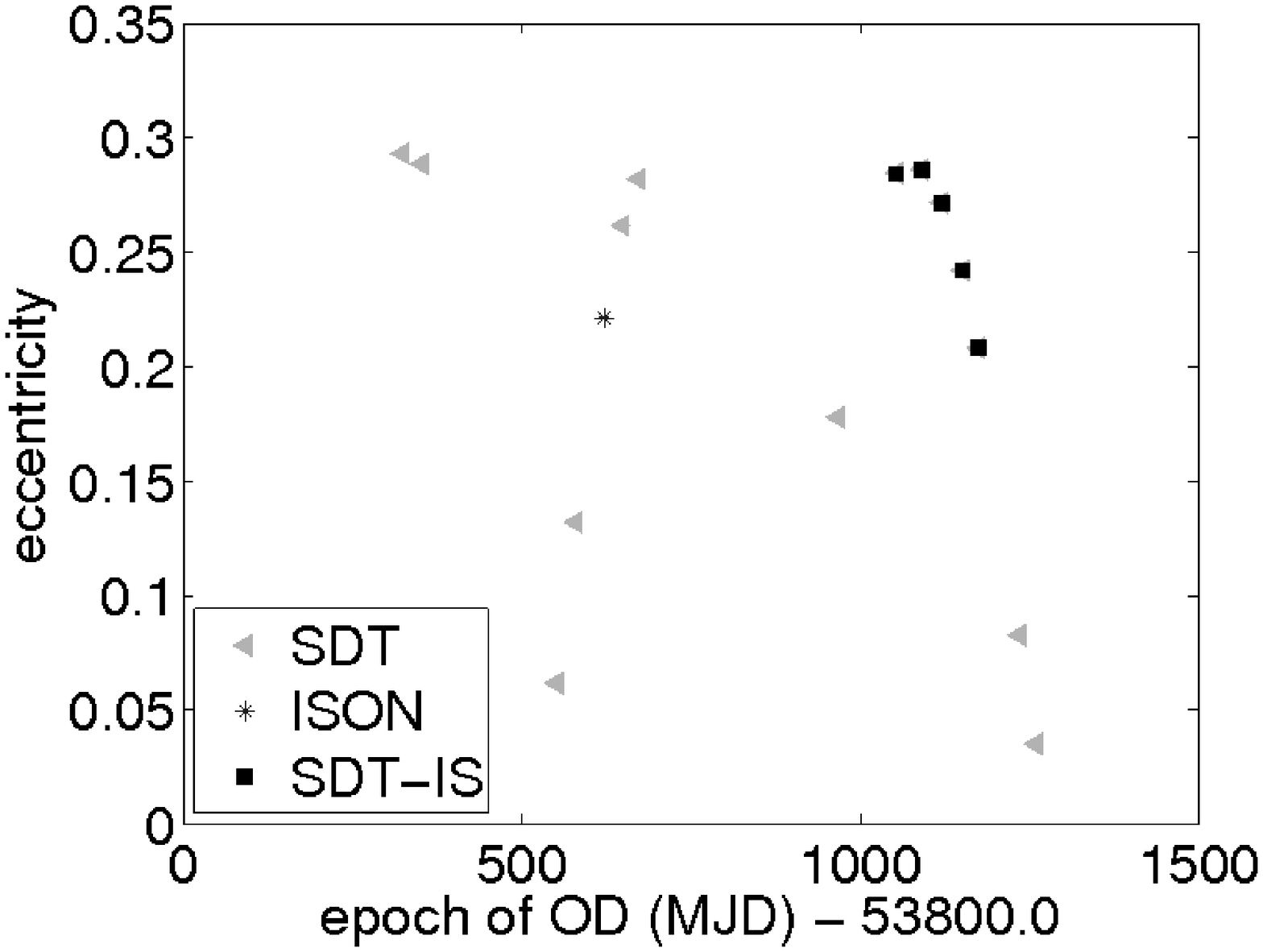}}
  \caption{Eccentricity as a function of time for orbits of the object (a) E08241A, (b) E06321D, (c) E07194A, (d) E07308B, (e) E06293A.}
  \label{ecc}
\end{figure*}

\section{Investigation of HAMR Objects in Sparse Data Setup}
The dynamical properties of HAMR objects were studied in the normalized sparse data setup
established in the previous section. Orbits are determined with two
observation sets only. The sets consist of four to eight observations
each. The observations are required to span at least a time interval of 1.2 hours
within the sets and need to be well spread over the anomaly for the objects in
orbits with a high eccentricity. The total fit interval for orbit
determination ranges between 10 and 120 days. As shown in the previous section
the comparability of the orbits do not seem to be dependent on these ranges. \\
\\
The orbits were first determined with observations from one observation site only, then with observations from different sites in the setup mentioned above. The observations used in this investigation stem from the ESASDT, ZIMLAT, and from several telescopes of the ISON network.\\
\\
\subsection{Selected Objects}
Five objects were selected for a detailed investigation. All objects were
discovered and first detected by the AIUB and are not listed in the USSTRATCOM
catalogue. All objects are faint debris objects. They were tracked successfully
over several years, and no maneuvers were detected. A set of osculating
orbital elements and an average value for the apparent magnitudes are listed
in Tab.\,\ref{obj}. The two objects with the lowest AMR values, E08241A and
E06321D, which were
used in the investigation of the sparse data orbit determination, are used here
again.\\
\\
\setlength{\arrayrulewidth}{0.5pt}
\setlength{\doublerulesep}{0.6pt} 
\begin{table}
\begin{center}
\caption{\bf Investigated HAMR objects: Internal name, epoch (MJD), eccentricity, inclination (deg), semi-major axis (km), area to mass ratio ($m^2/kg$) and apparent magnitude (mag).}
\begin{tabular}{lllllll}
\small NAME &\small Epoch &\small \textit{a} &\small \textit{e} &\small \textit{i} &\small AMR &\small Mag\\
\vspace{-0.4cm}\\
\hline
\vspace{-0.3cm}\\
\small 
\small E08241A&55213.0&41600&0.041&13.26&1.24&16.1\\
\small E06321D&55275.9&41400&0.035&7.00&2.29&15.3\\
\small E07194A&54877.0&40900&0.005&7.31&3.37&16.8\\
\small E07308B&54416.0&35600&0.264&7.63&8.83&15.8\\
\small E06293A&54951.0&40200&0.245&11.06&15.41&16.8\\
\end{tabular}
\label{obj}
\vspace{-0.5cm}
\end{center}
\end{table}
\begin{figure*}\textcolor{white}{[ht]}
  \centering
  \subfloat[]{\includegraphics[width=0.25\textwidth, height=0.13\textheight]{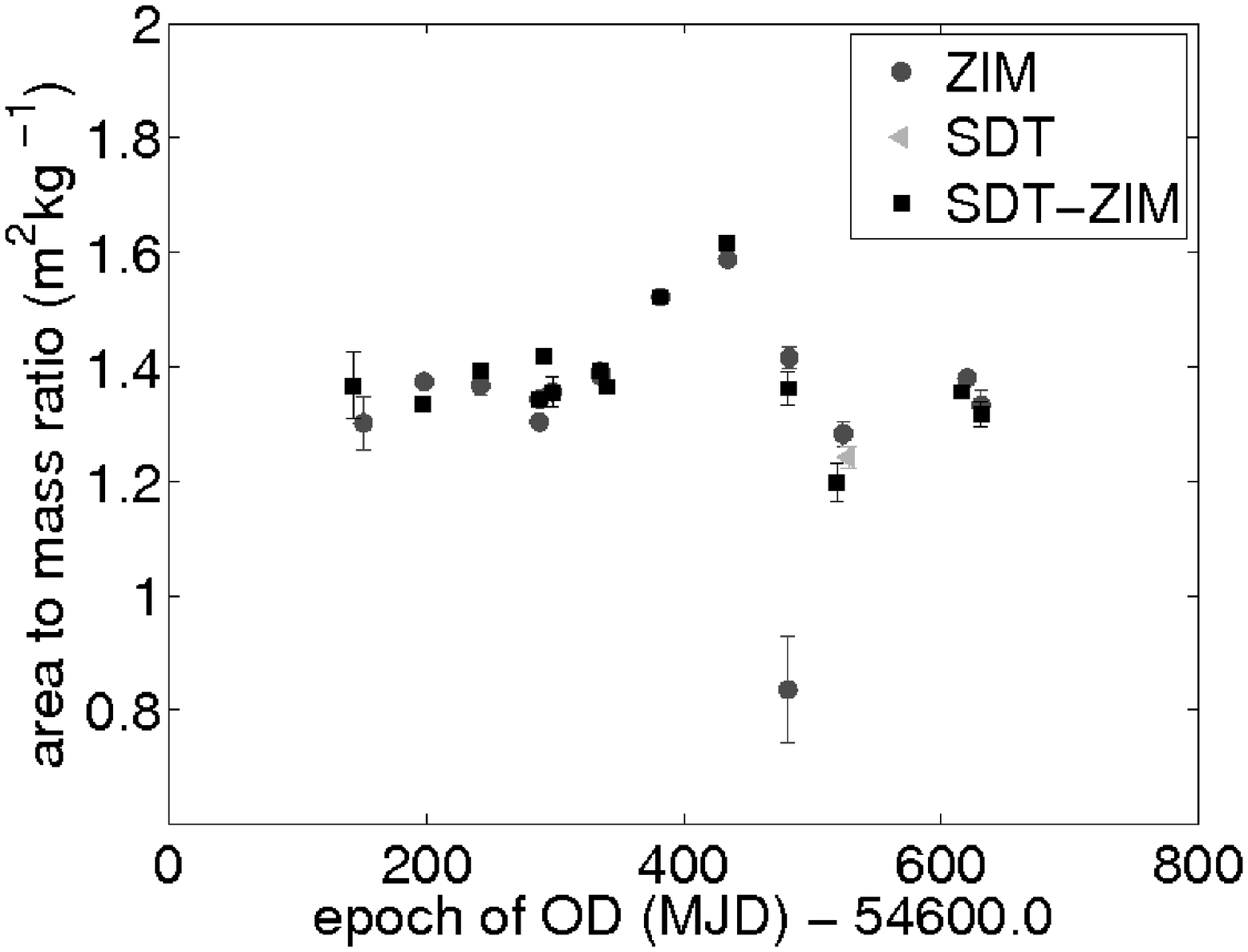}}  
  \subfloat[]{\includegraphics[width=0.25\textwidth, height=0.13\textheight]{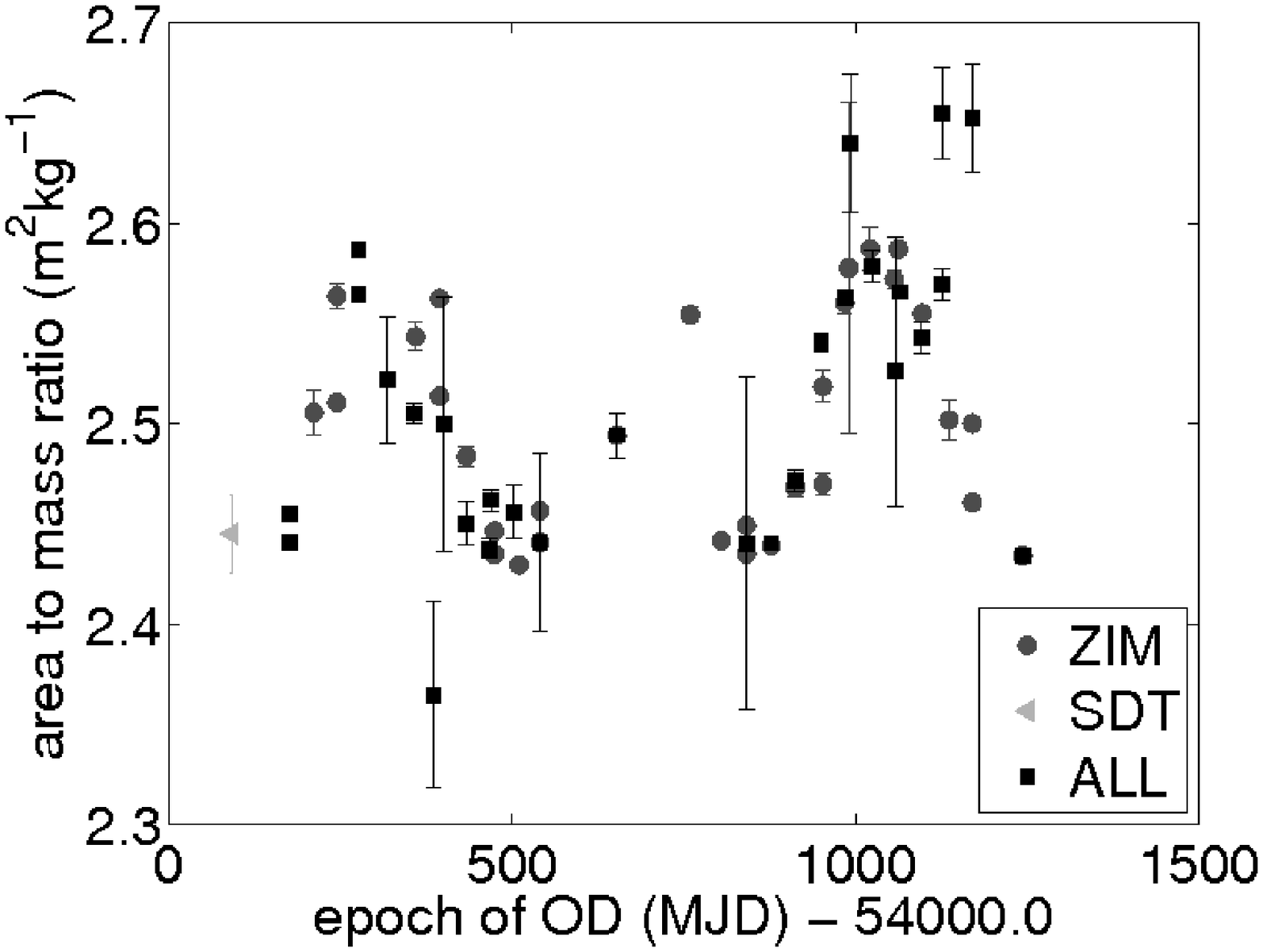}}
  \subfloat[]{\includegraphics[width=0.25\textwidth, height=0.13\textheight]{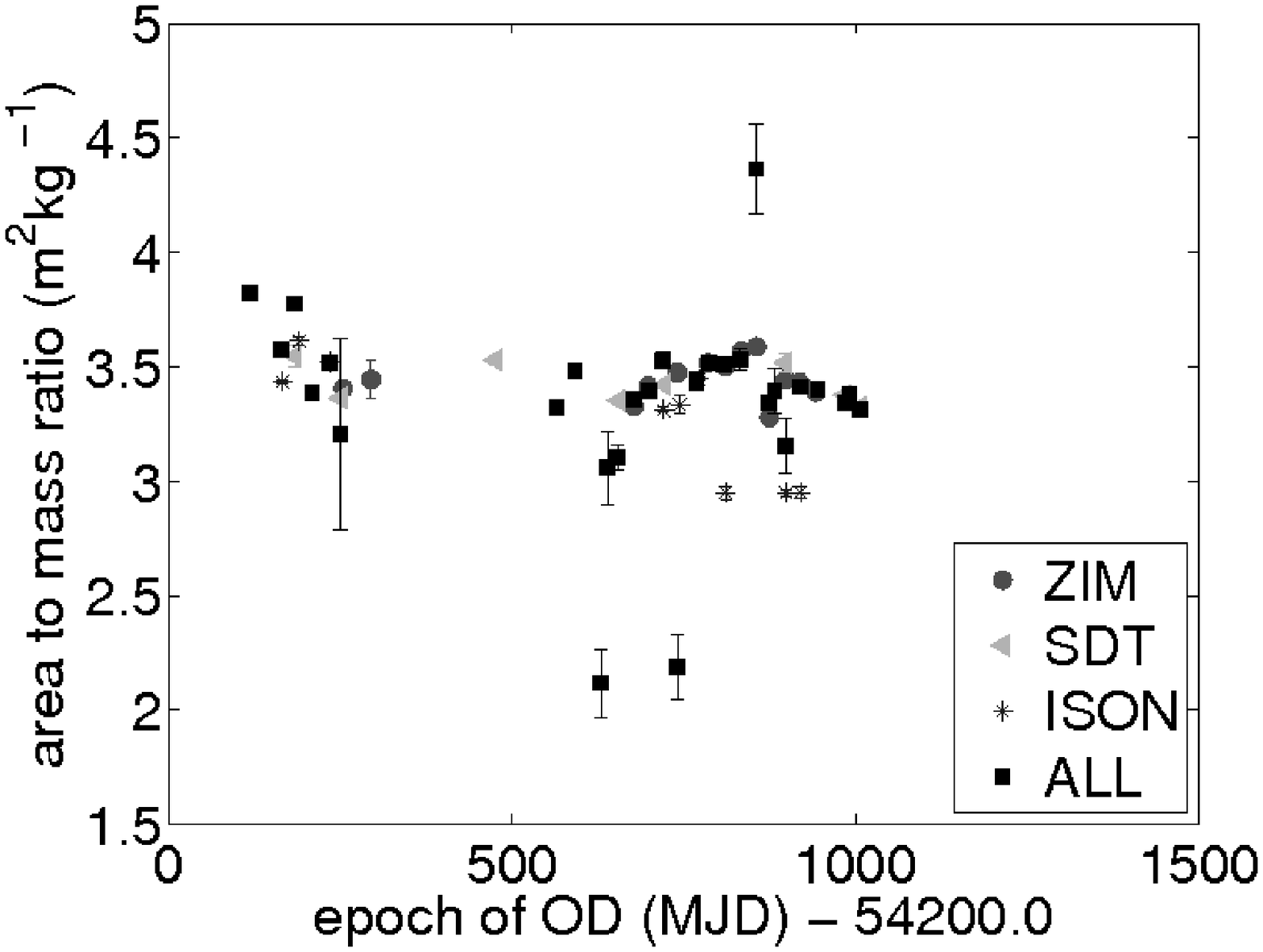}}\\
  \subfloat[]{\includegraphics[width=0.25\textwidth, height=0.13\textheight]{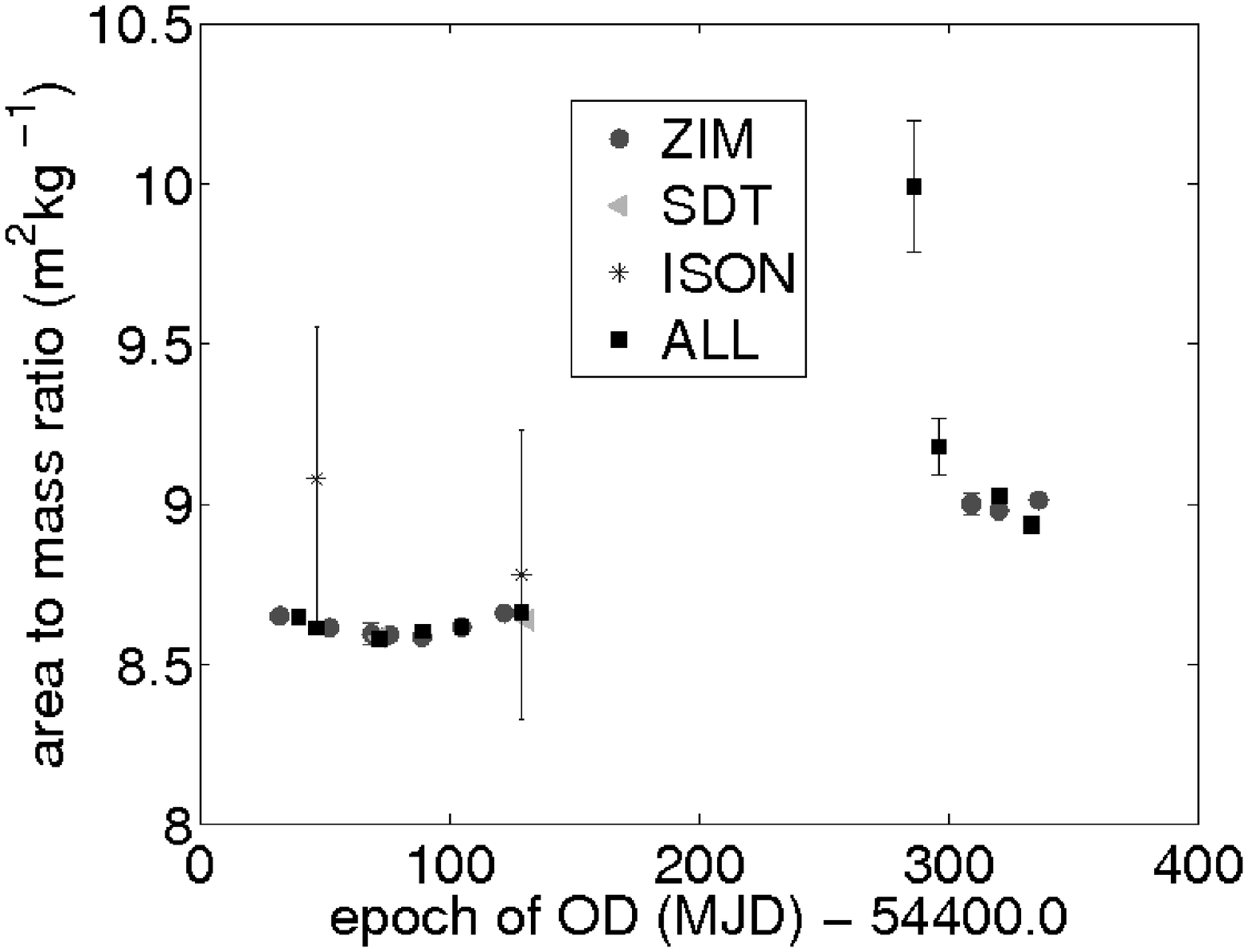}}
  \subfloat[]{\includegraphics[width=0.25\textwidth, height=0.13\textheight]{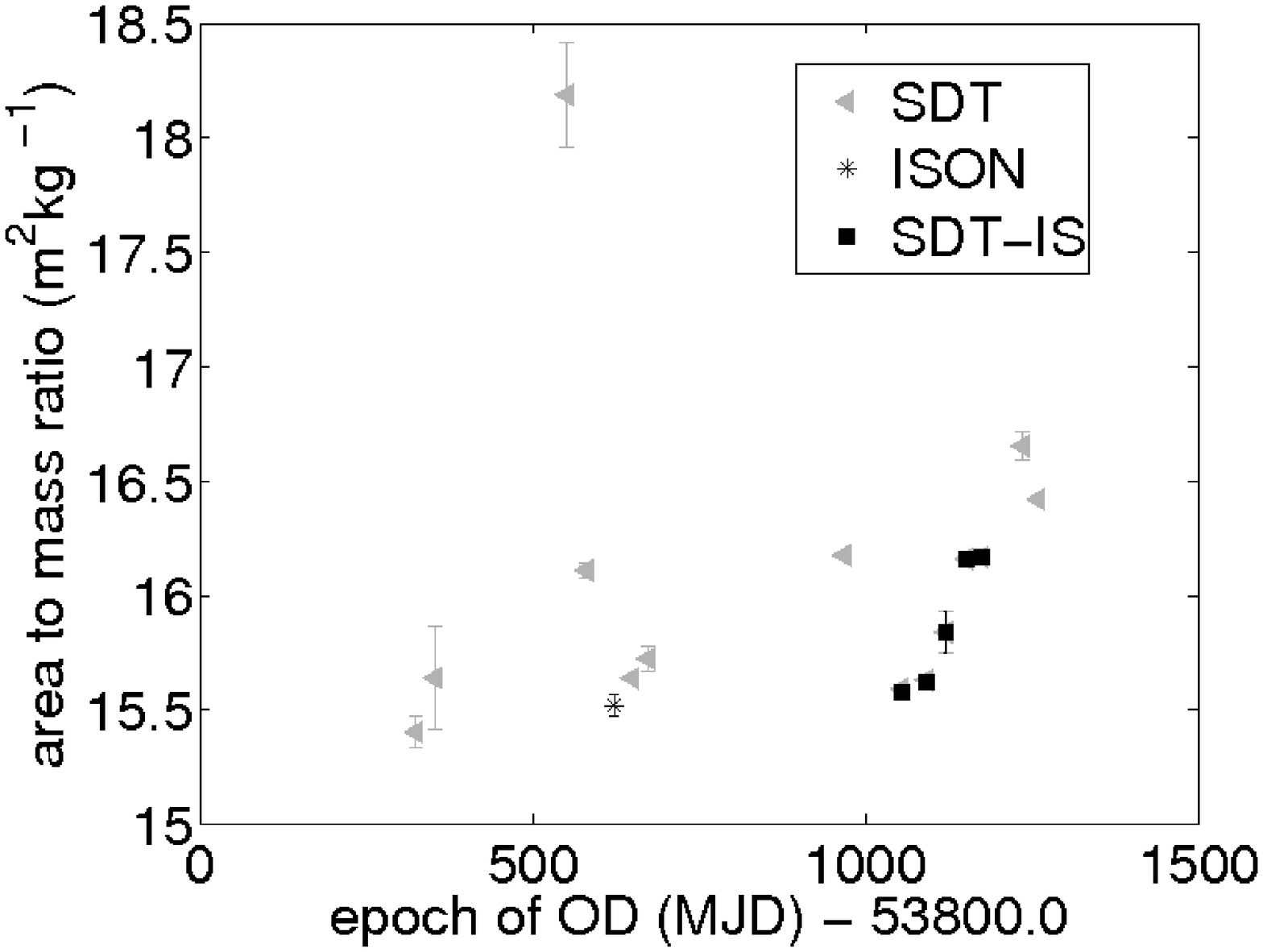}}
  \caption{AMR as a function of time for orbits of the object (a) E08241A, (b) E06321D, (c) E07194A, (d) E07308B, (e) E06293A.}
  \label{AMR}
\end{figure*} 
\begin{figure*}\textcolor{white}{[h]}
  \centering
  \subfloat[]{\includegraphics[width=0.4\textwidth]{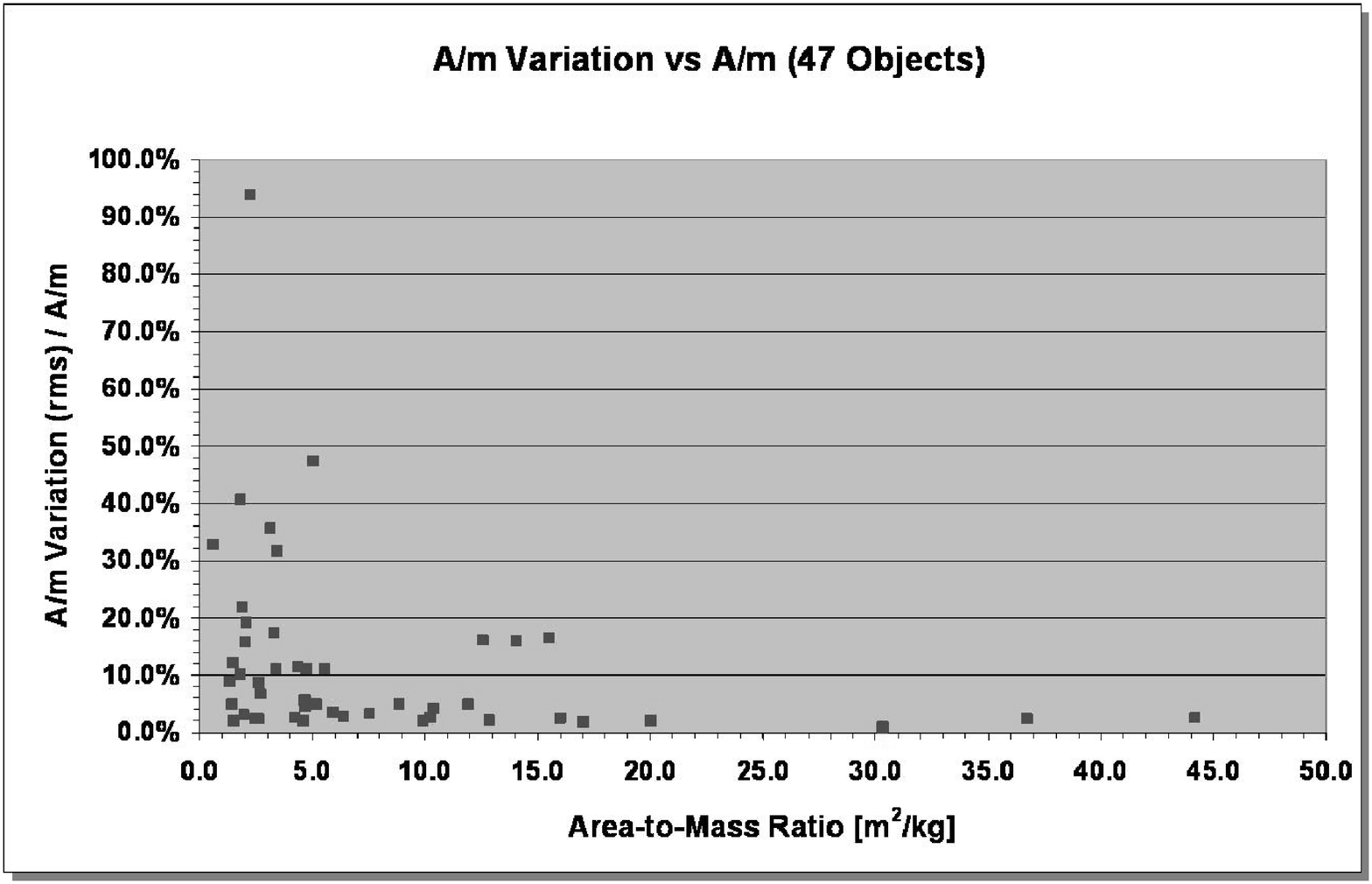}}
\hspace{0.3cm}  
  \subfloat[]{\includegraphics[width=0.4\textwidth]{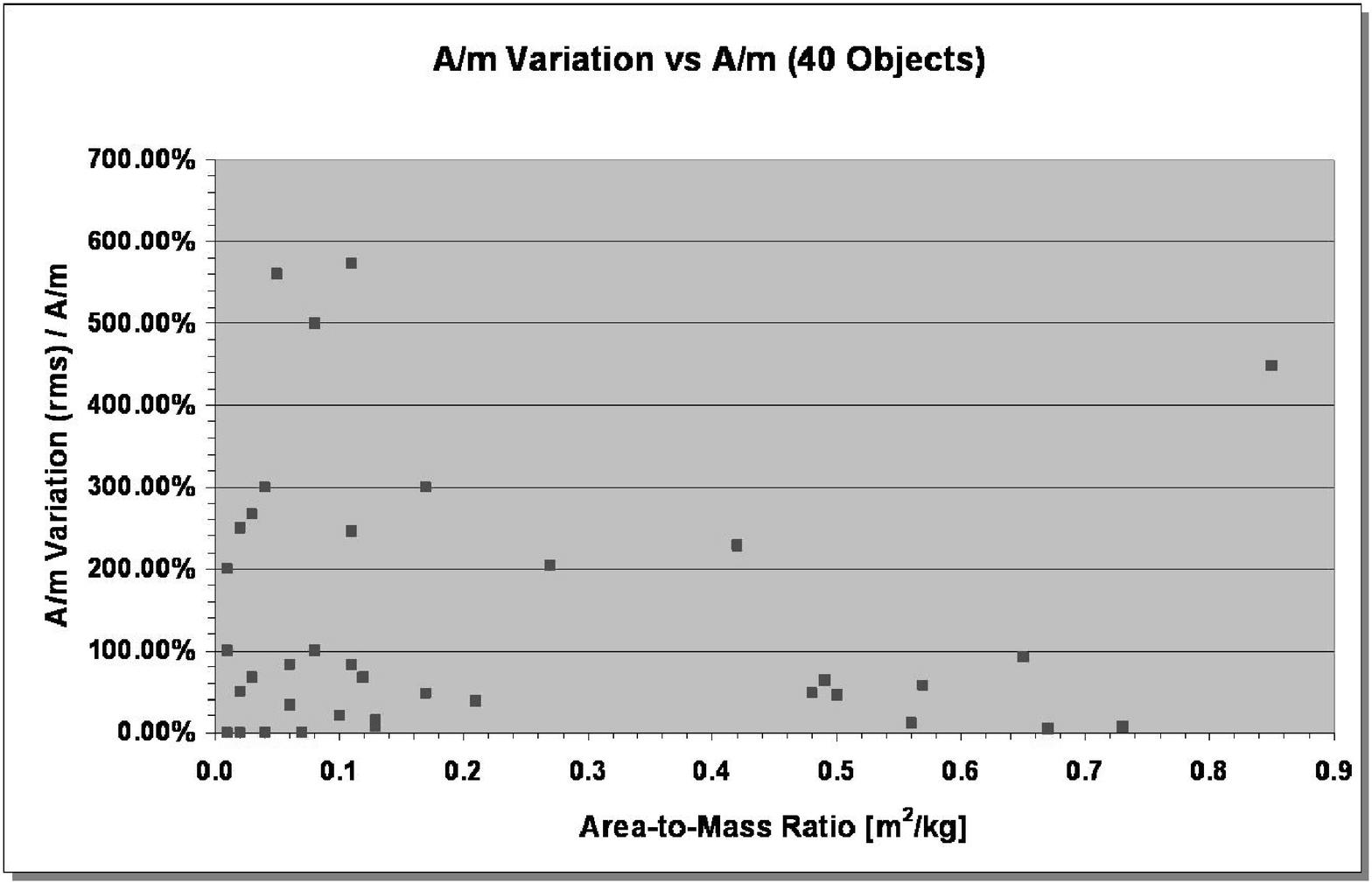}}
  \caption{Relative variation of AMR value as a function of the absolute AMR value of (a) 47 HAMR and (b) LAMR objects.}
  \label{LAMR-HAMR}
\end{figure*}
\begin{figure*}\textcolor{white}{[h]}
  \centering
  \subfloat[]{\includegraphics[width=0.25\textwidth, height=0.13\textheight]{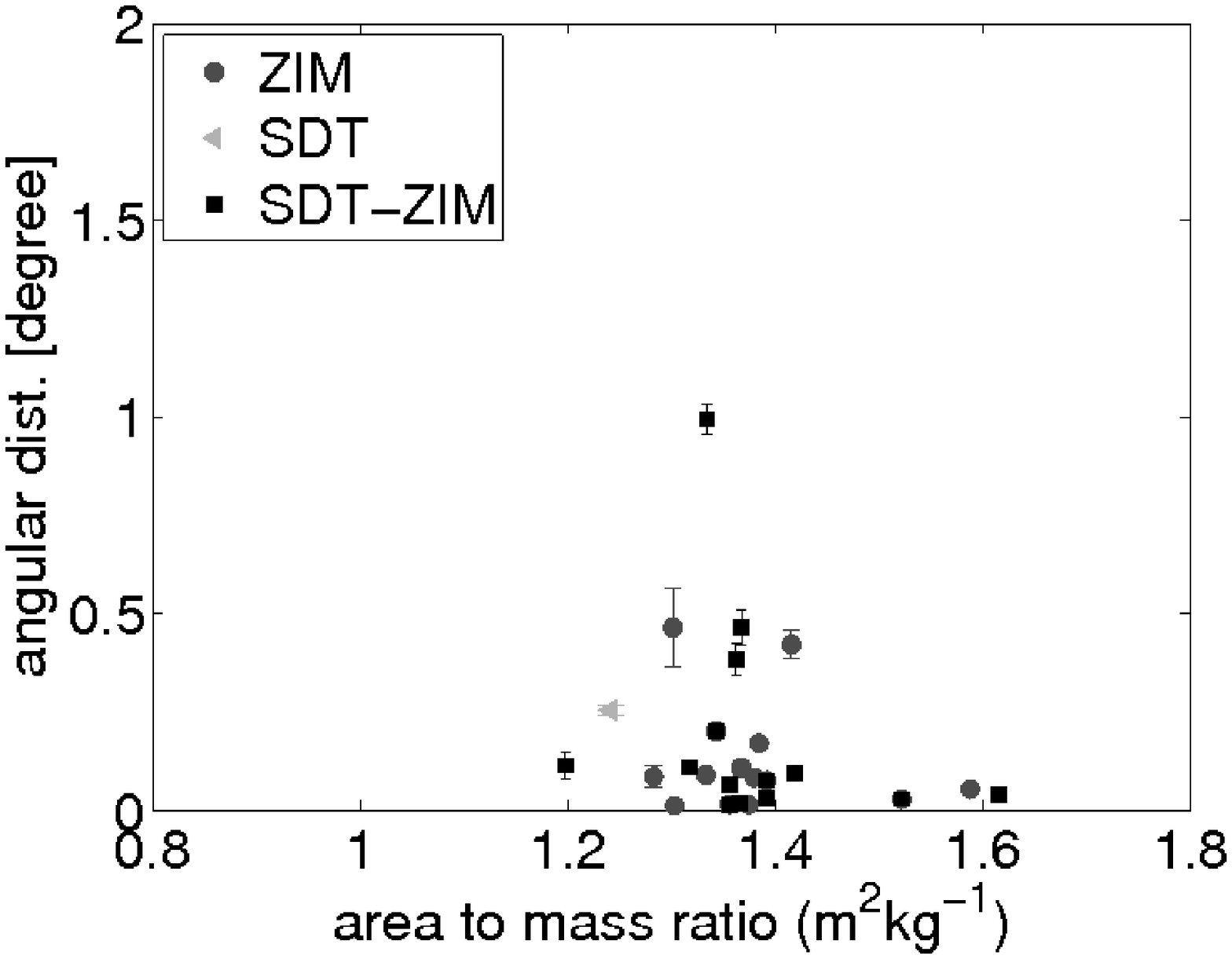}}  
  \subfloat[]{\includegraphics[width=0.25\textwidth, height=0.13\textheight]{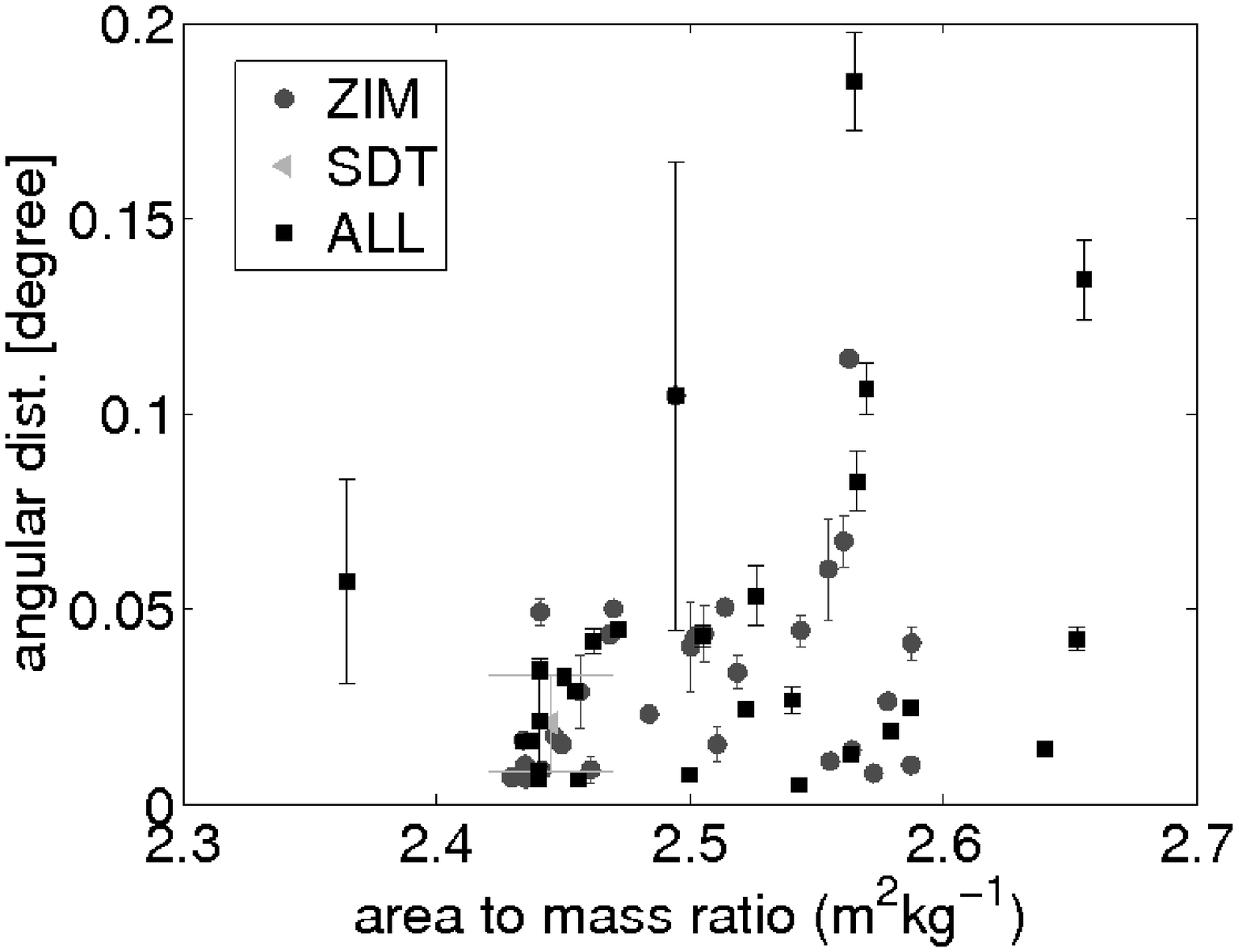}}
  \subfloat[]{\includegraphics[width=0.25\textwidth, height=0.13\textheight]{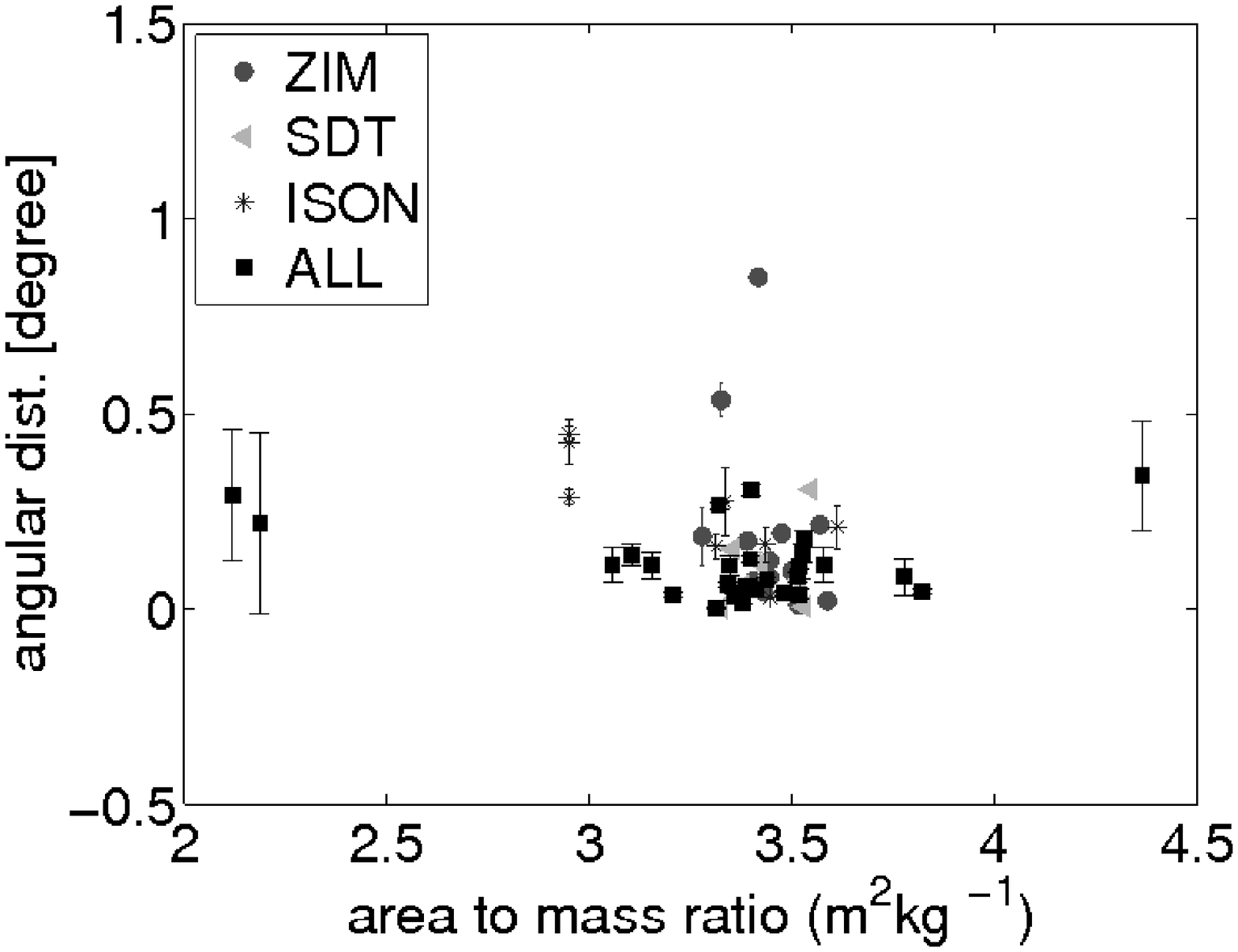}}\\
  \subfloat[]{\includegraphics[width=0.25\textwidth, height=0.13\textheight]{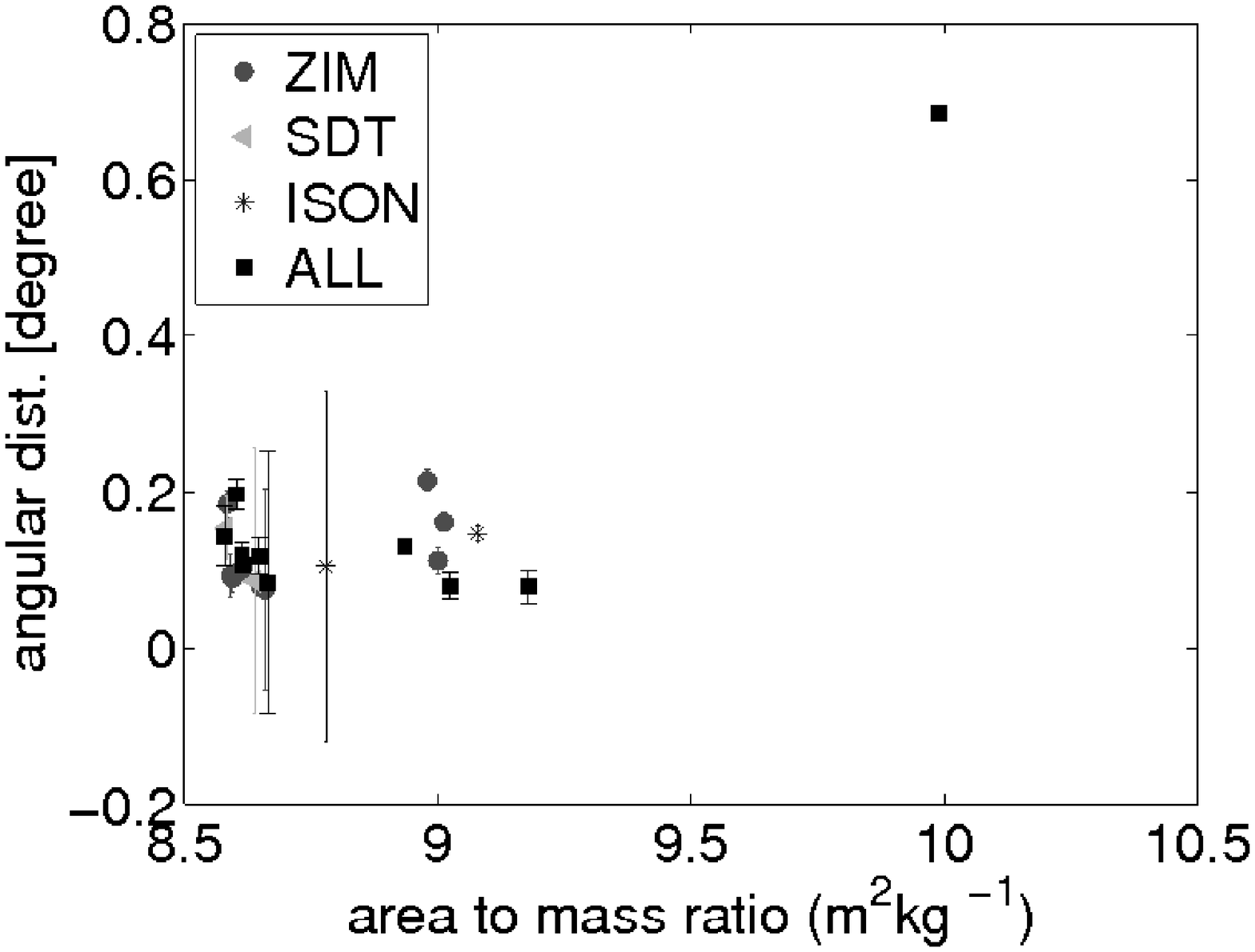}}
  \subfloat[]{\includegraphics[width=0.25\textwidth, height=0.13\textheight]{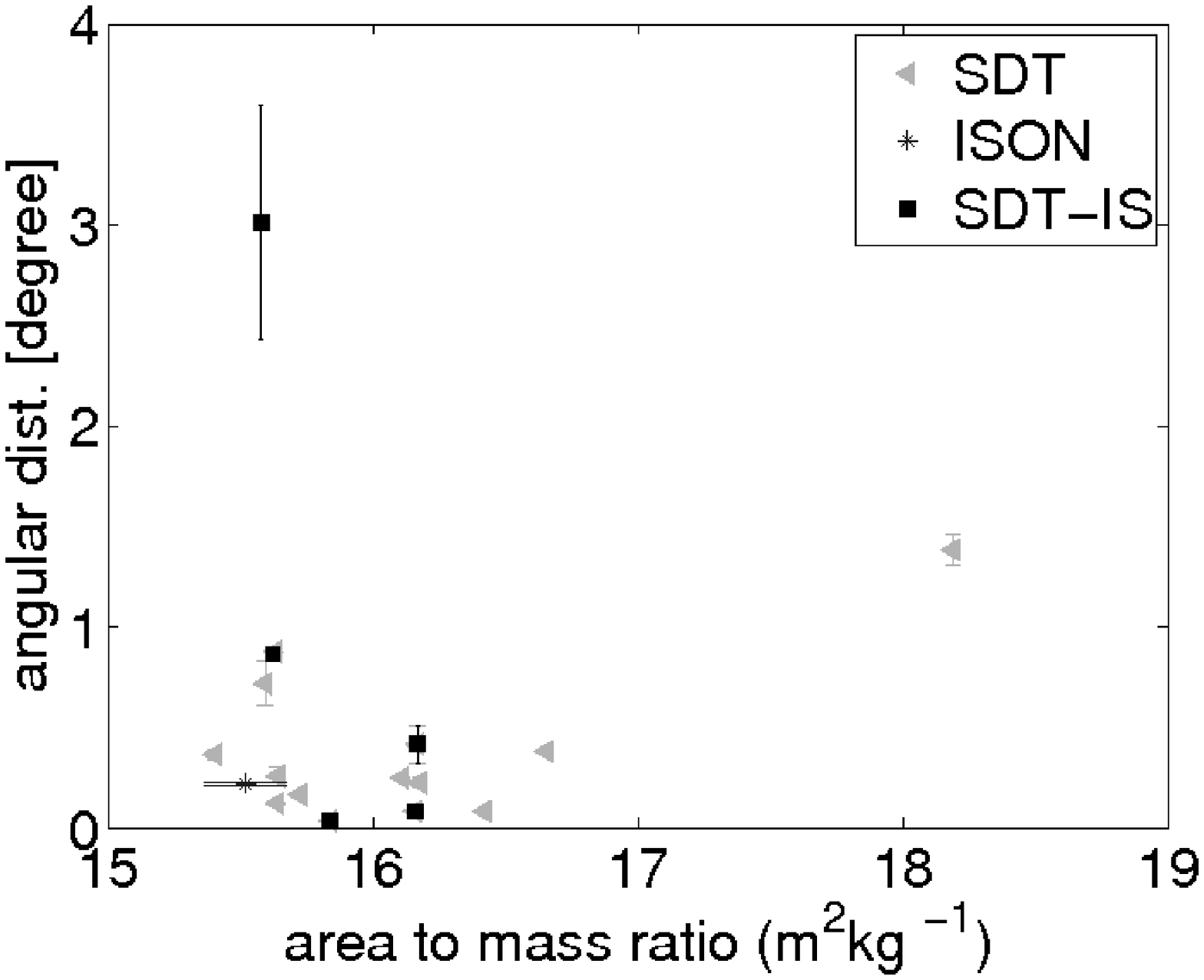}}
  \caption{Angular distances of predicted orbits on the celestial sphere as a function of AMR for orbits of the object (a) E08241A, (b) E06321D, (c) E07194A, (d) E07308B, (e) E06293A.}
  \label{distances}
\end{figure*} 
\begin{figure*}\textcolor{white}{[h]}
  \centering
  \subfloat[]{\includegraphics[width=0.25\textwidth, height=0.13\textheight]{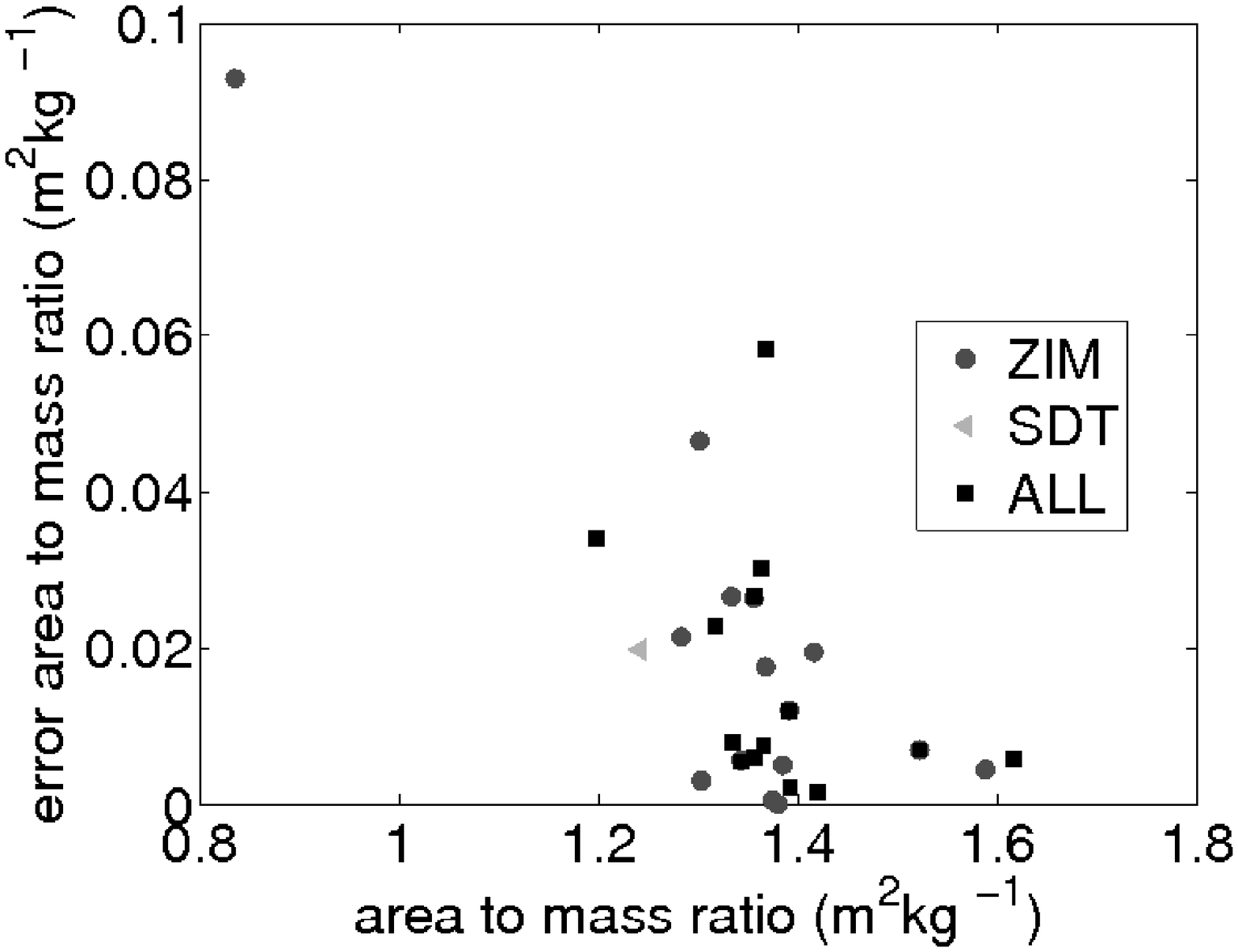}}  
  \subfloat[]{\includegraphics[width=0.25\textwidth, height=0.13\textheight]{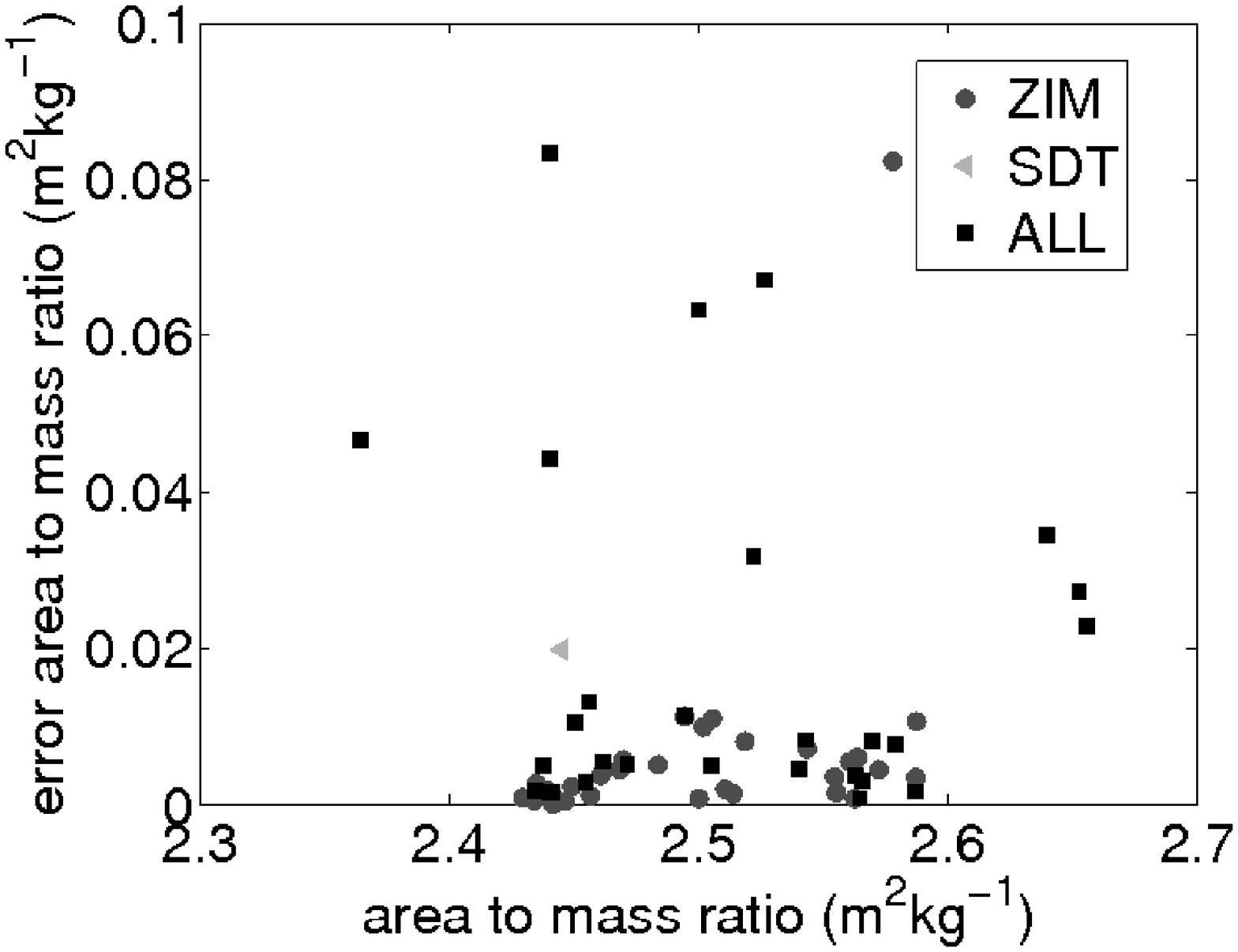}}
  \subfloat[]{\includegraphics[width=0.25\textwidth, height=0.13\textheight]{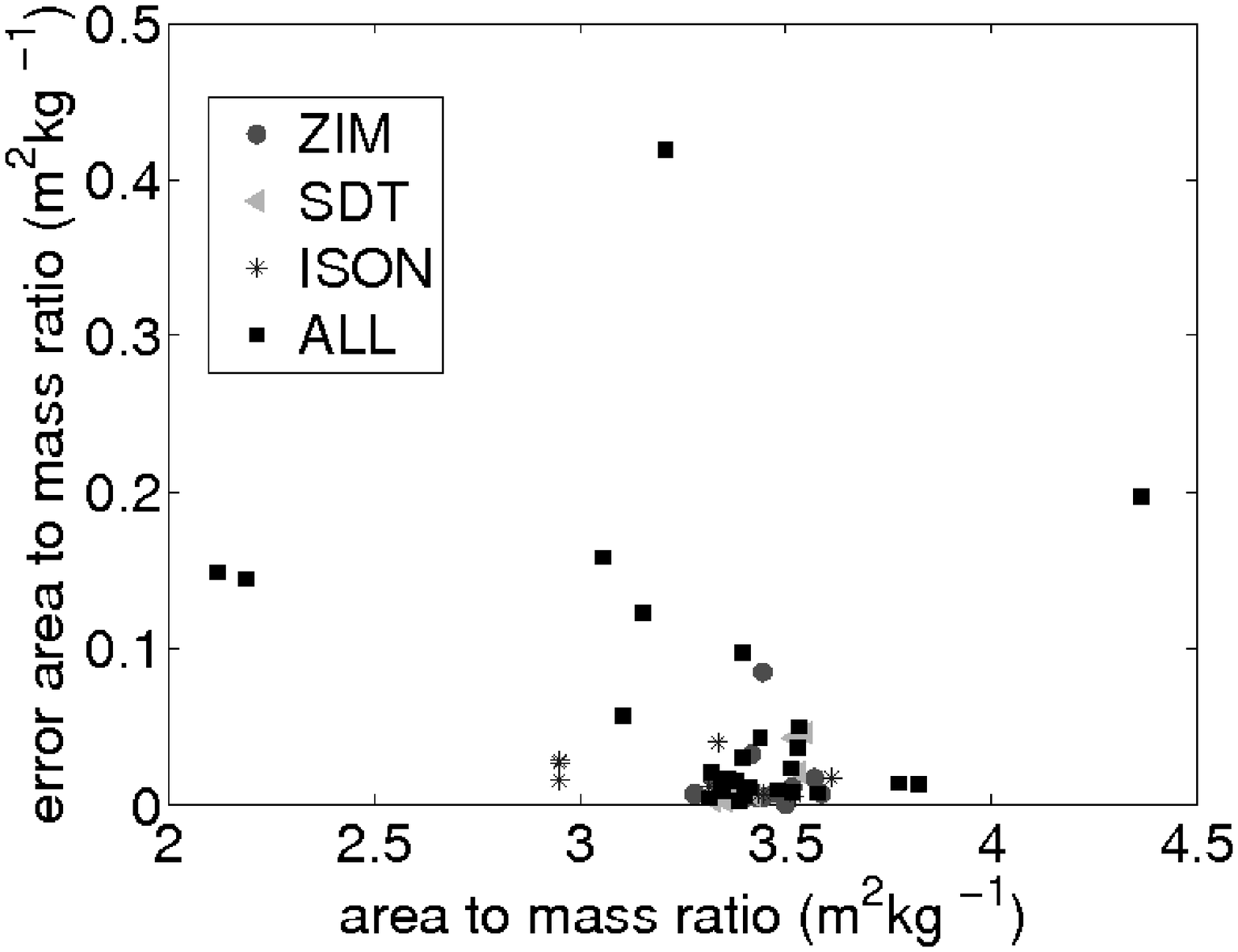}}\\
  \subfloat[]{\includegraphics[width=0.25\textwidth, height=0.13\textheight]{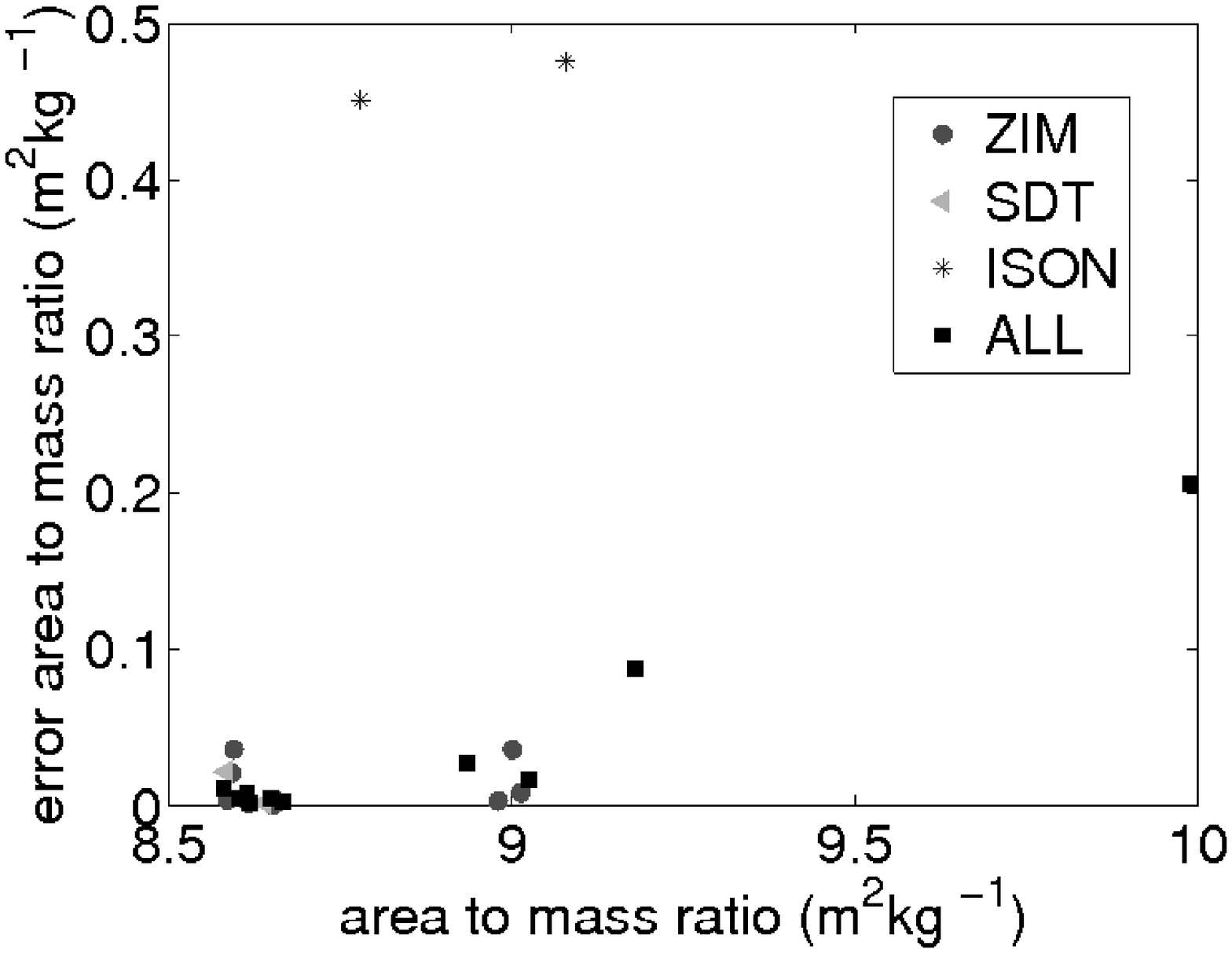}}
  \subfloat[]{\includegraphics[width=0.25\textwidth, height=0.13\textheight]{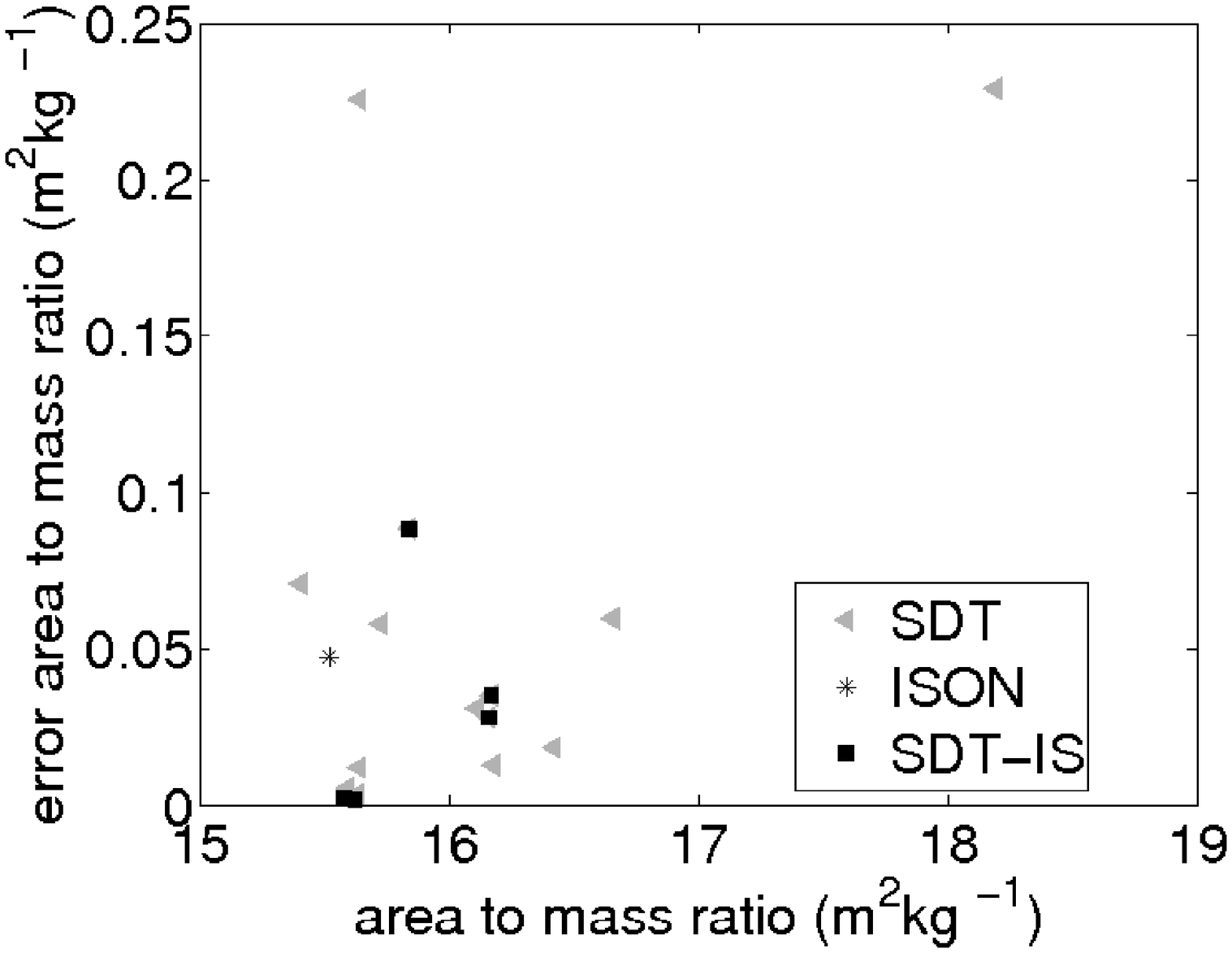}}
  \caption{Error of the AMR value as a function of AMR as estimated in orbits of the object (a) E08241A, (b) E06321D, (c) E07194A, (d) E07308B, (e) E06293A.}
  \label{DRP}
\end{figure*} 
\begin{figure*}\textcolor{white}{[h]}
  \centering
  \subfloat[]{\includegraphics[width=0.25\textwidth, height=0.13\textheight]{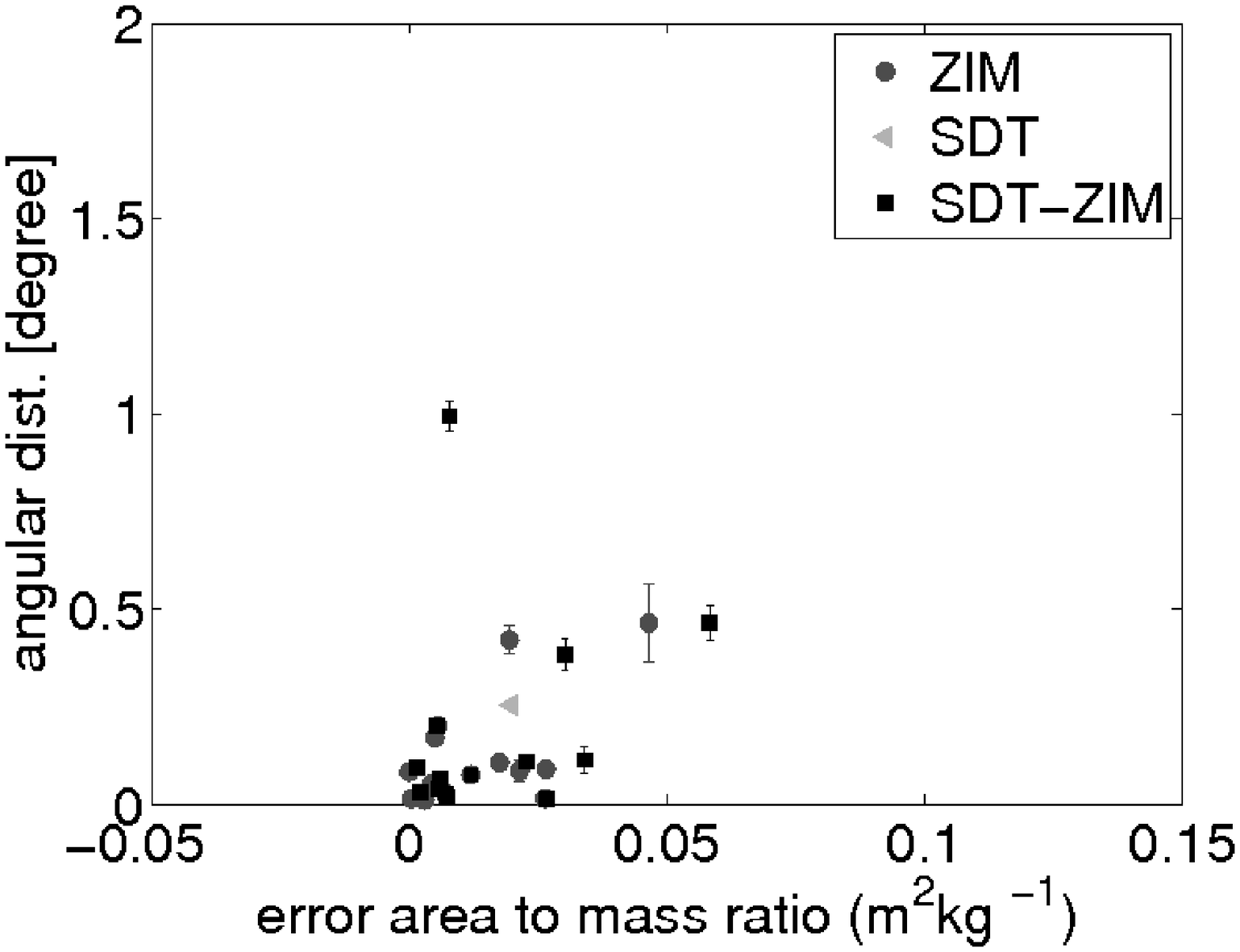}}  
  \subfloat[]{\includegraphics[width=0.25\textwidth, height=0.13\textheight]{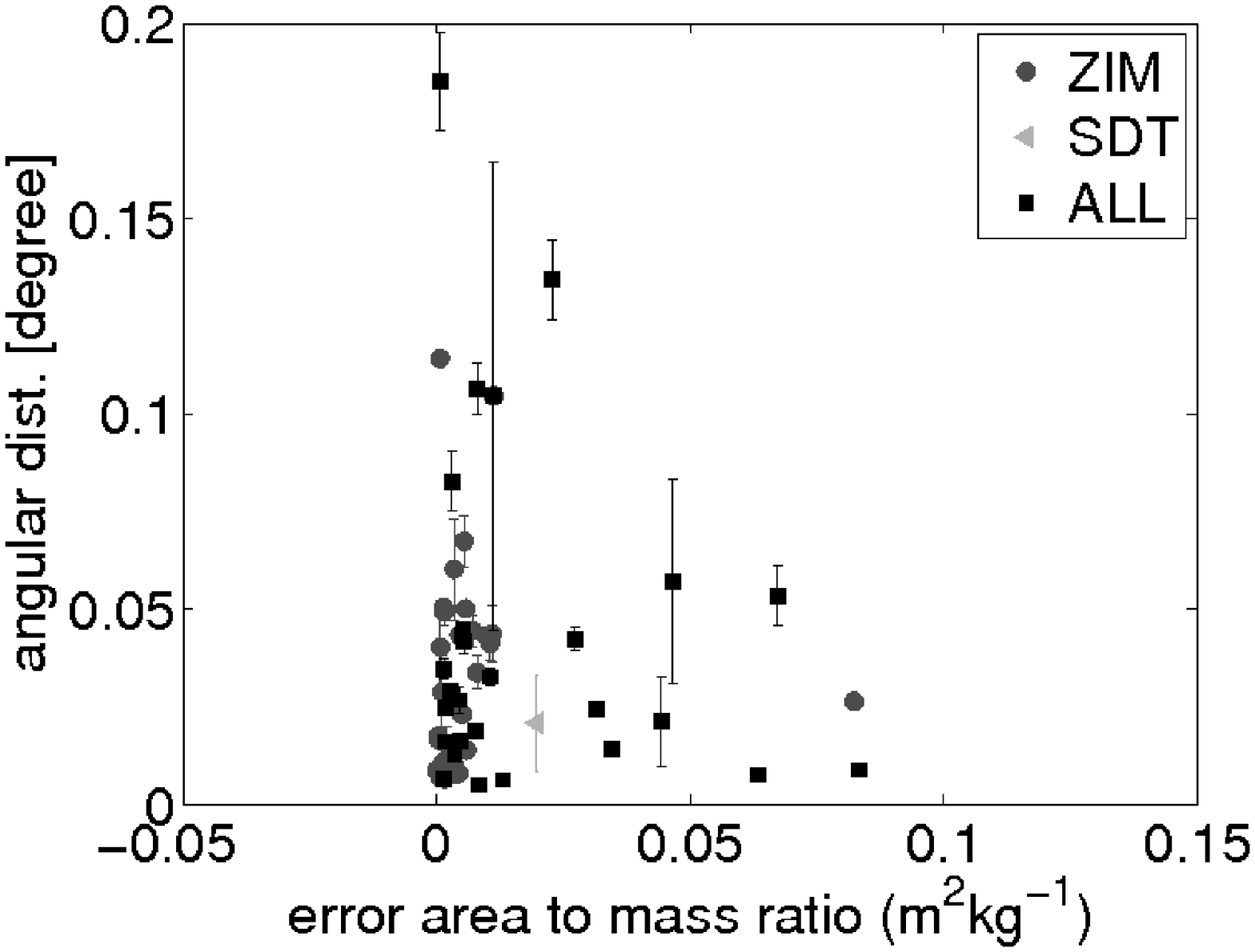}}
  \subfloat[]{\includegraphics[width=0.25\textwidth, height=0.13\textheight]{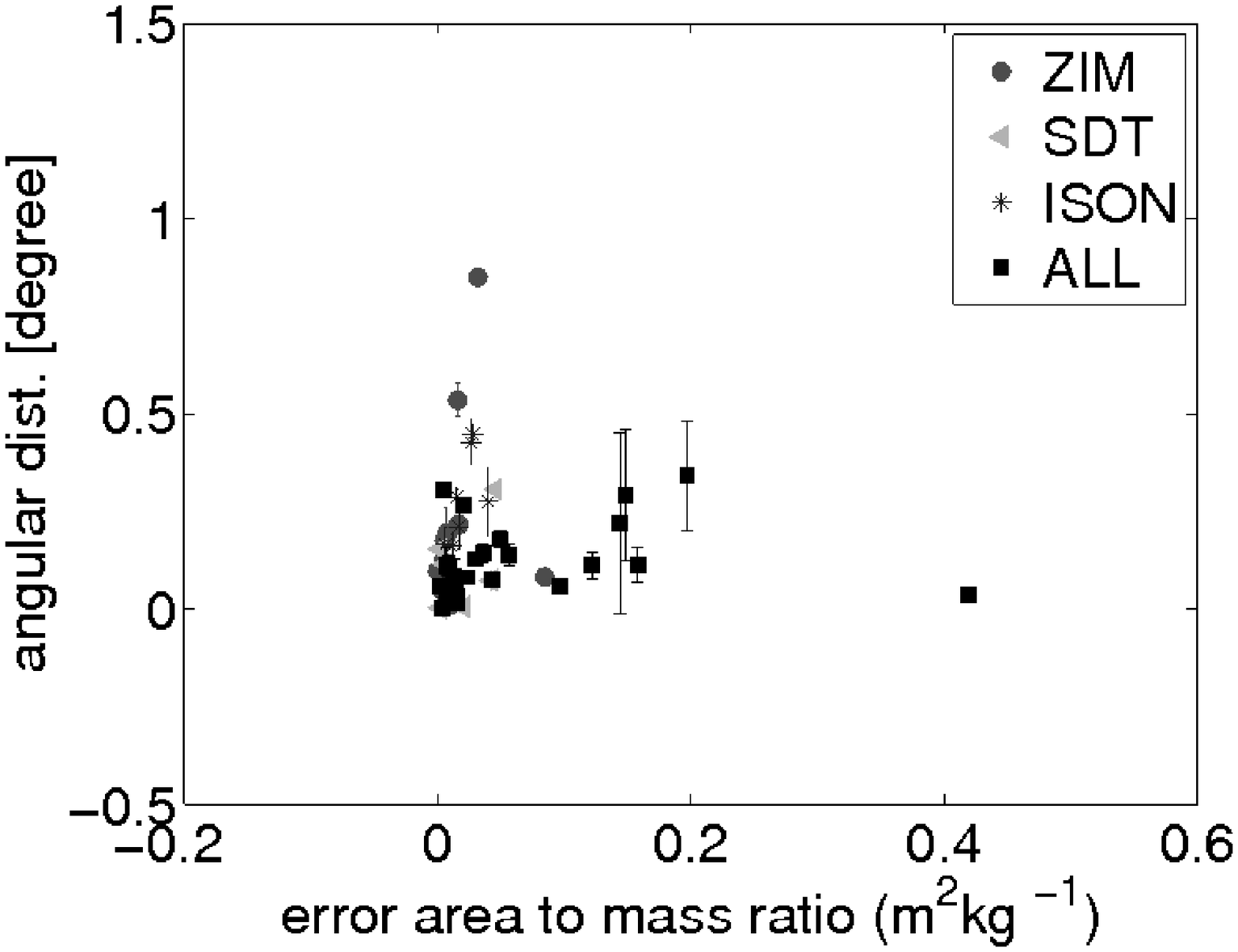}}\\
  \subfloat[]{\includegraphics[width=0.25\textwidth, height=0.13\textheight]{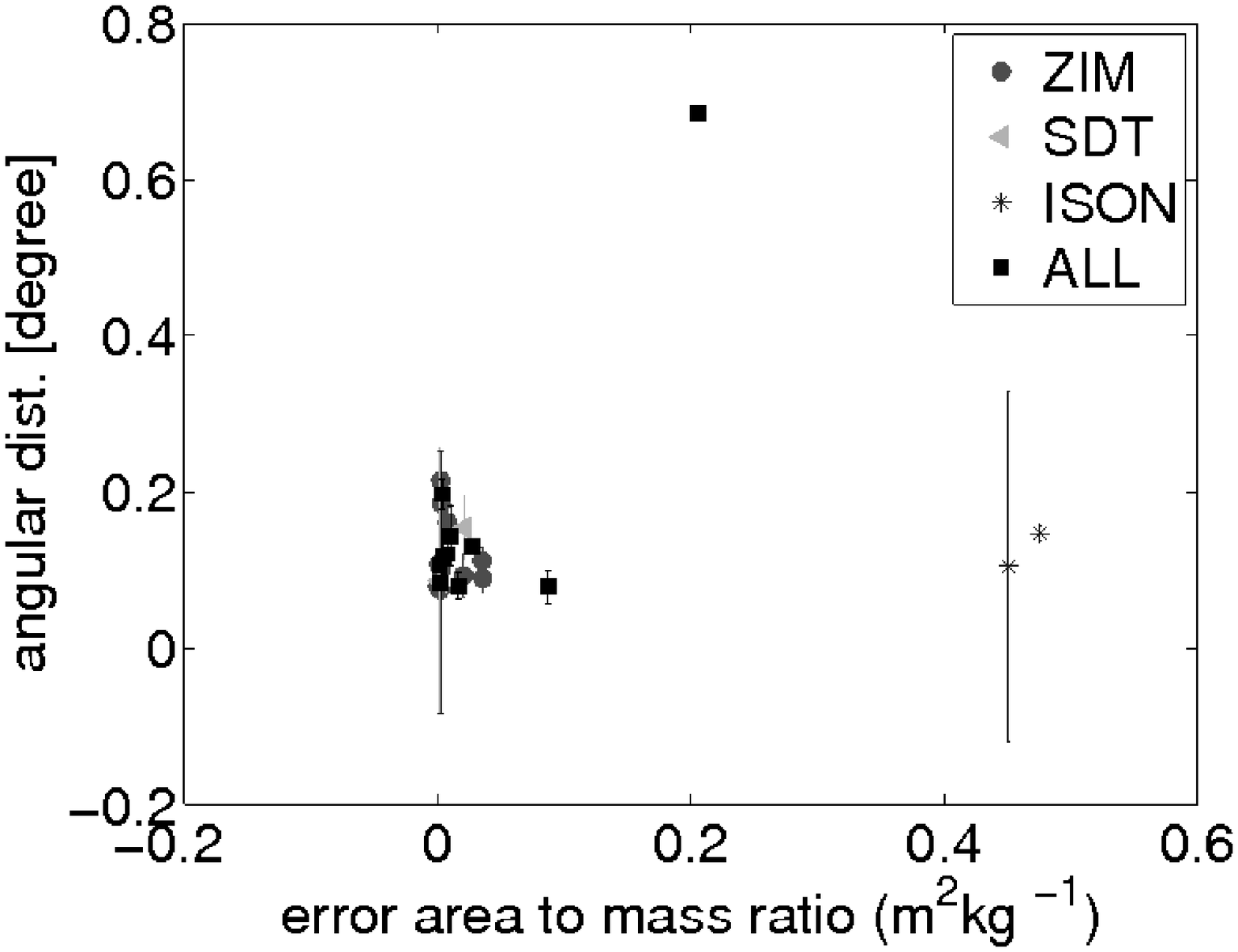}}
  \subfloat[]{\includegraphics[width=0.25\textwidth, height=0.13\textheight]{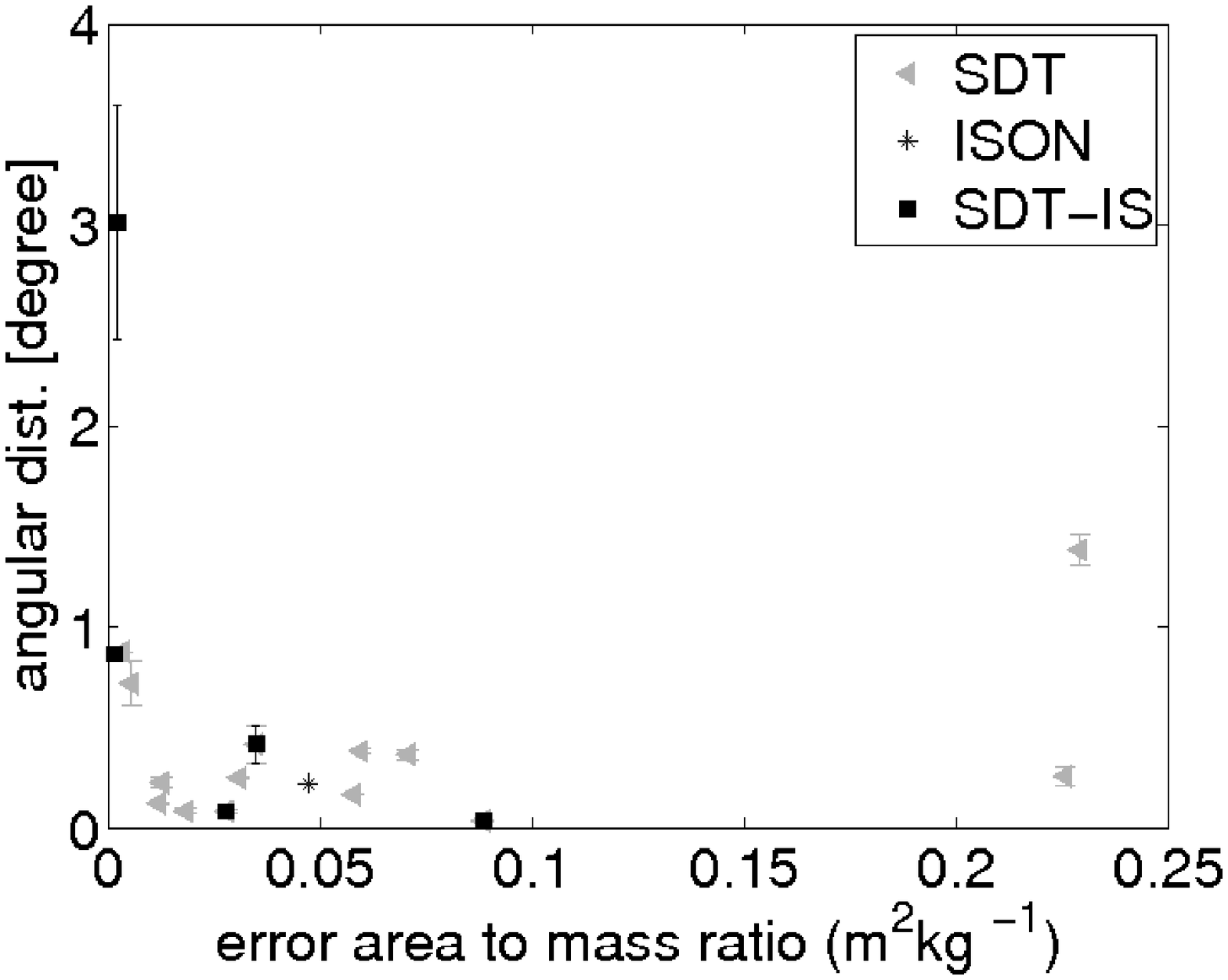}}
  \caption{Absolute values and standard deviations of the angular distances as a function of the error of the AMR value as found in orbit determination of the object (a) E08241A, (b) E06321D, (c) E07194A, (d) E07308B, (e) E06293A.}
  \label{DRP_arc}
\end{figure*}

\subsection{Evolution of Orbital Elements}
The evolution of the orbital elements over time is inspected in a first
step. Figure\,\ref{inc} shows the development of the inclination and its errors in inclination, of the five objects. The
error bars are too small, to be visible in the plot in most cases. The inclination values of the different orbits are closely aligned to each other and mark a consistent evolution, only in the case of
object E08241A in Fig.\,\ref{inc} a wider spread in the inclination values can be observed. The
orbits determined with observations from the different observation sites produce almost identical
results. For object E07308B and E06293A, which
have the highest AMR values, the inclination seems not to follow a steady
increase over time, but some smaller periodic substructure seems to be
superimposed. These may very well be the perturbations with a period of one
nodal year, which are well known for objects with high AMR, see e.g.,
\citet*{Liou2005}, \citet{Schildknecht05c}. \\
 \\
Figure \ref{ecc} shows the evolution of the eccentricity values and its errors estimated in orbit determination for the different objects. Periodic variations can be observed for all objects. The different orbits with observations from one site only or from different sites result in the same eccentricities.

\subsection{Evolution of Area to Mass Ratio Value}
Figure \ref{AMR} shows the AMR values as a function of time for the objects
listed in Tab.\,\ref{obj}. In all cases, the values for the AMR do not show
clear and obvious common trends, see Fig.\,\ref{inc} and \ref{ecc}. \\
\\
For object E08241A, the AMR values vary around a mean value of 1.4\,$\text{m}^2\text{kg}^{-1}$ with no obvious trend or periodic signal, see Fig.\,\ref{AMR}a.\\
\\
For object E06321D (see Fig.\,\ref{AMR}b), the AMR value seems to vary
periodically with a period of about one year around a value of
2.5\,$\text{m}^2\text{kg}^{-1}$, but also values of
2.35\,$\text{m}^2\text{kg}^{-1}$ and 2.65\,$\text{m}^2\text{kg}^{-1}$
occur. Similar results were obtained by \citet*{retoIAC08}, for the same
object, in different orbit determination setups. The AMR value of object
E07194A (see Fig.\,\ref{AMR}c) varies around 3.5\,$\text{m}^2\text{kg}^{-1}$,
but in the orbits determined with combined observations from all the sites,
so-called \textit{outliers} of 4.5\,$\text{m}^2\text{kg}^{-1}$ and
2.3\,$\text{m}^2\text{kg}^{-1}$ occur as well. These have, however large error
values.\\
\\
Object E07308B (see Fig.\,\ref{AMR}d) seems to generally increase its AMR value over time from a value of 8.5\,$\text{m}^2\text{kg}^{-1}$ up to 9.0\,$\text{m}^2\text{kg}^{-1}$. But single orbits also show AMR values of i.e. 10\,$\text{m}^2\text{kg}^{-1}$. \\
\\
Figure \ref{AMR}e shows that object E06293A, which is the object with the largest AMR value regarded here, has significant data gaps. A general trend of the AMR value in time, increasing from 15.5\,$\text{m}^2\text{kg}^{-1}$ to 16.5\,$\text{m}^2\text{kg}^{-1}$ cannot be excluded. But one orbit determined with ESASDT data also shows a value of 18.2\,$\text{m}^2\text{kg}^{-1}$, with a small formal error.\\
\\
No general correlation between the AMR value itself and the variations of the
AMR value could be
determined, no general trend is visible. A study on the
variation of AMR values was conducted by \citet{thomascospar10}. The variations of the AMR values of 47 HAMR objects
were investigated and compared to the AMR variations of orbits of 40 low AMR
(LAMR) value objects. No normalized or sparse data setup orbit determination setup was chosen. The AMR values in that analysis were determined in the standard orbit determination procedure at the AIUB, with fit arcs as long as possible for a successful, that is defined as leading to a small rms error, orbit determination. The results are illustrated in Fig.\,\ref{LAMR-HAMR}. No general trend in the AMR variations could be determined for either HAMR or LAMR objects. The relative variations of the AMR values of the LAMR objects were larger, than the AMR variations of the HAMR values. The AMR variations of the LAMR objects were of the order of several 100 percent. \\
\\
All orbits were predicted and compared to additional observations of the same object, which were not used for orbit determination. The additional observations were all checked via orbit determination, to ensure that they belong to the same object. Figure\,\ref{distances} shows the angular distances between the predicted ephemeris and observations. The values are averaged over all distances 50\,days after orbit determination and their standard deviations serve as error bars.\\
\\
Figure\,\ref{distances}a shows that for object E08241A, one orbit produces the largest distances of one degree. This orbit does not show up prominently in the orbital parameter plots (see Fig.\,\ref{ecc}a and \ref{inc}a) or AMR value plots (see Fig.\,\ref{AMR}a). The orbit with ZIMLAT data, which produced the \textit{outlier} AMR value of 0.82$\text{m}^2\text{kg}^{-1}$, does not show up prominently in the distance plot (Fig.\,\ref{distances}a). \\
\\
The mean value of all angular distances of object E06321D are well below
0.2\,degrees, but four orbits show large standard deviations in the angular
distance, as Fig.\,\ref{distances}b shows. All of them have been determined
with combined observations from ZIMLAT, ESASDT, and ISON observations. Their AMR values are 2.36\,$m^2kg^{-1}$, 2.50\,$\text{m}^2\text{kg}^{-1}$, 2.57\,$\text{m}^2\text{kg}^{-1}$,
and 2.66\,$\text{m}^2\text{kg}^{-1}$. The orbits with the AMR value of 2.36\,$\text{m}^2\text{kg}^{-1}$ does show up also in a group of \textit{outlier} AMR values, which do not seem to follow the periodic variation in the evolution of the AMR values. The other orbits, with large standard variations in the angular distance do not show up prominently (Fig.\,\ref{AMR}b). Those orbits with the largest standard variation in angular distance do not show the largest error in the AMR values either, as Fig.\,\ref{DRP_arc} shows.\\
\\
Figure\,\ref{distances}c shows for object E07194A three angular distances with
large standard deviations. The orbits were determined with observations from
all sites. They have AMR values of 2.12\,$\text{m}^2\text{kg}^{-1}$,
2.21\,$\text{m}^2\text{kg}^{-1}$, and 4.46\,$\text{m}^2\text{kg}^{-1}$. Those
are the smallest and largest AMR values in the determined orbits for
E07194A. These three values do also show up as outliers in
Fig.\,\ref{AMR}c. For objects E07308B and E06293A, the angular distances with a
large standard variation (see Fig.\,\ref{distances}d and e), do not show
significant outlier AMR values in Fig.\,\ref{AMR}d and e. For object E07308B,
the orbit with an AMR value of 10.15\,$\text{m}^2\text{kg}^{-1}$ shows the
largest mean value in the angular distance of almost 0.7 degrees but has a
small standard deviation in this distance (Fig.\,\ref{distances}d). This value
is significantly different compared to the other determined AMR values, see Fig.\,\ref{AMR}d.\\
\\
The dependency of the AMR value on the error of the AMR, as it was found in orbit determination, is investigated in the final step. No clear correlation could be determined between the AMR value and its rms value (Fig.\,\ref{DRP}).\\
\\
Figure\,\ref{DRP_arc} shows the angular distance distances on the celestial
sphere as a function of the error of the AMR value. As expected, for none of
the objects a clear correlation between the error of the AMR value and the
absolute value of the distances or the standard deviation of the distances
could be determined.\\
\\
All investigated objects show variations in the AMR value, but not a
common characteristic in these variations. It has to be noted that the result
may be affected by the relatively simple shadowing model that was used in
orbit determination; as it was shown \citet*{anselmo}, \citet{valk1}, shadowing
effects have a significant influence on the long term evolution of orbits of
HAMR objects. Investigation of simulated orbits with numerical and semi-analytical
methods, e.g. by \citet*{valk1}, \citet{valk2} also showed the existence of
irregular chaotic orbits and the significant influence of secondary resonances
on the orbits of HAMR objects. However, these simulations did assume a
constant AMR value. Complex attitude motion, irregular shapes, and/or
deformation of the actual objects, could lead to an actual change in the AMR
value itself over time, which may not be averaged out over the fit interval of orbit determination.

\section{Conclusions}
A sparse data setup was established to create comparable orbits over longer
time intervals. Orbits with two data sets only do produce small differences
between the propagated ephemerides and further observations, as long as 1.2
hours are covered within the sets. Other factors, such as that the
observations stem from different sites or the time interval between the sets,
are found to be negligible.\\
\\
The orbits of high area-to-mass ratio (HAMR) objects were analyzed in this setup. The
AMR value, that is the scaling factor of the direct radiation pressure (DRP) parameter, varies over
time. The order of magnitude of the variation of the area-to-mass ratio (AMR) value was not correlated with the order of magnitude of its error. \\
\\
The variation of the AMR is not averaged out in the fit interval of orbit
determination. In the evolution of the AMR value over time, no common
characteristic could be determined for different HAMR objects. Further work on
the orbits of HAMR objects is needed, to improve the radiation pressure model,
to determine possible attitude motion or deformations and to understand also
resonance effects and the existence of chaotic regions. 

\section*{Acknowledgments}

Special thank goes to ISON and the Keldysh Institute of Applied Mathematics, Moscow, for the supplementary observations.\\
The work was supported by the Swiss National Science Foundation through grants 200020-109527 and 200020-122070.\\
The observations from the ESASDT were acquired under ESA/ESOC contracts
15836/01/D/HK and 17835/03/D/HK.
\\
The authors thank the reviewer for useful hints. 
 \bibliographystyle{authordate1}
\bibliography{frueh_crop}

\label{lastpage}

\end{document}